\documentclass{article}
\usepackage[dvips]{graphicx}
\usepackage{epsfig}

\begin{document}
\title{Analytical Approach to the One-Dimensional Disordered Exclusion Process with Open Boundaries and Random Sequential Dynamics}
\author{M. Loulidi \footnote{regular associate of ICTP}\\
Laboratoire de Magn\'etisme et de Physique des Hautes Energies,\\
D\'epartement de Physique. Facult\'e des Sciences, B. P. 1014,\\
Rabat, Morocco\\}
\date{}
\maketitle
\begin{flushleft}
\large\bf{Abstract}
\end{flushleft}
\hspace*{1.1cm}A one dimensional disordered particle hopping rate asymmetric exclusion process (ASEP) with open boundaries and a random sequential dynamics is studied analytically. Combining the exact results of the steady states in the pure case with a perturbative mean field-like approach the broken particle-hole symmetry is highlighted and the phase diagram is studied in the parameter space $(\alpha,\beta)$, where $\alpha$ and $\beta$ represent respectively the injection rate and the extraction rate of particles. The model displays, as in the pure case, high-density, low-density and maximum-current phases. All critical lines are determined analytically showing that the high-density low-density first order phase transition occurs at $\alpha \neq \beta$. We show that the maximum-current phase extends its stability region as the disorder is increased and the usual $1/\sqrt{\ell }$-decay of the density profile in this phase is universal. Assuming that some exact results for the disordered model on a ring hold for a system with open boundaries, we derive some analytical results for platoon phase transition within the low-density phase and we give an analytical expression of its corresponding critical injection rate $\alpha^*$. As it was observed numerically$^{(19)}$, we show that the quenched disorder induces a cusp in the current-density relation at maximum flow in a certain region of parameter space and determine the analytical expression of its slope. The results of numerical simulations we develop agree with the analytical ones.\\
\vspace*{0.1cm}
\noindent \textbf{Key Words}:Disordered asymmetric exclusion model; steady state; boundary induced phase transitions; platoon phase.\\

\newpage
\section{Introduction}
\hspace*{0.5cm}It is well known that a set of equivalent problems including driven diffusion[1], 1d turbulence[2], growth of interfaces[3] and directed polymers in a random medium in 1+1 dimension[4] is considered as the same problem described by the noisy Burgers equation[2]. All those systems exhibit a none equilibrium stationary states and show a wealth of interesting phenomena that don't occur in thermal equilibrium. The standard asymmetric exclusion process(ASEP)[1,5] is relevant to a variety of phenomena in physics[1]. It is the one which describes a driven lattice gaz with hard core exclusion. Under suitable scaling the evolution of its macroscopic density is governed by a noisy Burgers equation, which is equivalent to the KPZ equation[6], in 1d. The ASEP with open boundary conditions is related to growth models with a defect or inhomogeneity[7] that introduces various types of phase transitions such as boundary induced phase transitions[8]. On the other hand it has a natural interpretation as a simplistic description of traffic flow on a single-lane high way and indeed forms the basis for more sophisticated traffic flow models[9].\\
\hspace*{0.5cm}The ASEP describes systems of interacting particles with hard core exclusion for double occupancy and with hopping rates differing for different directions. The bulk dynamics may be deterministic or stochastic. Open systems are coupled to reservoirs at both ends through stochastic boundaries. New methods for obtaining exact solutions for open 1d ASEP systems have been derived for random-sequential update[10] and were solved exactly for any values of injecting and extracting rates at the boundaries[11]. At the same time a very elegant solution using a matrix product ansatz for the weights of the stationary states[12] was given. Therefore one can calculate bulk properties, phase diagrams, density profiles of boundary layers and interfaces between coexisting bulk phases as well as correlation functions. A domain wall (shock) approach was given[13] to understand quantitatively the system dynamics, stationary states, the nature and the location of phase transitions. The collective velocity and the domain wall velocity are the crucial ingredients of this picture.\\ The different ways of updating sites are an essential part of the model. They affect the structure of phase diagram as well as the correlation function. The dynamics of updating may be applied in a random sequential order, parallel, i.e. fully synchronous for all sites or with sublattice parallel updating. For a detailed analytic results for ASEP we refer the reader to ref 14, 15 and references therein. The correlations are weakest for the random sequential updating, intermediate for ordered sequential and sublattice parallel updating and strongest for parallel updating. As a matter of fact, wealth of interesting exact results were obtained for random sequential and sublattice parallel dynamics, the fully parallel updating offers some difficulties.\\
As far as a great deal of results are known about the ASEP with a deterministic jumps rates, a little is known analytically about systems with quenched disorder jump rates[16]. The random sequential version of the disordered model on a ring has been studied[17]. It was shown that a transition occurs between inhomogeneous phase of low-density(jammed phase), where a traffic jam forms behind the slowest particle, and a high density congested phase (laminar phase), where all particles have to move more slowly than there preferred speed. It was shown that this transition persists in both dynamics parallel and ordered sequential updating and it is analogous to Bose condensation[18]. The disordered ASEP with open boundaries has been studied using numerical simulations[19]. A hopping parameter $\Delta t$ was introduced to interpolate between fully parallel ($\Delta t=1$) and random sequential ($\Delta t\rightarrow 0$) dynamics. It was found that the phase diagram is similar to that obtained in the pure case a part the shift of the first order transition line describing the coexistence of the low and high-density phases. The numerical results in the maximum current phase are consistent with a power law-decay $1/\sqrt{\ell }$ of the density profile as in the pure case.\\
In this paper we present an analytical study of a disordered ASEP with open boundary conditions. Particles jump in each time step to the right to vacant nearest neighbor sites with a probability $p_{\mu }$ associated with particle $\mu$ . The probability of injected (extracted) particles at the right (left) boundary is $\alpha (\beta )$ . The model is defined in sec. 2 and studied analytically after developing the equation of the steady state probabilities[10] of a system of length $N$ for a random sequential updating. Using a mean field-like approach we give an analytical solution for arbitrary $\alpha$ and $\beta$. The disorder doesn't affect dramatically the topology of the phase diagram. We show in sec. 3 that the phase diagram exhibits three different regions: the low-density and high-density phases, which split into two regions characterized by different behavior of the density profile, and the maximum-current phase. The density profile in the various phases and the critical lines are obtained analytically and studied in sec. 4. Thus, an analytical expression of the shift induced by the broken particle-hole symmetry of the first order critical line between low and high density phases is obtained. The platoon phase transition is studied in sec. 5 where we discuss in detail the phase diagram and derive analytical expressions for bulk densities in low and high density phases. According to a power law distribution of particle hopping probabilities $p_{\mu}$, we establish the analytical expression of the critical injection rate $\alpha*$ that corresponds to the platoon phase transition. We show that for the maximum current density, the current at this transition has gotten a negative slope whose analytical expression is calculated. To support our analytical results numerical simulations were performed in sec. 6. Our conclusion is given in sec 7.
\section{Model and Method}
\hspace*{0.5cm}We consider a one-dimensional lattice of $N$ sites. Each site $i$, $1\leq i\leq N$ , is either occupied by one particle or is empty. A configuration of the system is characterized by binary variables $\tau _{i}$, where $\tau _{i}=0$ ($\tau _{i}=1$) if site $i$ is empty (full). A quenched random rate $p_{\mu}$ is associated with each particle $\mu$. The dynamics of the system is governed by the following rule: at each time step $t\rightarrow t+1$ one chooses at random an integer $0\leq i \leq N$ with a probability $1/(N+1)$, then the particle $\mu$ on site $i$ hops to site $(i+1)$ if it is empty with a random probability $p_{\mu}$ such that:
\begin{equation}
\begin{array}{cc} 
\tau _{i}(t+1)=1 & \mbox{with probability}\hspace*{0.5cm}p_{\mu }\tau _{i}(t)\tau_{i+1}(t)+(1-p_{\mu })\tau _{i}(t)\\
\tau _{i+1}(t+1)=1 & \mbox{with probability}\hspace*{0.5cm}\tau _{i+1}(t)+p_{\mu }(1-\tau _{i+1}(t))\tau _{i}(t)
\end{array}	
\end{equation}
At the left boundary the site $1$ remains occupied at time $t+1$ if it was occupied at time $t$, and it gets occupied with probability $\alpha $ if it was empty at time $t$. Therefore:
\begin{equation}
\begin{array}{cc}
\tau _{1}(t+1)=1 & \mbox{with probability}\hspace*{0.5cm}\tau _{1}(t)+\alpha
(1-\tau _{1}(t))\\
\tau _{1}(t+1)=0 & \mbox{with probability}\hspace*{0.5cm}(1-\alpha )(1-\tau_{1}(t))  
\end{array}    	
\end{equation}
Similarly, the site $N$ remains empty at time $t+1$ if it was empty at time $t$, and it gets empty with probability $\beta $ if it was occupied at time $t$. So:
\begin{equation}
\begin{array}{cc}
\tau _{N}(t+1)=1 & \mbox{with probability}\hspace*{0.5cm}(1-\beta )\tau
_{N}(t)) \\
\tau _{N}(t+1)=0 & \hspace*{1.3cm}\mbox{with probability}\hspace*{0.5cm}1-(1-\beta )(1-\tau_{N}(t))
\end{array}        	
\end{equation}
The steady state defined from the stochastic dynamics (1)-(3) is given in terms of the probabilities $P_{N}(\{\tau _{i}\},\{p_{\mu }\})$ of finding the specific configuration of particles represented by the set of the occupation numbers $\{\tau _{i}\}$ and jumping rate probabilities $\{p_{\mu}\}$ on the chain with $N$ sites. Thus, following the dynamics rule described above it is easy to show that the probability $P_{N}(\{\tau_{i}\},\{p_{\mu }\})$ satisfies the relation:
\begin{eqnarray}
&&P_{N}(\{\tau _{i}\},\{p_{\mu }\})=\frac{1-\alpha }{N+1}P_{N}(\{\tau_{i}\},\{p_{\mu }\})+\frac{\alpha }{N+1}\tau _{1}[P_{N}(0,\{\tau _{i\neq
1}\},\{p_{\mu }\})+P_{N}(1,\{\tau _{i\neq 1}\},\{p_{\mu }\})] \nonumber \\
&&+\frac{1}{N+1}[P_{N}(\{\tau_{i}\},\{p_{\mu }\})+p_{\lambda}(\tau _{2}-\tau _{1})P_{N}(1,0,\{\tau _{i\neq 1,2}\},\{p_{\mu}\})] +...+
\frac{1}{N+1}[P_{N}(\{\tau _{i}\},\{p_{\mu }\}) \nonumber \\
&&+p_{\nu }(\tau _{N}-\tau _{N-1})P_{N}(\{\tau _{i\neq(N-1),N}\},1,0,\{p_{\mu }\})]+\frac{1-\beta }{N+1}P_{N}(\{\tau_{i}\},\{p_{\mu }\})  \nonumber \\
&&+\frac{\beta }{N+1}(1-\tau _{N})[P_{N}(\{\tau _{i\neq N}\},0,\{p_{\mu}\})+P_{N}(\{\tau _{i\neq N}\},1,\{p_{\mu }\})],  
\end{eqnarray}
which can be written as:
\begin{eqnarray}
&&\alpha (2\tau _{1}-1)P_{N}(0,\{\tau _{i\neq 1}\},\{p_{\mu }\})+(\tau_{2}-\tau _{1})p_{\lambda}P_{N}(1,0,\{\tau _{i\neq 1,2}\},\{p_{\mu }\})+...\\	\nonumber
&&+(\tau _{N}-\tau _{N-1})p_{\nu }P_{N}(\{\tau _{i\neq (N-1),N}\},1,0,\{p_{\mu}\})+\beta (1-2\tau _{N})P_{N}(\{\tau _{i\neq N}\},1,\{p_{\mu }\})=0	,
\end{eqnarray}
where $p_{\lambda}$ and $p_{\nu}$ are the hopping rate probabilities of particles $\lambda$ and $\nu$ located respectively at site $i=1$ and site $i=N-1$.\\
It turns out to be more convenient to write the recursion relation with unnormalized probabilities $f_{N}(\{\tau _{i}\},\{p_{\mu }\})$
related to $P_{N}(\{\tau _{i}\},\{p_{\mu }\})$ by:
\begin{equation}
P_{N}(\{\tau _{i}\},\{p_{\mu }\})=\frac{f_{N}(\{\tau _{i}\},\{p_{\mu }\})}{Z_{N}(\{p_{\mu }\})}  
\end{equation}
where 
\[
Z_{N}(\{p_{\mu }\})=\sum_{\{\tau _{i}=0,1\}}f_{N}(\{\tau_{i}\},\{p_{\mu }\})  
\]
In contrast to the pure case,i.e $p_{\mu}=p$, we are not able to find any recursion relation for $f_{N}(\{\tau _{i}\},\{p_{\mu }\})$ which allows to calculate exactly the average of any physical quantity as the occupation number, $<\tau_i>$. But, we will use a perturbative development based on a mean field-like approach to determine analytically such quantities. Indeed, if we consider the unnormalized probability $f_{N}(\{\tau _{i}\},p_m)$ for the pure case, where $p_m$ is the averaged value of $p_{\mu}$ according to a common distribution $\varphi(p)$ with a support on the the interval $[c,1] $, we may write:
\[
P_N(\{\tau_i\},\{p_{\mu}\})=\frac{e^{H_{N}^{0}}e^{V_N}}{Z_{N}(\{p_{\mu }\})},   
\]
where $H_N=ln(f_N(\{\tau _{i}\},\{p_{\mu }\}))$, $H_{N}^{0}=ln(f_{N}(\{\tau _{i}\},p_m))$ and $V_N= H_N-H_{N}^{0}$. Using a perturbative development of the potential $V_N$ that we assume to be weak for jumping rate probabilities $p_{\mu }$, we can determine the average occupation number $<\tau_i>$ as follow:
\[
<\tau_i>=\frac{\sum_{\{\tau _{i}=0,1\}}\tau_i e^{H_{N}^{0}} e^{V_N}}{\sum_{\{\tau _{i}=0,1\}}e^{H_{N}^{0}} e^{V_N}}
\]
By introducing the partition function $Z_N(p_m)$ for the pure case, the above equation can be written as:
\[
<\tau_i>=\frac{<\tau_i e^{V_{N}}>^0}{<e^{V_{N}}>^0},
\]
where the symbol $<.>^0$ indicates that the average is made using the unnormalized probability $f_N(\{\tau _{i}\},p_m)$.\\
Finally, using a perturbative development of the $e^{V_N}$ we obtain
\[
<\tau_i>_N=(1-<V_N>^0)(<\tau_i>^0+<\tau_i V_N>^0).
\]
Knowing that for any real function $f(x_1,x_2,...,x_n)$ we may write, using the the differential operator $\partial_{x_i}$:
\[
f(\lambda_1, \lambda_2,..., \lambda_n)=e^{\sum_{i=1}^{n}\lambda_i\partial_{x_i}}f(x_1,x_2,...,x_n)/_{\{x_i\}=0},
\]
where $ \lambda_1, \lambda_2,..., \lambda_n$ are real values, the potential $V_N(\{\tau_i\},\{p_{\mu}\})=ln(\frac{f_N(\{\tau _{i}\},\{p_{\mu }\})}{f_N(\{\tau _{i}\},p_m)})$ is developed as follow:
\begin{eqnarray}
&&V_N(\{\tau_i\},\{p_{\mu}\})=e^{\sum_{i=1}^{N}\tau_i \partial{x_i}} V_N(\{x_i\},\{p_{\mu}\})/_{\{x_i=0\}}\\  \nonumber
&&\hspace*{2cm}=\prod_{i=1}^{N}e^{\tau_i \partial{x_i}} V_N(\{x_i\},\{p_{\mu}\})/_{\{x_i=0\}}	 
\end{eqnarray}
Since $\tau_i=0,1$ the identity $e^{\lambda \tau_i}=1+\tau_i(e^{\lambda}-1)$ may be introduced in eq. 7 to obtain:
\begin{eqnarray}
&&V_N(\{\tau_i\},\{p_{\mu}\})=\prod_{i=1}^{N}[1+\tau_i (e^{\partial_{x_i}}-1)]V_N(\{x_i\},\{p_{\mu}\})/_{\{x_i=0\}}\\ \nonumber
&& =[1+\sum_i\tau_i(e^{\partial_{x_i}}-1)+\sum_{i,j}\tau_i \tau_j (e^{\partial_{x_i}}-1)(e^{\partial_{x_j}}-1)+...]V_N(\{x_i\},\{p_{\mu}\})/_{\{x_i=0\}} 	
\end{eqnarray}
In order to evaluate the average $<V_N(\{\tau_i\},\{p_{\mu}\})>^0$ we use the decoupling approximation $<\tau_i\tau_j\tau_{k}...\tau_l>^0 \simeq <\tau_i>^0<\tau_j>^0<\tau_{k}>^0...<\tau_l>^0$ and neglect the higher order of the development, i.e $(<\tau_i>^0~)^m\ll 1$ for $m>1$. Then, if we restrict ourselves to the first order we get from eq.8
\[
<V_N(\{\tau_i\},\{p_{\mu}\})>^0=V_N(0,\{p_{\mu}\})+\sum_i<\tau_i(e^{\partial_{x_i}}-1)V_N(\{x_i\},\{p_{\mu}\})/_{\{x_i=0\}}>^0
\]
and then the average of the occupation number is given by:
\begin{equation}
<\tau_i>_N=<\tau_i>_{N}^{0}[1-V_N(0,\{p_{\mu}\})-\sum_i<\tau_i>_{N}^{0}(e^{\partial_{x_i}}-1)V_N(\{x_i\},\{p_{\mu}\})/_{\{x_i=0\}}]  
\end{equation}
\section{The density profile of the system}
\hspace*{0.5cm}In order to study the density profile of the system we compute from the recursion relation (6) the average occupation number
$\overline{<\tau_{i}>}_{N}$ such that:
\begin{equation}
\overline{<\tau _{i}>}_{N}=\overline{\left( \frac{T_{N,i}(\{p_{\mu}\})}{Z_{N}(\{p_{\mu}\})}
\right) }  
\end{equation}
where
\begin{equation}
T_{N,i}(\{p_{\mu}\})=\sum_{\{\tau _{i}\}=0,1}\tau _{i}f_{N}(\tau_{1},\tau _{2},...,\tau _{N},\{p_{\mu}\})  
\end{equation}
The symbol $<.>$ denotes the average over the configurations $ (\tau _{1},\tau _{2},...,\tau _{N})$ while the bar indicates the average on the quenched disorder. In order to analyze the density profile of the system, we will study its discrete derivative defined in ref 11. It is given for a set of values ${p_{\mu}}$ by:
\begin{equation}
t_{N}^{\ell}(\{p_{\mu}\})=\frac{T_{N,\ell+1}(\{p_{\mu}\})-T_{N,\ell}(\{p_{\mu}\})}{Z_{N}(\{p_{\mu}\})}  
\end{equation}
This quantity can be developed using eq. 10 and the perturbative development of eq. 9:
\[
t_{N}^{\ell}(\{p_{\mu}\})=t_{N}^{\ell}(p_m)[1-V_N(0,\{p_{\mu}\})+\sum_{i=1}^{N}<\tau_i>_{N}^{0}(e^{\partial_{x_i}}-1)V_N(\{x_i\},\{p_{\mu}\})/_{\{x_i=0\}}]
\]
where $t_{N}^{\ell}(p_m)$ is the derivative of the density profile in the pure case for the averaged value $\overline{p_{\mu}}=p_m$. From the expression of the potential $V_N(0,\{p_{\mu}\})$ we obtain
\[
1-V_N(0,\{p_{\mu}\})=\frac{f_N(0,p_m)}{f_N(0,\{p_{\mu}\})}
\]
and
\[
(e^{\partial_{x_i}}-1)V_N(\{x_i\},\{p_{\mu}\})=\frac{f_N(0,p_m)}{f_N(0,\{p_{\mu}\})}\frac{\partial}{\partial_{x_i}}
\left(\frac{f_N(\{x_i\},p_m)}{f_N(\{x_i\},\{p_{\mu}\})}\right)_{\{x_i=0\}}
\]
Finally we get the expression of $t_{N}^{\ell}(\{p_{\mu}\})$:
\begin{equation}
t_{N}^{\ell}(\{p_{\mu}\})=t_{N}^{\ell}(p_m)\frac{f_N(0,p_m)}{f_N(0,\{p_{\mu}\})}\left[1-\sum_{i=1}^{N}<\tau_i>_{N}^{0}\frac{\partial}{\partial_{x_i}}\left(\frac{f_N(\{x_i\},p_m)}{f_N(\{x_i\},\{p_{\mu}\})}\right)_{\{x_i=0\}}\right] 	
\end{equation}
\hspace*{0.5cm}In order to analyze the density profile of the system we will study the average value of its discrete derivative
$\overline{t_{N}^{l}(\{p_{\mu}\})}$ obtained in eq. 13. However, to give the explicit form of $t_{N}^{\ell}(\{p_{\mu}\})$ we have to determine the ones of $<\tau_i>_{N}^{0}$, $t_{N}^{\ell}(p_m)$
and the ratio $f_N(0,p_m)/f_N(0,\{p_{\mu}\})$ of unnormalized probabilities. The discrete derivative $t_{N}^{\ell}(p_m)$ is obtained from eq. 12 since we may calculate the exact expression of $Z_N(p_m)$ and $T_N^{\ell}(p_m)$ in the same manner as in ref 11. Thus, we obtain:
\[
t_{N}^{l}(p_m)=Z_{N}^{-1}(p_m)(1-\frac{\alpha }{p_m}-\frac{\beta }{p_m})(\beta p_m)^{l}G_{N-l,N-l}^{N-l}(\beta /p_m)(\alpha p_m)^{N-l}
G_{l,l}^{l}(\alpha /p_m).
\]
It can be written more conveniently by using a simple rescaling $\alpha\rightarrow\alpha/p_m$ and $\beta\rightarrow\beta/p_m$ as:
\begin{equation}
t_{N}^{\ell }(p_m)=\widetilde{Z}_{N}^{-1}(p_m)F_{\ell }(\alpha /p_m)F_{N-\ell}(\beta /p_m)  
\end{equation}
with
\begin{equation}
F_{N}(x)=x^{-N-1}G_{N,N}^{N}(x)  
\end{equation}
and
\begin{equation}
\widetilde{Z}_{N}(p_m)=\frac{Z_{N}(p_m)}{p_{m}^{2N}(1-\alpha /p_m-\beta /p_m)(\beta/p_m)^{N+1}(\alpha /p_m)^{N+1}}  
\end{equation}
$\widetilde{Z}_{N}(p_m)$ can be given explicitly from the function $F_{N}(x)$:
\begin{equation}
\widetilde{Z}_{N}(p_m)=\left\{\begin{array}{cc}
\frac{F_{N}(\beta /p_m)-F_{N}(\alpha /p_m)}{(\alpha /p_m)(1-\alpha /p_m)-(\beta/p_m)(1-\beta /p_m)}
& \mbox{for $\alpha \neq \beta ,(p_m-\beta )$} \\
-\frac{F^{^{\prime }}(\beta /p_m)}{1-2\beta /p_m} & \mbox{for $\alpha =\beta \neq p_m/2$}
\end{array}
\right.  		
\end{equation}
where $F^{\prime}(x)$ is the derivative with $x$. For the definition of the functions $G_{N,K}^M(x)$ we refer the reader to ref 11.\\ 
On the other hand, the recursion relation of $f_N(\{\tau_i\},p_m)$ solution of eq. 5 can be constructed by the same way as in ref 10. We obtain:
\begin{eqnarray}
&&f_{N}(\tau_{1},\tau_{2},...,\tau_{N},p_m)=\alpha p_m\tau_{N}f_{N-1}(\tau_{1},\tau _{2},...,\tau_{N},p_m)+\beta p_m(1-\tau_{N})...(1-\tau_{1})f_{N-1}(\tau_{1},\tau_{2},...,\tau_{N},p_m)\nonumber\\
&&+\alpha \beta \sum_{k=1}^{N-1}(1-\tau_{N})...
(1-\tau_{k+1})\tau _{k}[f_{N-1}(\tau_{1},...,\tau_{k-1},1,\tau_{k+1},...,\tau _{N},p_m)+f_{N-1}(\tau_{1},...,\tau_{k-1},0,\tau_{k+1},...,\tau_{N},p_m)]\nonumber,
\end{eqnarray}
which leads to the unnormalized probability $f_N(0,p_m)=\beta (\beta p_m)^{N-1}$. Although we are not able to give such recursion relation in the disordered case, we may derive from equation 4 the following recursion relations:
\[
f_N(0,\{p_{\mu}\})=\beta p_{\nu}f_{N-1}(0,\{p_{\mu}\}) \hspace*{1cm} \mbox{and} \hspace*{1cm} f_N(0,..,0,1,\{p_{\mu}\})=\alpha p_{\nu} f_{N-1}(0,\{p_{\mu}\}),
\]
where $p_{\nu}$ denotes the hopping probability of the particle $\nu$ located at site $N-1$. These recursion relations solve the steady state equation $\alpha f_N(0,\{p_{\mu}\})=\beta f_N(0,..,0,1,\{p_{\mu}\})$, obtained immediately from equation 4, and easily lead to $f_N(0,\{p_{\mu}\})=\beta \prod_{\mu=1}^{N-1} \beta p_{\mu}$.\\
Up to the first order of the development in eq. 13, the discrete derivative of the average occupation is written as:
\begin{equation}
\overline{t_{N}^{l}(\{p_{\mu}\})}\sim t_{N}^{l}(p_m)\prod_{\mu =1}^{N}\overline{\left (\frac{p_m}{p_{\mu}}\right )} 
\end{equation}
By averaging over the disordered jumping rate probabilities it becomes:
\begin{equation}
\overline{t_{N}^{l}(p\{_{\mu}\})}=\widetilde{Z}_{N}^{-1}(p_m)F_{\ell }^{d}(\alpha /p_m)F_{N-\ell}^{d}(\beta /p_m) 
\end{equation}
where
\[
F_{N}^{d}(x)=\left(p_m\int\limits_{c}^{1}\frac{\varphi(p)}{p}dp\right)^N F_N(x)   
\]
\section{Density profile in the hydrodynamic limit and phase diagram}
\hspace*{0.5cm}The behavior of the density profile is discussed in the hydrodynamic limit $N\rightarrow \infty $ at large distances from both ends, i.e., $\ell >>1$ and $r=(N-\ell )>>1$.
The behavior of the average quantity $\overline{t_{N}^{\ell }(\{p_{\mu}\})}$ allows us to localize the critical lines and then give the structure of the phase diagram. Using the asymptotic expression of $F_{N}(\sigma /p_m)$ for large $N$[11]:
\begin{equation}
F_{N}(\sigma /p_m)=\left\{\begin{array}{cc}
\frac{1-2\sigma /p_m}{[(\sigma /p_m)(1-\sigma /p_m)]^{N+1}}& \mbox{if $\sigma <p_m/2$} \\
\frac{4^{N}}{\sqrt{\pi }(1-2\sigma /p_m)^{2}N^{3/2}}& \mbox{if $\sigma >p_m/2$}
\end{array}
\right. 
\end{equation}
the shape of the density profile is computed from eq(14-19).\\
Since the method we use is a mean-field like approximation, it doesn't give the exact behavior near critical lines. But, it may be considered as a good approximation far from them as it is illustrated by the numerical results presented in what follows in sec.6(Fig 2). Thus, based on the approached density profile given in eq. 18 we will deduce the shape of the phase diagram from the divergence of scale lengths defined below. \\ 
\hspace*{0.5cm}The high-density phase HD$_1$ is located within the region $\beta <\alpha <p_m/2$. From the results obtained above the density profile decays exponentially such that:
\begin{equation}
\overline{t_{N}^{l}(\{p_{\mu}\})}=\left(\frac{p_m-2\alpha}{p_m}\right)\left(1-\frac{\beta(p_m-\beta)}
{\alpha(p_m-\alpha)}\right)e^{-\ell /\xi} 
\end{equation}
where the length scale $\xi$ is given by:
\begin{equation}
\xi^{-1}=ln\left[\frac{\beta(p_m-\beta)}{\alpha(p_m-\alpha)}f(c)\right]
\end{equation}
with $f(c)=p_m\int\limits_{c}^{1}\frac{\varphi(p)}{p}dp$.\\
We remark that the length scale $\xi^{-1}$, which is rather different than the one obtained in the pure case, has gotten an additional factor that is responsible for the particle-hole symmetry breaking. The low-density phase LD$_1$ is bounded by the region defined by $\alpha <p_m/2$ and $\beta <p_m/2$ with $\alpha <\beta $. As for HD$_1$, the derivative of the density profile $\overline{t_{N}^{\ell }(\{p_{\mu}\})}$ may be obtained from eq. 14-19 such that:
\begin{equation}
\overline{t_{N}^{l}(\{p_{\mu}\})}=\left(\frac{p_m-2\beta}{p_m}\right)\left(1-\frac{\alpha(p_m-\alpha)}{\beta(p_m-\beta)}\right)e^{-r/\xi}     
\end{equation}
where $r=N-\ell$ and the length scale $\xi^{-1}=ln\left[\frac{\alpha(p_m-\alpha)}{f(c)\beta(p_m-\beta)}\right]$. The current for high(low) density phase HD$_1$(LD$_1$) is given by $j=\beta \rho_{bulk}$($j=\alpha(1-\rho_{bulk}))$ where $\rho_{bulk}=\overline{<\tau_N>}$.\\
The coexistence line LD$_1$ $\leftrightarrow$ HD$_1$ is determined by the divergence of the length scale $\xi$ defined in eq. 22. It is given by:
\begin{equation}
\beta=\frac{p_m}{2}-\left(\frac{p_{m}^{2}}{4}-\frac{\alpha (p_m-\alpha)}{f(c)}\right)^{1/2}
\end{equation}
\hspace*{0.5cm}The high-density phase HD$_2$ is located within the region defined by $\alpha >p_m/2$ and $\beta <p_m/2$. The behavior of the derivative $\overline{t_{N}^{\ell }(p)}$ changes to the form:
\begin{equation}
\overline{t_{N}^{l}(\{p_{\mu}\})}=\left(\frac{(p_m-\alpha-\beta)(\alpha-\beta)}{\sqrt{\pi}(p_m-2\alpha)^2}\right)
\ell^{(-3/2)}e^{-\ell/\xi_{\beta}}     
\end{equation}
where the length scale $\xi_{\beta}$ is defined as:
\begin{equation}
\xi_{\beta}^{-1}=ln\left[4\frac{\beta(p_m-\beta)}{p_m^2}f(c)\right]       
\end{equation}
The current and the boundary values of the density are given by the same expressions as in the high- density phase HD$_1$: $\overline{<\tau _{N}>}=\rho _{bulk}$ and $\overline{<\tau _{1}>}=1-\frac{\beta }{\alpha }$ $\rho _{bulk}$. The profile of the low-density phase LD$_2$($\alpha <p_m/2$ , $\beta >p_m/2$) may be obtained using the same calculations as above. The density and the current are the same as in LD$_1$ phase but the density profile is given by:
\begin{equation}
\overline{t_{N}^{l}(\{p_{\mu}\})}=\frac{(p_m-\beta-\alpha)(\beta-\alpha)}{\sqrt{\pi}(p_m-2\beta)^2}r^{(-3/2)}e^{-r/\xi_{\alpha}} 
\end{equation}
where $\xi_{\alpha}^-1=ln\left[4f(c)\frac{\alpha(p_m-\alpha)}{p_m^2}\right]$.\\
\hspace*{0,5cm}The maximum-current phase is located in the region defined by $\alpha >p_m/2$ and $\beta > p_m/2$. The derivative $\overline{t_{N}^{\ell }(p)}$ depends neither on $\alpha$ nor on $\beta $. It takes the same expression as in the pure case independently of the disorder distribution:
\begin{equation}
\overline{t_{N}^{l}(\{p_{\mu}\})}=-\frac{1}{4\sqrt{\pi}}(1-\frac{\ell}{N})^{-3/2}\ell^{-3/2}			
\end{equation}
However, the density approaches its bulk value $\rho_{bulk}=1/2$ as $r^{-1/2}$ with the distance $r=\ell $ from the origin from above and $r=N-\ell $ from the boundary from below. The current depends on the average of the hopping rate parameter and it takes its maximal value $\overline{j(\{p_{\mu}\})}=p_m/4$. Therefore one obtains the boundary values $\overline{<\tau _{N}>}=\frac{p_m}{4\beta }$ and $\overline{<\tau _{1}>}=1-\frac{p_m}{4\alpha }$.\\	
The phase transition between HD$_2$ and the maximum-current phase MC, which is of second order, depends on the value of $c$ for a given distribution $\varphi (p_\mu)$. It occurs when the length scale $\xi _{\beta }$ defined in eq. 26 diverges and then is located at the critical value $\beta_c$:
\begin{equation}
\beta_c=\frac{p_m}{2}(1-\sqrt{1-1/f(c)})	
\end{equation}
The density profile and the current should be continuous at the critical line.\\
In order to determine the phase transition between LD$_1$ and LD$_2$ we consider, for low values of $\alpha$ ($\alpha<\alpha_c$), the transition between HD$_2$ and MC phases on the line $\beta = \beta_c$. In the MC phase the bulk density $\rho_{bulk}= 1/2$ and the way in which it is approached doesn't depend on $\beta$, whereas in the high density phase HD$_2$, $\beta$ determines completely $\rho_{bulk}$ and the way the profile decays to it. This is due to the fact that for $\beta > \beta_c$, the particles close to the boundary don't block each other and are quickly extracted. As a result, any perturbation corresponding to a change in the extraction rate $\beta$ doesn't spread into the system. As this description of the effect of $\beta$ increasing beyond $\beta_c$ on the transition from the high density phase HD$_2$ to MC phase doesn't depend on the injection rate $\alpha$ at the origin, we conclude that the transition between low density LD$_1$ and LD$_2$ phases should be located at the same critical line $\beta = \beta_c$. Based on the divergence of the correlation length $\xi_{\alpha}$ (eq.27) one can find that the critical line LD$_2$ $\leftrightarrow$ MC is located at $\alpha=\alpha_c=\beta_c$. It is of second order. The behavior of the profile at the critical line between high density phases HD$_1$ and HD$_2$ may be determined using the same arguments as above based on the transition line LD$_2$ $\leftrightarrow$ MC. Consequently, the critical line between high density phases HD$_1$ and HD$_2$ should be at $\alpha=\alpha_c$ since the transition from the low density phase LD$_2$ to MC phase doesn't depend on the extraction rate $\beta$. Since all critical lines should meet at the critical point $(\alpha_c,\beta_c)$ the coexistence line of LD$_1$ $\leftrightarrow$ HD$_1$ described by the equation 24 should give the critical value $\beta_c$ for $\alpha=\alpha_c$. Unfortunately it is not the case. This discrepancy is due to the fact that our perturbative mean field-like approach doesn't give the exact value of the correlation length $\xi$ as high correlations are neglected. As a consequence, the thermal average and the quenched disorder are, up to the first order of the development, decoupled (eq. 18). We think that if we take into account of correlations we will obtain the exact value of $\xi$ and then the discrepancy will be discarded. But, it is not obvious to develop any analytical result of the model using our approach without the decoupling approximation. In order to improve the expression of $\beta$ that gives the correct critical line we remark that if we multiply the second term of eq. 24 by $\frac{1}{\sqrt{1+1/f(c)}}$ the critical value $\beta=\beta_c$ is found for $\alpha=\alpha_c$. We note that by defining a new correlation length:
\[
(\xi^{'})^{-1}=ln\left[\frac{\beta^{'}(p_m^{'}-\beta^{'})}{\alpha^{'}(p_m^{'}-\alpha^{'})}f(c)\right]
\]
with the transformation $\beta^{'}=\beta-\frac{p_m}{2}(1-\frac{1}{\sqrt{1+1/f(c)}})$, $\alpha^{'}=\frac{\alpha}{\sqrt{1+1/f(c)}}$ and $p_m^{'}=\frac{p_m}{\sqrt{1+1/f(c)}}$, we recover that all lines meet at the critical point $(\alpha_c,\beta_c)$. Such correlation length may be thought as resulting from the coupling of thermal average and quenched disorder since the new parameters $\alpha^{'}$ and $\beta^{'}$ depend on the disorder through $f(c)$.\\
On the other hand, $\beta$ should vanish for $\alpha \rightarrow 0$ while in the limit of the pure case, i.e $f(c)\rightarrow 1$ we should obtain $\beta=\alpha$. Consequently, a perturbative development to higher order should lead to the following expression of the coexistence critical line LD$_1$ $\leftrightarrow$ HD$_1$:
 
\begin{equation}
\beta=\kappa(\alpha)=\frac{p_m}{2}-\frac{1}{\sqrt{1+1/f(c)}}\left(\frac{p_m^2}{4}-\frac{\alpha(p_m-\alpha)}{f(c)}\right)^{1/2}+\frac{p_m}{2}\left(\frac{\alpha}{\alpha_c}-1\right)\left(1-\frac{1}{\sqrt{1+1/f(c)}}\right)
\end{equation}

\section{Discussion of results}
\hspace*{0.5cm}In this section we will discuss in detail the phase diagram, derive analytical expressions of bulk densities, $\rho_{bulk}$, in high-density and low density phases and present some analytical results for platoon transition.
\subsection{Phase diagram and bulk densities}
\hspace*{0.5cm}The phase diagram (Fig.1) of the disordered ASEP presents the usual phases namely the high-density, low-density and maximum-current phases.
The high-density phase is split into two regions. The argument of the logarithm of the length scale $\xi$ in the first region ($\alpha < \alpha_c$, $\beta < \beta_c$) depends on the distribution of the particle jumping rate through $f(c)=p_m\overline{(1/p_{\mu})}$ that is responsible for the particle-hole symmetry breaking. When $\xi$ diverges, a phase transition between high and low density phases occurs at $\alpha \neq \beta$. The transition line which is of first order is given by eq. 48. Using the perturbative development of the bulk density $\rho_{bulk}=\overline{<\tau_N>}$ we obtain, up to a second order, for high density region:
\begin{equation}
\rho_{bulk}=(1-\frac{\beta}{p_m})g(\alpha,\beta)  	
\end{equation}
where
\[
g(\alpha,\beta)=\overline{\prod_{\mu =1}^{N}\left(\frac{p_m}{p_{\mu}}\right)}-\sum_{i=1}^{N}<\tau_i>^0_N\overline{\prod_{\mu =1}^{N}\left(\frac{p_m}{p_{\mu}}\right)\frac{\partial}{\partial x_i}
\left(\frac{f_N(x_i,p_{\mu})}{f_N(x_i,p_m)}\right )}_{\{x_i=0\}}
\]
Since we can not determine explicitly the expression of the function $f_N(\{x_i\},\{p_{\mu}\})$ we will give, based on some physical arguments, an approached form of the function $g(\alpha,\beta)$ after having shown that it doesn't depend on the injection rate $\alpha$.\\
It was shown[17] that as the probability distribution of the gap (the number of vacant sites) in front of the $\mu^{th}$ particle satisfies the stationary condition one provides $p_{\mu}\alpha_{\mu}=constant=v$, where $\alpha_{\mu}$ is the probability that the $\mu^{th}$ particle has a vacant site in front of it and $v$ is the average velocity of particles. Averaging over the quenched disorder we may obtain an approximated form of the velocity $v(\rho)$ such that $v(\rho)=p_m\overline{\alpha_{\mu}}=p_m(1-\rho)$. Thus, the expression of the current of our disordered model may be written as $j=p_m\rho(1-\rho)$. The local density $\rho(x,t)$ of the driven lattice gaz we study satisfies the continuity equation $\frac{\partial \rho(x,t)}{\partial t}-\frac{\partial j(x,t)}{\partial x}=0$. Consequently, the collective velocity may be given from the exact non equilibrium fluctuation dissipation theorem[13]:
\[
V^{coll}=\frac{\partial j(\rho)}{\partial \rho}.
\]

\begin{figure}[h]
\begin{center}
\includegraphics[width=8cm,height=6cm,angle=-90]{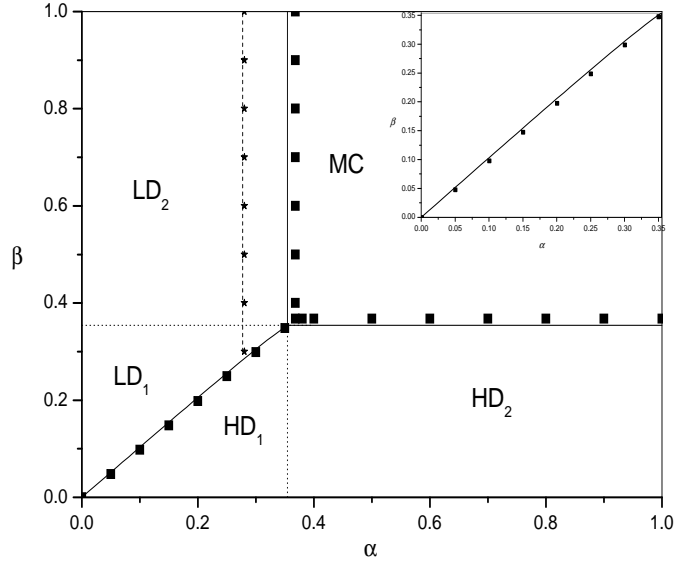}
\caption{The fundamental diagram of the ASEP. The continuous lines present the analytical result of the critical lines while the squares show the numerical ones for $L=2000$, $c=0.5$ and $n=1$. The dashed line(stars)localizes the platoon transition obtained analytically (numerically). The dotted lines delimit the different behaviors of low and high density regions. The inset represents an enlarging of the first critical line LD$_1$-HD$_1$.}
\end{center}
\end{figure}
One finds $V^{coll}=p_m(1-2\rho)$. It changes sign at $\rho=1/2$ where the current takes its maximal value $j=p_m/4$. In high density regime and for a fixed $\beta$, any small perturbation causes incoming particles to pile up behind the perturbation rather to spread into the bulk. As a result, the collective velocity of the center of mass of the perturbation is negative and thereafter the bulk density, $\rho_{bulk}$, is independent on the injection rate $\alpha$. Consequently, $g(\alpha,\beta)\equiv g(\beta)$. On the other hand, one should recover the limiting cases namely that for $c=1$ (pure case) $g(\beta)\rightarrow 1$ while in the limit $\beta \rightarrow 0$ the bulk density reaches its maximal value $\rho_{bulk}=1$. By increasing $\beta$ for a fixed value of $\alpha > \alpha_c$, $\rho_{bulk}$ should decrease going to $\rho_{bulk}=1/2$ at $\beta=\beta_c$ where a continuous phase transition to the MC phase occurs. Following these arguments the explicit expression of the bulk density in high density phase may be given by:
\[
\rho_{bulk}=\left(1-\frac{\beta}{p_m}\right)\left(1-\frac{\beta(p_m-2\beta_c)}{2\beta_c(p_m-\beta)}\right)  
\]
which may be simply written as:
\begin{equation}
\rho_{bulk}= 1-\frac{\beta}{2\beta_c}
\end{equation}
\hspace*{0.5cm}As for the pure case, the domain wall(shock)velocity is derived from the usual continuity equation:
\begin{equation}
v_{sh}=\frac{j_+-j_-}{\rho_+-\rho_-}	
\end{equation}
where $j_{\pm}$, $\rho_{\pm}$ are the bulk stationary state values of the current and the density in the left (-) and the right (+) parts of
the domain wall. Since $j_{\pm}=p_m\rho_{\pm}(1-\rho_{\pm})$ we obtain
\begin{equation}
v_{sh}=p_m(1-\rho_+-\rho_-)	
\end{equation}

The expression of the domain wall velocity $v_{sh}$ changes its functional form depending on the phase transition lines where it should vanish. The bulk density in the low-density phase may be obtained by assuming that the domain wall picture[13] remains still valid for the disordered case. On the one hand, it should depend only on the injection rate $\alpha$ since any perturbation of the stationary state, for a fixed value of $\alpha$, doesn't penetrate in the bulk. On the other hand, for the low-density/high-density domain wall $(0|1)$ $v_{sh}$ should vanish at the first order critical line LD$_1$ $\leftrightarrow$ HD$_1$. As $\rho_+$ is given by eq. 32, we obtain the bulk density in the low-density phase $\rho_{bulk}=\frac{\kappa(\alpha)}{2\alpha_c}$. For the maximum current/high density domain wall $(m|1)$ we have $v_{sh}=\frac{p_m}{2}(\frac{\beta}{\beta_c}-1)$,($\rho_-=1/2$). It vanishes exactly at $\beta=\beta_c$. For the low density/maximum current domain wall $(0|m)$ the domain wall velocity, $v_{sh}=\frac{p_m}{2}(\frac{\kappa(\alpha)}{\alpha_c}-1)$, vanishes at $\alpha=\alpha_c$, since $\kappa(\alpha_c)=\alpha_c$.\\
\hspace*{0.5cm}The shape of the phase diagram doesn't depend on the distribution of the particle jumping rate and MC phase gains more space by decreasing $c$ since $\alpha_c$ decreases. Indeed, it's easy to show that $1\le \mid f(c) \mid < p_m lnc^{-1}$, which implies that $f(c)>1$ and then $\alpha_c \rightarrow \frac{p_m}{4f(c)}$ for $c<<1$.\\
It's worthwhile to note that the critical point $(\alpha_c,\beta_c)$ where the three phases meet should be located in between c/2 and 1/2, which correspond respectively to the pure case $p_{\mu}=c$ and $p_{\mu}=1$. For the common distribution
\begin{equation}
\varphi(p)=\frac{n+1}{(1-c)^{n+1}}(p-c)^n 
\end{equation}
we have $\alpha_c=(1+n+c)\frac{(1-\sqrt{1-1/f(c)})}{n+2}$, which agrees with the value expected in ref 19. As it was argued numerically[19] the density in the MC phase decays algebraically and the deviation from its bulk value $\rho_{bulk}=1/2$ decays asymptotically as $\delta \rho_{\ell}=\mid<\tau_{\ell}>-\rho_{bulk}\mid=a\ell ^{-1/2}$. It is insensitive to the order of the maximum current. On the other hand, Hager et al[20] have conjunctured, using scaling arguments that this power law behavior holds for any one dimensional lattice gaz model. Based on the height difference correlation function of the one dimensional KPZ equation they have shown that the constant $a$, which was calculated for $n >1$, is proportional to the compressibility. This result is independent on the shape of the distribution $\varphi (p)$, the dynamics of updating[19,20], on the symmetry of the exclusion model[20] and neither on $\alpha $ nor $\beta$. It is a universal behavior.
\subsection{Platoon phase}
\hspace*{0.5cm}The exact current-density relation in the disordered ASEP can be obtained analytically in the hydrodynamic limit $N\rightarrow \infty $ from an implicit equation for the stationary state particle velocity $v(\rho)$[17,19]:
\begin{equation}
\rho =\left( 1+v\int\limits_{c}^{1}\frac{\varphi (p)}{p-v}dp\right) ^{-1}	
\end{equation}
This relation holds for densities $\rho ^{*}<\rho <1$ , where $\rho ^{*}$ is the critical density for the onset of platoon formation[17,19]. For $\rho <\rho ^{*}$ the overall speed will be set by the slowest particle and then $v(\rho )=c$ while for $\rho >\rho ^{*},$ $v$ is determined from eq. 36. Consequently, the value of $\rho ^{*}$ is found by setting $v=c$ in the rhs of this equation and the current-density relation becomes:
\begin{equation}
j(\rho )=\left\{ \begin{array}{cc}
\rho v(\rho )& \mbox{if $\rho >\rho ^{*}$} \\ 
\rho c & \mbox{if $\rho <\rho ^{*}$}
\end{array}
\right.  		
\end{equation}
Since in the low-density phase we have $\rho _{bulk}=\kappa(\alpha)/2\alpha_c$ the critical value $\alpha^{*}$ for the platoon transition (The dashed line in fig. 1) is given by:
\begin{equation}
\frac{2\alpha_c}{\kappa(\alpha^*)}=\int\limits_{c}^{1}\frac{p}{p-c}\varphi(p)dp
\end{equation}
Consequently, the critical line separating the inhomogeneous phase ($\rho <\rho ^{*}$) from the congested phase ($\rho >\rho ^{*}$) is independent of $ \beta $ as it was shown numerically[19]. It depends only on the choice of the jumping rate probability distribution $\varphi (p)$. According to the common distribution (35) we obtain
\[
\kappa(\alpha^*)=\frac {n(1-c)}{(n+c)}p_m(1-\sqrt{1-1/f(c)}),
\]
in agreement with the expression obtained in ref 19 since $\rho^*=\frac{\kappa(\alpha^*)}{2 \alpha^*}=\frac{n(1-c)}{(n+c)}$.\\
It was shown that the platoon phase transition is of second order, in the sense of ref 17, for $0<n\leq 1$. In this regime the velocity $v(\rho )$ is continuous at $\rho =\rho ^{*}$ while it is discontinuous for $n>1$. The current-density relation(eq. 37) leads to a shape with a quadratic maximum at $\rho _{\max }>\rho ^{*}$, as it was shown in ref 19. For $n>1$ the platoon transition becomes of first order and it is possible to give, within our approach, the value of $c$ under which $\rho _{\max }=\rho ^{*}$. As we have $\rho _{\max }=1/2$ we get for $n>1,$ $c=\frac{n}{1+2n}$.\\
From the expressions of the bulk density and the current given above, the derivative of $\overline{j(\{p_{\mu}\})}$ with $\rho $ is calculated at the platoon transition. We obtain for $\rho^*=\rho_{max}$,
\begin{equation}
\frac{\partial \overline{j}}{\partial \rho })_{\rho _{\max }=\rho
^{*}}=\left\{ \begin{array}{cc}
c & \mbox{if $\rho <\rho ^{*}=\rho _{\max }$} \\ 
c- \frac{1}{\vartheta(c)} & \mbox{if $\rho >\rho ^{*}=\rho _{\max }$}
\end{array}
\right.			
\end{equation}
where $\vartheta(c)=\rho_{max}\int\limits_{c}^{1}\frac{p}{(p-c)^2}\varphi(p)dp$.\\
According to the distribution 35 one obtain $\vartheta(c)=\frac{(n+1)(n-1+c)}{2n(n-1)(1-c)^2}$. It's easy to see that within the region for which $\rho^*=\rho_{max}$ one has $1/\vartheta(c)>c$. Thus, the current presents a negative local slope for $\rho > \rho^*=\rho_{max}$. Consequently, the disorder induces a cusp at the maximum of $j(\rho )$ at the first order platoon transition. This result was observed in a previous numerical study[19].

\section{Numerical simulations}
\hspace*{0.5cm}In order to support our analytical approach we have performed numerical simulations of the disordered ASEP on a lattice of sizes $L=2000-5\times10^4$ and random initial distribution of particles. Once the system reaches the stationary state we compute the average $<Q>$ of any physical quantity $Q(t)$ over $4\times 10^5-10^8$ time steps. We note that, in contrast to the case of ring, a separate disorder average is not necessary in the case of open boundaries since new particles are constantly injected into the system. In what follows we will just give some numerical results that we compare with the analytical ones, namely the phase diagram, the density profile in the high density and MC phases and the platoon transition.\\
\hspace*{0.5cm}The phase diagram(Fig. 1) is determined by computing the bulk density $\rho_{bulk}$ for different values of $\alpha$ and $\beta$. For a fixed value of $\alpha < \alpha_c$, where $\alpha_c$ is the critical value which separates the two high density regions HD$_1$ and HD$_2$, $\rho_{bulk}$ decreases when increasing $\beta$ until a critical value $\beta_{cr}(\alpha)$ where a first-order transition characterized by a discontinuity of $\rho_{bulk}$ brings the system into the low-density phase where the density becomes independent of $\beta$. The first-order transition high density-low density phase occurs at a critical line $\beta_{cr}(\alpha) \neq \alpha$. This is an effect of the disorder induced in the model which breaks the particle-hole symmetry. We note that the critical line obtained analytically is in good agreement with the numerical results. For $\alpha > \alpha_c$ the system exhibits, at a critical value $\beta_c$, a second-order phase transition to the MC phase where the bulk density and the current are independent of $\alpha$ and $\beta$. The numerical and analytical values of $\beta_c$ are in good agreement. For a fixed value of $\beta > \beta_c$ the bulk density increases when increasing $\alpha$ until the critical value $\alpha_c$, where we find the MC phase. In order to check the validity of the analytical expressions of the density in the high density (eq. 32) and low density phases we have shown in fig. 2 the variation of $\rho_{bulk}$ in high(low) density with the extraction(injection)rate $\beta$($\alpha$). We may consider that the analytical forms present a good fit of the numerical values especially for high values of $c$.

\setlength{\unitlength}{0.240900pt}
\ifx\plotpoint\undefined\newsavebox{\plotpoint}\fi
\sbox{\plotpoint}{\rule[-0.200pt]{0.400pt}{0.400pt}}%
\begin{picture}(1500,900)(0,0)
\sbox{\plotpoint}{\rule[-0.200pt]{0.400pt}{0.400pt}}%
\put(201.0,123.0){\rule[-0.200pt]{4.818pt}{0.400pt}}
\put(181,123){\makebox(0,0)[r]{ 0.6}}
\put(1419.0,123.0){\rule[-0.200pt]{4.818pt}{0.400pt}}
\put(201.0,205.0){\rule[-0.200pt]{4.818pt}{0.400pt}}
\put(181,205){\makebox(0,0)[r]{ 0.65}}
\put(1419.0,205.0){\rule[-0.200pt]{4.818pt}{0.400pt}}
\put(201.0,287.0){\rule[-0.200pt]{4.818pt}{0.400pt}}
\put(181,287){\makebox(0,0)[r]{ 0.7}}
\put(1419.0,287.0){\rule[-0.200pt]{4.818pt}{0.400pt}}
\put(201.0,368.0){\rule[-0.200pt]{4.818pt}{0.400pt}}
\put(181,368){\makebox(0,0)[r]{ 0.75}}
\put(1419.0,368.0){\rule[-0.200pt]{4.818pt}{0.400pt}}
\put(201.0,450.0){\rule[-0.200pt]{4.818pt}{0.400pt}}
\put(181,450){\makebox(0,0)[r]{ 0.8}}
\put(1419.0,450.0){\rule[-0.200pt]{4.818pt}{0.400pt}}
\put(201.0,532.0){\rule[-0.200pt]{4.818pt}{0.400pt}}
\put(181,532){\makebox(0,0)[r]{ 0.85}}
\put(1419.0,532.0){\rule[-0.200pt]{4.818pt}{0.400pt}}
\put(201.0,614.0){\rule[-0.200pt]{4.818pt}{0.400pt}}
\put(181,614){\makebox(0,0)[r]{ 0.9}}
\put(1419.0,614.0){\rule[-0.200pt]{4.818pt}{0.400pt}}
\put(201.0,695.0){\rule[-0.200pt]{4.818pt}{0.400pt}}
\put(181,695){\makebox(0,0)[r]{ 0.95}}
\put(1419.0,695.0){\rule[-0.200pt]{4.818pt}{0.400pt}}
\put(201.0,777.0){\rule[-0.200pt]{4.818pt}{0.400pt}}
\put(181,777){\makebox(0,0)[r]{ 1}}
\put(1419.0,777.0){\rule[-0.200pt]{4.818pt}{0.400pt}}
\put(201.0,123.0){\rule[-0.200pt]{0.400pt}{4.818pt}}
\put(201,82){\makebox(0,0){ 0}}
\put(201.0,757.0){\rule[-0.200pt]{0.400pt}{4.818pt}}
\put(378.0,123.0){\rule[-0.200pt]{0.400pt}{4.818pt}}
\put(378,82){\makebox(0,0){ 0.05}}
\put(378.0,757.0){\rule[-0.200pt]{0.400pt}{4.818pt}}
\put(555.0,123.0){\rule[-0.200pt]{0.400pt}{4.818pt}}
\put(555,82){\makebox(0,0){ 0.1}}
\put(555.0,757.0){\rule[-0.200pt]{0.400pt}{4.818pt}}
\put(732.0,123.0){\rule[-0.200pt]{0.400pt}{4.818pt}}
\put(732,82){\makebox(0,0){ 0.15}}
\put(732.0,757.0){\rule[-0.200pt]{0.400pt}{4.818pt}}
\put(908.0,123.0){\rule[-0.200pt]{0.400pt}{4.818pt}}
\put(908,82){\makebox(0,0){ 0.2}}
\put(908.0,757.0){\rule[-0.200pt]{0.400pt}{4.818pt}}
\put(1085.0,123.0){\rule[-0.200pt]{0.400pt}{4.818pt}}
\put(1085,82){\makebox(0,0){ 0.25}}
\put(1085.0,757.0){\rule[-0.200pt]{0.400pt}{4.818pt}}
\put(1262.0,123.0){\rule[-0.200pt]{0.400pt}{4.818pt}}
\put(1262,82){\makebox(0,0){ 0.3}}
\put(1262.0,757.0){\rule[-0.200pt]{0.400pt}{4.818pt}}
\put(1439.0,123.0){\rule[-0.200pt]{0.400pt}{4.818pt}}
\put(1439,82){\makebox(0,0){ 0.35}}
\put(1439.0,757.0){\rule[-0.200pt]{0.400pt}{4.818pt}}
\put(201.0,123.0){\rule[-0.200pt]{298.234pt}{0.400pt}}
\put(1439.0,123.0){\rule[-0.200pt]{0.400pt}{157.549pt}}
\put(201.0,777.0){\rule[-0.200pt]{298.234pt}{0.400pt}}
\put(201.0,123.0){\rule[-0.200pt]{0.400pt}{157.549pt}}
\put(40,450){\makebox(0,0){$\rho$}}
\put(820,21){\makebox(0,0){$\beta$}}
\put(820,839){\makebox(0,0){Fig 2a}}
\put(201,777){\raisebox{-.8pt}{\makebox(0,0){$+$}}}
\put(236,758){\raisebox{-.8pt}{\makebox(0,0){$+$}}}
\put(272,735){\raisebox{-.8pt}{\makebox(0,0){$+$}}}
\put(307,722){\raisebox{-.8pt}{\makebox(0,0){$+$}}}
\put(342,703){\raisebox{-.8pt}{\makebox(0,0){$+$}}}
\put(378,688){\raisebox{-.8pt}{\makebox(0,0){$+$}}}
\put(413,660){\raisebox{-.8pt}{\makebox(0,0){$+$}}}
\put(449,652){\raisebox{-.8pt}{\makebox(0,0){$+$}}}
\put(484,641){\raisebox{-.8pt}{\makebox(0,0){$+$}}}
\put(519,620){\raisebox{-.8pt}{\makebox(0,0){$+$}}}
\put(555,605){\raisebox{-.8pt}{\makebox(0,0){$+$}}}
\put(590,593){\raisebox{-.8pt}{\makebox(0,0){$+$}}}
\put(625,554){\raisebox{-.8pt}{\makebox(0,0){$+$}}}
\put(661,548){\raisebox{-.8pt}{\makebox(0,0){$+$}}}
\put(696,543){\raisebox{-.8pt}{\makebox(0,0){$+$}}}
\put(732,513){\raisebox{-.8pt}{\makebox(0,0){$+$}}}
\put(767,497){\raisebox{-.8pt}{\makebox(0,0){$+$}}}
\put(802,455){\raisebox{-.8pt}{\makebox(0,0){$+$}}}
\put(838,467){\raisebox{-.8pt}{\makebox(0,0){$+$}}}
\put(873,447){\raisebox{-.8pt}{\makebox(0,0){$+$}}}
\put(908,385){\raisebox{-.8pt}{\makebox(0,0){$+$}}}
\put(944,402){\raisebox{-.8pt}{\makebox(0,0){$+$}}}
\put(979,368){\raisebox{-.8pt}{\makebox(0,0){$+$}}}
\put(1015,394){\raisebox{-.8pt}{\makebox(0,0){$+$}}}
\put(1050,354){\raisebox{-.8pt}{\makebox(0,0){$+$}}}
\put(1085,367){\raisebox{-.8pt}{\makebox(0,0){$+$}}}
\put(1121,311){\raisebox{-.8pt}{\makebox(0,0){$+$}}}
\put(1156,278){\raisebox{-.8pt}{\makebox(0,0){$+$}}}
\put(1191,282){\raisebox{-.8pt}{\makebox(0,0){$+$}}}
\put(1227,272){\raisebox{-.8pt}{\makebox(0,0){$+$}}}
\put(1262,227){\raisebox{-.8pt}{\makebox(0,0){$+$}}}
\put(1298,227){\raisebox{-.8pt}{\makebox(0,0){$+$}}}
\put(1333,212){\raisebox{-.8pt}{\makebox(0,0){$+$}}}
\put(1368,175){\raisebox{-.8pt}{\makebox(0,0){$+$}}}
\put(1404,205){\raisebox{-.8pt}{\makebox(0,0){$+$}}}
\put(1439,177){\raisebox{-.8pt}{\makebox(0,0){$+$}}}
\put(201,777){\makebox(0,0){$\diamond$}}
\put(236,759){\makebox(0,0){$\diamond$}}
\put(272,740){\makebox(0,0){$\diamond$}}
\put(307,722){\makebox(0,0){$\diamond$}}
\put(342,703){\makebox(0,0){$\diamond$}}
\put(378,685){\makebox(0,0){$\diamond$}}
\put(413,666){\makebox(0,0){$\diamond$}}
\put(449,648){\makebox(0,0){$\diamond$}}
\put(484,629){\makebox(0,0){$\diamond$}}
\put(519,611){\makebox(0,0){$\diamond$}}
\put(555,592){\makebox(0,0){$\diamond$}}
\put(590,574){\makebox(0,0){$\diamond$}}
\put(625,555){\makebox(0,0){$\diamond$}}
\put(661,537){\makebox(0,0){$\diamond$}}
\put(696,518){\makebox(0,0){$\diamond$}}
\put(732,500){\makebox(0,0){$\diamond$}}
\put(767,482){\makebox(0,0){$\diamond$}}
\put(802,463){\makebox(0,0){$\diamond$}}
\put(838,445){\makebox(0,0){$\diamond$}}
\put(873,426){\makebox(0,0){$\diamond$}}
\put(908,408){\makebox(0,0){$\diamond$}}
\put(944,389){\makebox(0,0){$\diamond$}}
\put(979,371){\makebox(0,0){$\diamond$}}
\put(1015,352){\makebox(0,0){$\diamond$}}
\put(1050,334){\makebox(0,0){$\diamond$}}
\put(1085,315){\makebox(0,0){$\diamond$}}
\put(1121,297){\makebox(0,0){$\diamond$}}
\put(1156,278){\makebox(0,0){$\diamond$}}
\put(1191,260){\makebox(0,0){$\diamond$}}
\put(1227,241){\makebox(0,0){$\diamond$}}
\put(1262,223){\makebox(0,0){$\diamond$}}
\put(1298,205){\makebox(0,0){$\diamond$}}
\put(1333,186){\makebox(0,0){$\diamond$}}
\put(1368,168){\makebox(0,0){$\diamond$}}
\put(1404,149){\makebox(0,0){$\diamond$}}
\put(1439,131){\makebox(0,0){$\diamond$}}
\put(201.0,123.0){\rule[-0.200pt]{298.234pt}{0.400pt}}
\put(1439.0,123.0){\rule[-0.200pt]{0.400pt}{157.549pt}}
\put(201.0,777.0){\rule[-0.200pt]{298.234pt}{0.400pt}}
\put(201.0,123.0){\rule[-0.200pt]{0.400pt}{157.549pt}}
\end{picture}

\setlength{\unitlength}{0.240900pt}
\ifx\plotpoint\undefined\newsavebox{\plotpoint}\fi
\sbox{\plotpoint}{\rule[-0.200pt]{0.400pt}{0.400pt}}%
\begin{picture}(1500,900)(0,0)
\font\gnuplot=cmr10 at 10pt
\gnuplot
\sbox{\plotpoint}{\rule[-0.200pt]{0.400pt}{0.400pt}}%
\put(201.0,123.0){\rule[-0.200pt]{4.818pt}{0.400pt}}
\put(181,123){\makebox(0,0)[r]{ 0}}
\put(1419.0,123.0){\rule[-0.200pt]{4.818pt}{0.400pt}}
\put(201.0,205.0){\rule[-0.200pt]{4.818pt}{0.400pt}}
\put(181,205){\makebox(0,0)[r]{ 0.05}}
\put(1419.0,205.0){\rule[-0.200pt]{4.818pt}{0.400pt}}
\put(201.0,287.0){\rule[-0.200pt]{4.818pt}{0.400pt}}
\put(181,287){\makebox(0,0)[r]{ 0.1}}
\put(1419.0,287.0){\rule[-0.200pt]{4.818pt}{0.400pt}}
\put(201.0,368.0){\rule[-0.200pt]{4.818pt}{0.400pt}}
\put(181,368){\makebox(0,0)[r]{ 0.15}}
\put(1419.0,368.0){\rule[-0.200pt]{4.818pt}{0.400pt}}
\put(201.0,450.0){\rule[-0.200pt]{4.818pt}{0.400pt}}
\put(181,450){\makebox(0,0)[r]{ 0.2}}
\put(1419.0,450.0){\rule[-0.200pt]{4.818pt}{0.400pt}}
\put(201.0,532.0){\rule[-0.200pt]{4.818pt}{0.400pt}}
\put(181,532){\makebox(0,0)[r]{ 0.25}}
\put(1419.0,532.0){\rule[-0.200pt]{4.818pt}{0.400pt}}
\put(201.0,613.0){\rule[-0.200pt]{4.818pt}{0.400pt}}
\put(181,613){\makebox(0,0)[r]{ 0.3}}
\put(1419.0,613.0){\rule[-0.200pt]{4.818pt}{0.400pt}}
\put(201.0,695.0){\rule[-0.200pt]{4.818pt}{0.400pt}}
\put(181,695){\makebox(0,0)[r]{ 0.35}}
\put(1419.0,695.0){\rule[-0.200pt]{4.818pt}{0.400pt}}
\put(201.0,777.0){\rule[-0.200pt]{4.818pt}{0.400pt}}
\put(181,777){\makebox(0,0)[r]{ 0.4}}
\put(1419.0,777.0){\rule[-0.200pt]{4.818pt}{0.400pt}}
\put(201.0,123.0){\rule[-0.200pt]{0.400pt}{4.818pt}}
\put(201,82){\makebox(0,0){ 0}}
\put(201.0,757.0){\rule[-0.200pt]{0.400pt}{4.818pt}}
\put(378.0,123.0){\rule[-0.200pt]{0.400pt}{4.818pt}}
\put(378,82){\makebox(0,0){ 0.05}}
\put(378.0,757.0){\rule[-0.200pt]{0.400pt}{4.818pt}}
\put(555.0,123.0){\rule[-0.200pt]{0.400pt}{4.818pt}}
\put(555,82){\makebox(0,0){ 0.1}}
\put(555.0,757.0){\rule[-0.200pt]{0.400pt}{4.818pt}}
\put(732.0,123.0){\rule[-0.200pt]{0.400pt}{4.818pt}}
\put(732,82){\makebox(0,0){ 0.15}}
\put(732.0,757.0){\rule[-0.200pt]{0.400pt}{4.818pt}}
\put(908.0,123.0){\rule[-0.200pt]{0.400pt}{4.818pt}}
\put(908,82){\makebox(0,0){ 0.2}}
\put(908.0,757.0){\rule[-0.200pt]{0.400pt}{4.818pt}}
\put(1085.0,123.0){\rule[-0.200pt]{0.400pt}{4.818pt}}
\put(1085,82){\makebox(0,0){ 0.25}}
\put(1085.0,757.0){\rule[-0.200pt]{0.400pt}{4.818pt}}
\put(1262.0,123.0){\rule[-0.200pt]{0.400pt}{4.818pt}}
\put(1262,82){\makebox(0,0){ 0.3}}
\put(1262.0,757.0){\rule[-0.200pt]{0.400pt}{4.818pt}}
\put(1439.0,123.0){\rule[-0.200pt]{0.400pt}{4.818pt}}
\put(1439,82){\makebox(0,0){ 0.35}}
\put(1439.0,757.0){\rule[-0.200pt]{0.400pt}{4.818pt}}
\put(201.0,123.0){\rule[-0.200pt]{298.234pt}{0.400pt}}
\put(1439.0,123.0){\rule[-0.200pt]{0.400pt}{157.549pt}}
\put(201.0,777.0){\rule[-0.200pt]{298.234pt}{0.400pt}}
\put(40,450){\makebox(0,0){$\rho$}}
\put(820,21){\makebox(0,0){$\alpha$}}
\put(820,839){\makebox(0,0){Fig 2b}}
\put(201.0,123.0){\rule[-0.200pt]{0.400pt}{157.549pt}}
\put(201,123){\raisebox{-.8pt}{\makebox(0,0){$\diamond$}}}
\put(236,142){\raisebox{-.8pt}{\makebox(0,0){$\diamond$}}}
\put(272,160){\raisebox{-.8pt}{\makebox(0,0){$\diamond$}}}
\put(307,179){\raisebox{-.8pt}{\makebox(0,0){$\diamond$}}}
\put(342,198){\raisebox{-.8pt}{\makebox(0,0){$\diamond$}}}
\put(378,217){\raisebox{-.8pt}{\makebox(0,0){$\diamond$}}}
\put(413,235){\raisebox{-.8pt}{\makebox(0,0){$\diamond$}}}
\put(449,254){\raisebox{-.8pt}{\makebox(0,0){$\diamond$}}}
\put(484,273){\raisebox{-.8pt}{\makebox(0,0){$\diamond$}}}
\put(519,291){\raisebox{-.8pt}{\makebox(0,0){$\diamond$}}}
\put(555,310){\raisebox{-.8pt}{\makebox(0,0){$\diamond$}}}
\put(590,329){\raisebox{-.8pt}{\makebox(0,0){$\diamond$}}}
\put(625,348){\raisebox{-.8pt}{\makebox(0,0){$\diamond$}}}
\put(661,366){\raisebox{-.8pt}{\makebox(0,0){$\diamond$}}}
\put(696,385){\raisebox{-.8pt}{\makebox(0,0){$\diamond$}}}
\put(732,404){\raisebox{-.8pt}{\makebox(0,0){$\diamond$}}}
\put(767,422){\raisebox{-.8pt}{\makebox(0,0){$\diamond$}}}
\put(802,441){\raisebox{-.8pt}{\makebox(0,0){$\diamond$}}}
\put(838,460){\raisebox{-.8pt}{\makebox(0,0){$\diamond$}}}
\put(873,479){\raisebox{-.8pt}{\makebox(0,0){$\diamond$}}}
\put(908,497){\raisebox{-.8pt}{\makebox(0,0){$\diamond$}}}
\put(944,516){\raisebox{-.8pt}{\makebox(0,0){$\diamond$}}}
\put(979,535){\raisebox{-.8pt}{\makebox(0,0){$\diamond$}}}
\put(1015,553){\raisebox{-.8pt}{\makebox(0,0){$\diamond$}}}
\put(1050,572){\raisebox{-.8pt}{\makebox(0,0){$\diamond$}}}
\put(1085,591){\raisebox{-.8pt}{\makebox(0,0){$\diamond$}}}
\put(1121,609){\raisebox{-.8pt}{\makebox(0,0){$\diamond$}}}
\put(1156,628){\raisebox{-.8pt}{\makebox(0,0){$\diamond$}}}
\put(1191,647){\raisebox{-.8pt}{\makebox(0,0){$\diamond$}}}
\put(1227,665){\raisebox{-.8pt}{\makebox(0,0){$\diamond$}}}
\put(1262,684){\raisebox{-.8pt}{\makebox(0,0){$\diamond$}}}
\put(1298,702){\raisebox{-.8pt}{\makebox(0,0){$\diamond$}}}
\put(1333,721){\raisebox{-.8pt}{\makebox(0,0){$\diamond$}}}
\put(1368,740){\raisebox{-.8pt}{\makebox(0,0){$\diamond$}}}
\put(201,123){\makebox(0,0){$+$}}
\put(236,142){\makebox(0,0){$+$}}
\put(272,160){\makebox(0,0){$+$}}
\put(307,178){\makebox(0,0){$+$}}
\put(342,200){\makebox(0,0){$+$}}
\put(378,215){\makebox(0,0){$+$}}
\put(413,238){\makebox(0,0){$+$}}
\put(449,247){\makebox(0,0){$+$}}
\put(484,274){\makebox(0,0){$+$}}
\put(519,283){\makebox(0,0){$+$}}
\put(555,309){\makebox(0,0){$+$}}
\put(590,331){\makebox(0,0){$+$}}
\put(625,337){\makebox(0,0){$+$}}
\put(661,359){\makebox(0,0){$+$}}
\put(696,389){\makebox(0,0){$+$}}
\put(732,395){\makebox(0,0){$+$}}
\put(767,405){\makebox(0,0){$+$}}
\put(802,438){\makebox(0,0){$+$}}
\put(838,434){\makebox(0,0){$+$}}
\put(873,449){\makebox(0,0){$+$}}
\put(908,473){\makebox(0,0){$+$}}
\put(944,480){\makebox(0,0){$+$}}
\put(979,543){\makebox(0,0){$+$}}
\put(1015,541){\makebox(0,0){$+$}}
\put(1050,564){\makebox(0,0){$+$}}
\put(1085,563){\makebox(0,0){$+$}}
\put(1121,601){\makebox(0,0){$+$}}
\put(1156,603){\makebox(0,0){$+$}}
\put(1191,597){\makebox(0,0){$+$}}
\put(1227,653){\makebox(0,0){$+$}}
\put(1262,644){\makebox(0,0){$+$}}
\put(1298,708){\makebox(0,0){$+$}}
\put(1333,710){\makebox(0,0){$+$}}
\put(1368,705){\makebox(0,0){$+$}}
\end{picture}

\noindent Figure 2: The variation of the bulk density vs $\beta$ and $\alpha$ respectively in (a) high density and (b) low density regions for $c=0.75$ and $n=1$.$\diamond$ presents the analytical expression and + denotes the numerical simulation result for $L=2000$.\\

The regions of the high density phase and low density phase exhibit different density profiles. In fig. 3 we restrected ourself to present the density profiles of HD$_1$ and HD$_2$ phases. They present different behaviors far from the bulk from below as we have shown using our analytical approach.

\setlength{\unitlength}{0.240900pt}
\ifx\plotpoint\undefined\newsavebox{\plotpoint}\fi
\sbox{\plotpoint}{\rule[-0.200pt]{0.400pt}{0.400pt}}%
\begin{picture}(1500,900)(0,0)
\sbox{\plotpoint}{\rule[-0.200pt]{0.400pt}{0.400pt}}%
\put(221.0,143.0){\rule[-0.200pt]{4.818pt}{0.400pt}}
\put(201,143){\makebox(0,0)[r]{ 0.8}}
\put(1419.0,143.0){\rule[-0.200pt]{4.818pt}{0.400pt}}
\put(221.0,215.0){\rule[-0.200pt]{4.818pt}{0.400pt}}
\put(201,215){\makebox(0,0)[r]{ 0.81}}
\put(1419.0,215.0){\rule[-0.200pt]{4.818pt}{0.400pt}}
\put(221.0,286.0){\rule[-0.200pt]{4.818pt}{0.400pt}}
\put(201,286){\makebox(0,0)[r]{ 0.82}}
\put(1419.0,286.0){\rule[-0.200pt]{4.818pt}{0.400pt}}
\put(221.0,358.0){\rule[-0.200pt]{4.818pt}{0.400pt}}
\put(201,358){\makebox(0,0)[r]{ 0.83}}
\put(1419.0,358.0){\rule[-0.200pt]{4.818pt}{0.400pt}}
\put(221.0,430.0){\rule[-0.200pt]{4.818pt}{0.400pt}}
\put(201,430){\makebox(0,0)[r]{ 0.84}}
\put(1419.0,430.0){\rule[-0.200pt]{4.818pt}{0.400pt}}
\put(221.0,502.0){\rule[-0.200pt]{4.818pt}{0.400pt}}
\put(201,502){\makebox(0,0)[r]{ 0.85}}
\put(1419.0,502.0){\rule[-0.200pt]{4.818pt}{0.400pt}}
\put(221.0,573.0){\rule[-0.200pt]{4.818pt}{0.400pt}}
\put(201,573){\makebox(0,0)[r]{ 0.86}}
\put(1419.0,573.0){\rule[-0.200pt]{4.818pt}{0.400pt}}
\put(221.0,645.0){\rule[-0.200pt]{4.818pt}{0.400pt}}
\put(201,645){\makebox(0,0)[r]{ 0.87}}
\put(1419.0,645.0){\rule[-0.200pt]{4.818pt}{0.400pt}}
\put(221.0,717.0){\rule[-0.200pt]{4.818pt}{0.400pt}}
\put(201,717){\makebox(0,0)[r]{ 0.88}}
\put(1419.0,717.0){\rule[-0.200pt]{4.818pt}{0.400pt}}
\put(221.0,788.0){\rule[-0.200pt]{4.818pt}{0.400pt}}
\put(201,788){\makebox(0,0)[r]{ 0.89}}
\put(1419.0,788.0){\rule[-0.200pt]{4.818pt}{0.400pt}}
\put(221.0,860.0){\rule[-0.200pt]{4.818pt}{0.400pt}}
\put(201,860){\makebox(0,0)[r]{ 0.9}}
\put(1419.0,860.0){\rule[-0.200pt]{4.818pt}{0.400pt}}
\put(221.0,143.0){\rule[-0.200pt]{0.400pt}{4.818pt}}
\put(221,102){\makebox(0,0){ 0}}
\put(221.0,840.0){\rule[-0.200pt]{0.400pt}{4.818pt}}
\put(465.0,143.0){\rule[-0.200pt]{0.400pt}{4.818pt}}
\put(465,102){\makebox(0,0){ 20}}
\put(465.0,840.0){\rule[-0.200pt]{0.400pt}{4.818pt}}
\put(708.0,143.0){\rule[-0.200pt]{0.400pt}{4.818pt}}
\put(708,102){\makebox(0,0){ 40}}
\put(708.0,840.0){\rule[-0.200pt]{0.400pt}{4.818pt}}
\put(952.0,143.0){\rule[-0.200pt]{0.400pt}{4.818pt}}
\put(952,102){\makebox(0,0){ 60}}
\put(952.0,840.0){\rule[-0.200pt]{0.400pt}{4.818pt}}
\put(1195.0,143.0){\rule[-0.200pt]{0.400pt}{4.818pt}}
\put(1195,102){\makebox(0,0){ 80}}
\put(1195.0,840.0){\rule[-0.200pt]{0.400pt}{4.818pt}}
\put(1439.0,143.0){\rule[-0.200pt]{0.400pt}{4.818pt}}
\put(1439,102){\makebox(0,0){ 100}}
\put(1439.0,840.0){\rule[-0.200pt]{0.400pt}{4.818pt}}
\put(221.0,143.0){\rule[-0.200pt]{293.416pt}{0.400pt}}
\put(1439.0,143.0){\rule[-0.200pt]{0.400pt}{172.725pt}}
\put(221.0,860.0){\rule[-0.200pt]{293.416pt}{0.400pt}}
\put(221.0,143.0){\rule[-0.200pt]{0.400pt}{172.725pt}}
\put(40,501){\makebox(0,0){density}}
\put(830,21){\makebox(0,0){site i}}
\put(343,752){\makebox(0,0){HD$_2$}}
\put(465,573){\makebox(0,0){HD$_1$}}
\put(306,246){\raisebox{-.8pt}{\makebox(0,0){$\diamond$}}}
\put(318,373){\raisebox{-.8pt}{\makebox(0,0){$\diamond$}}}
\put(331,455){\raisebox{-.8pt}{\makebox(0,0){$\diamond$}}}
\put(343,515){\raisebox{-.8pt}{\makebox(0,0){$\diamond$}}}
\put(355,559){\raisebox{-.8pt}{\makebox(0,0){$\diamond$}}}
\put(367,593){\raisebox{-.8pt}{\makebox(0,0){$\diamond$}}}
\put(379,625){\raisebox{-.8pt}{\makebox(0,0){$\diamond$}}}
\put(392,642){\raisebox{-.8pt}{\makebox(0,0){$\diamond$}}}
\put(404,667){\raisebox{-.8pt}{\makebox(0,0){$\diamond$}}}
\put(416,664){\raisebox{-.8pt}{\makebox(0,0){$\diamond$}}}
\put(428,682){\raisebox{-.8pt}{\makebox(0,0){$\diamond$}}}
\put(440,686){\raisebox{-.8pt}{\makebox(0,0){$\diamond$}}}
\put(452,688){\raisebox{-.8pt}{\makebox(0,0){$\diamond$}}}
\put(465,685){\raisebox{-.8pt}{\makebox(0,0){$\diamond$}}}
\put(477,687){\raisebox{-.8pt}{\makebox(0,0){$\diamond$}}}
\put(489,694){\raisebox{-.8pt}{\makebox(0,0){$\diamond$}}}
\put(501,704){\raisebox{-.8pt}{\makebox(0,0){$\diamond$}}}
\put(513,698){\raisebox{-.8pt}{\makebox(0,0){$\diamond$}}}
\put(525,690){\raisebox{-.8pt}{\makebox(0,0){$\diamond$}}}
\put(538,687){\raisebox{-.8pt}{\makebox(0,0){$\diamond$}}}
\put(550,701){\raisebox{-.8pt}{\makebox(0,0){$\diamond$}}}
\put(562,708){\raisebox{-.8pt}{\makebox(0,0){$\diamond$}}}
\put(574,698){\raisebox{-.8pt}{\makebox(0,0){$\diamond$}}}
\put(586,705){\raisebox{-.8pt}{\makebox(0,0){$\diamond$}}}
\put(599,697){\raisebox{-.8pt}{\makebox(0,0){$\diamond$}}}
\put(611,684){\raisebox{-.8pt}{\makebox(0,0){$\diamond$}}}
\put(623,681){\raisebox{-.8pt}{\makebox(0,0){$\diamond$}}}
\put(635,700){\raisebox{-.8pt}{\makebox(0,0){$\diamond$}}}
\put(647,708){\raisebox{-.8pt}{\makebox(0,0){$\diamond$}}}
\put(659,707){\raisebox{-.8pt}{\makebox(0,0){$\diamond$}}}
\put(672,698){\raisebox{-.8pt}{\makebox(0,0){$\diamond$}}}
\put(684,702){\raisebox{-.8pt}{\makebox(0,0){$\diamond$}}}
\put(696,697){\raisebox{-.8pt}{\makebox(0,0){$\diamond$}}}
\put(708,701){\raisebox{-.8pt}{\makebox(0,0){$\diamond$}}}
\put(720,694){\raisebox{-.8pt}{\makebox(0,0){$\diamond$}}}
\put(733,693){\raisebox{-.8pt}{\makebox(0,0){$\diamond$}}}
\put(745,685){\raisebox{-.8pt}{\makebox(0,0){$\diamond$}}}
\put(757,684){\raisebox{-.8pt}{\makebox(0,0){$\diamond$}}}
\put(769,687){\raisebox{-.8pt}{\makebox(0,0){$\diamond$}}}
\put(781,689){\raisebox{-.8pt}{\makebox(0,0){$\diamond$}}}
\put(793,693){\raisebox{-.8pt}{\makebox(0,0){$\diamond$}}}
\put(806,693){\raisebox{-.8pt}{\makebox(0,0){$\diamond$}}}
\put(818,687){\raisebox{-.8pt}{\makebox(0,0){$\diamond$}}}
\put(830,698){\raisebox{-.8pt}{\makebox(0,0){$\diamond$}}}
\put(842,691){\raisebox{-.8pt}{\makebox(0,0){$\diamond$}}}
\put(854,694){\raisebox{-.8pt}{\makebox(0,0){$\diamond$}}}
\put(867,685){\raisebox{-.8pt}{\makebox(0,0){$\diamond$}}}
\put(879,679){\raisebox{-.8pt}{\makebox(0,0){$\diamond$}}}
\put(891,692){\raisebox{-.8pt}{\makebox(0,0){$\diamond$}}}
\put(903,690){\raisebox{-.8pt}{\makebox(0,0){$\diamond$}}}
\put(915,694){\raisebox{-.8pt}{\makebox(0,0){$\diamond$}}}
\put(927,694){\raisebox{-.8pt}{\makebox(0,0){$\diamond$}}}
\put(940,688){\raisebox{-.8pt}{\makebox(0,0){$\diamond$}}}
\put(952,704){\raisebox{-.8pt}{\makebox(0,0){$\diamond$}}}
\put(964,697){\raisebox{-.8pt}{\makebox(0,0){$\diamond$}}}
\put(976,697){\raisebox{-.8pt}{\makebox(0,0){$\diamond$}}}
\put(988,695){\raisebox{-.8pt}{\makebox(0,0){$\diamond$}}}
\put(1001,706){\raisebox{-.8pt}{\makebox(0,0){$\diamond$}}}
\put(1013,693){\raisebox{-.8pt}{\makebox(0,0){$\diamond$}}}
\put(1025,690){\raisebox{-.8pt}{\makebox(0,0){$\diamond$}}}
\put(1037,694){\raisebox{-.8pt}{\makebox(0,0){$\diamond$}}}
\put(1049,691){\raisebox{-.8pt}{\makebox(0,0){$\diamond$}}}
\put(1061,699){\raisebox{-.8pt}{\makebox(0,0){$\diamond$}}}
\put(1074,695){\raisebox{-.8pt}{\makebox(0,0){$\diamond$}}}
\put(1086,703){\raisebox{-.8pt}{\makebox(0,0){$\diamond$}}}
\put(1098,692){\raisebox{-.8pt}{\makebox(0,0){$\diamond$}}}
\put(1110,691){\raisebox{-.8pt}{\makebox(0,0){$\diamond$}}}
\put(1122,687){\raisebox{-.8pt}{\makebox(0,0){$\diamond$}}}
\put(1134,688){\raisebox{-.8pt}{\makebox(0,0){$\diamond$}}}
\put(1147,694){\raisebox{-.8pt}{\makebox(0,0){$\diamond$}}}
\put(1159,700){\raisebox{-.8pt}{\makebox(0,0){$\diamond$}}}
\put(1171,694){\raisebox{-.8pt}{\makebox(0,0){$\diamond$}}}
\put(1183,682){\raisebox{-.8pt}{\makebox(0,0){$\diamond$}}}
\put(1195,700){\raisebox{-.8pt}{\makebox(0,0){$\diamond$}}}
\put(1208,690){\raisebox{-.8pt}{\makebox(0,0){$\diamond$}}}
\put(1220,686){\raisebox{-.8pt}{\makebox(0,0){$\diamond$}}}
\put(1232,684){\raisebox{-.8pt}{\makebox(0,0){$\diamond$}}}
\put(1244,695){\raisebox{-.8pt}{\makebox(0,0){$\diamond$}}}
\put(1256,704){\raisebox{-.8pt}{\makebox(0,0){$\diamond$}}}
\put(1268,690){\raisebox{-.8pt}{\makebox(0,0){$\diamond$}}}
\put(1281,692){\raisebox{-.8pt}{\makebox(0,0){$\diamond$}}}
\put(1293,693){\raisebox{-.8pt}{\makebox(0,0){$\diamond$}}}
\put(1305,685){\raisebox{-.8pt}{\makebox(0,0){$\diamond$}}}
\put(1317,695){\raisebox{-.8pt}{\makebox(0,0){$\diamond$}}}
\put(1329,697){\raisebox{-.8pt}{\makebox(0,0){$\diamond$}}}
\put(1342,694){\raisebox{-.8pt}{\makebox(0,0){$\diamond$}}}
\put(1354,686){\raisebox{-.8pt}{\makebox(0,0){$\diamond$}}}
\put(1366,690){\raisebox{-.8pt}{\makebox(0,0){$\diamond$}}}
\put(1378,695){\raisebox{-.8pt}{\makebox(0,0){$\diamond$}}}
\put(1390,694){\raisebox{-.8pt}{\makebox(0,0){$\diamond$}}}
\put(1402,696){\raisebox{-.8pt}{\makebox(0,0){$\diamond$}}}
\put(1415,685){\raisebox{-.8pt}{\makebox(0,0){$\diamond$}}}
\put(1427,678){\raisebox{-.8pt}{\makebox(0,0){$\diamond$}}}
\put(1439,694){\raisebox{-.8pt}{\makebox(0,0){$\diamond$}}}
\put(233,790){\makebox(0,0){$+$}}
\put(245,724){\makebox(0,0){$+$}}
\put(258,710){\makebox(0,0){$+$}}
\put(270,696){\makebox(0,0){$+$}}
\put(282,705){\makebox(0,0){$+$}}
\put(294,705){\makebox(0,0){$+$}}
\put(306,696){\makebox(0,0){$+$}}
\put(318,702){\makebox(0,0){$+$}}
\put(331,702){\makebox(0,0){$+$}}
\put(343,705){\makebox(0,0){$+$}}
\put(355,697){\makebox(0,0){$+$}}
\put(367,687){\makebox(0,0){$+$}}
\put(379,700){\makebox(0,0){$+$}}
\put(392,691){\makebox(0,0){$+$}}
\put(404,694){\makebox(0,0){$+$}}
\put(416,684){\makebox(0,0){$+$}}
\put(428,684){\makebox(0,0){$+$}}
\put(440,688){\makebox(0,0){$+$}}
\put(452,697){\makebox(0,0){$+$}}
\put(465,696){\makebox(0,0){$+$}}
\put(477,698){\makebox(0,0){$+$}}
\put(489,681){\makebox(0,0){$+$}}
\put(501,694){\makebox(0,0){$+$}}
\put(513,690){\makebox(0,0){$+$}}
\put(525,697){\makebox(0,0){$+$}}
\put(538,702){\makebox(0,0){$+$}}
\put(550,692){\makebox(0,0){$+$}}
\put(562,697){\makebox(0,0){$+$}}
\put(574,690){\makebox(0,0){$+$}}
\put(586,694){\makebox(0,0){$+$}}
\put(599,685){\makebox(0,0){$+$}}
\put(611,684){\makebox(0,0){$+$}}
\put(623,689){\makebox(0,0){$+$}}
\put(635,693){\makebox(0,0){$+$}}
\put(647,693){\makebox(0,0){$+$}}
\put(659,698){\makebox(0,0){$+$}}
\put(672,698){\makebox(0,0){$+$}}
\put(684,686){\makebox(0,0){$+$}}
\put(696,693){\makebox(0,0){$+$}}
\put(708,699){\makebox(0,0){$+$}}
\put(720,697){\makebox(0,0){$+$}}
\put(733,698){\makebox(0,0){$+$}}
\put(745,693){\makebox(0,0){$+$}}
\put(757,684){\makebox(0,0){$+$}}
\put(769,690){\makebox(0,0){$+$}}
\put(781,700){\makebox(0,0){$+$}}
\put(793,707){\makebox(0,0){$+$}}
\put(806,702){\makebox(0,0){$+$}}
\put(818,695){\makebox(0,0){$+$}}
\put(830,695){\makebox(0,0){$+$}}
\put(842,683){\makebox(0,0){$+$}}
\put(854,698){\makebox(0,0){$+$}}
\put(867,694){\makebox(0,0){$+$}}
\put(879,688){\makebox(0,0){$+$}}
\put(891,686){\makebox(0,0){$+$}}
\put(903,686){\makebox(0,0){$+$}}
\put(915,694){\makebox(0,0){$+$}}
\put(927,695){\makebox(0,0){$+$}}
\put(940,695){\makebox(0,0){$+$}}
\put(952,700){\makebox(0,0){$+$}}
\put(964,700){\makebox(0,0){$+$}}
\put(976,688){\makebox(0,0){$+$}}
\put(988,691){\makebox(0,0){$+$}}
\put(1001,706){\makebox(0,0){$+$}}
\put(1013,698){\makebox(0,0){$+$}}
\put(1025,700){\makebox(0,0){$+$}}
\put(1037,702){\makebox(0,0){$+$}}
\put(1049,693){\makebox(0,0){$+$}}
\put(1061,693){\makebox(0,0){$+$}}
\put(1074,695){\makebox(0,0){$+$}}
\put(1086,695){\makebox(0,0){$+$}}
\put(1098,705){\makebox(0,0){$+$}}
\put(1110,699){\makebox(0,0){$+$}}
\put(1122,690){\makebox(0,0){$+$}}
\put(1134,690){\makebox(0,0){$+$}}
\put(1147,702){\makebox(0,0){$+$}}
\put(1159,690){\makebox(0,0){$+$}}
\put(1171,701){\makebox(0,0){$+$}}
\put(1183,689){\makebox(0,0){$+$}}
\put(1195,703){\makebox(0,0){$+$}}
\put(1208,693){\makebox(0,0){$+$}}
\put(1220,697){\makebox(0,0){$+$}}
\put(1232,683){\makebox(0,0){$+$}}
\put(1244,698){\makebox(0,0){$+$}}
\put(1256,694){\makebox(0,0){$+$}}
\put(1268,694){\makebox(0,0){$+$}}
\put(1281,696){\makebox(0,0){$+$}}
\put(1293,700){\makebox(0,0){$+$}}
\put(1305,698){\makebox(0,0){$+$}}
\put(1317,691){\makebox(0,0){$+$}}
\put(1329,692){\makebox(0,0){$+$}}
\put(1342,700){\makebox(0,0){$+$}}
\put(1354,699){\makebox(0,0){$+$}}
\put(1366,694){\makebox(0,0){$+$}}
\put(1378,694){\makebox(0,0){$+$}}
\put(1390,706){\makebox(0,0){$+$}}
\put(1402,706){\makebox(0,0){$+$}}
\put(1415,693){\makebox(0,0){$+$}}
\put(1427,692){\makebox(0,0){$+$}}
\put(1439,684){\makebox(0,0){$+$}}
\put(221.0,143.0){\rule[-0.200pt]{293.416pt}{0.400pt}}
\put(1439.0,143.0){\rule[-0.200pt]{0.400pt}{172.725pt}}
\put(221.0,860.0){\rule[-0.200pt]{293.416pt}{0.400pt}}
\put(221.0,143.0){\rule[-0.200pt]{0.400pt}{172.725pt}}
\end{picture}

\noindent Figure 3: The density profile of HD$_1$($\alpha=0.2, \beta=0.1$) and $HD_2$($\alpha=0.6, \beta=0.1$) phases.\\

The density profile in the MC phase decays, within the mean field theory[8,21], as $\rho(x)\sim x^{1/(m-1)}$, where $m$ is an exponent that describes the behavior of $J(\rho)-J(\rho_{max})$ near the maximal density $\rho_{max}$. The deterministic limit $p_{\mu}=1$ of the totally ASEP, within a fully parallel dynamics, for which $m=1$ is the only exactly solved case and it is found[22] that the MC phase disappears altogether. In the disordered ASEP the current-density relation depends on the choice of the distribution and the dynamics of updating. But, it seems that the density profile is not[19,20]. In fig. 4 we have presented the density profile in the MC phase for the random sequential updating. The data are consistent with a $1/ \sqrt{\ell}$-decay as we have shown analytically and they don't depend neither on $n$ nor $c$.\\

\setlength{\unitlength}{0.240900pt}
\ifx\plotpoint\undefined\newsavebox{\plotpoint}\fi
\sbox{\plotpoint}{\rule[-0.200pt]{0.400pt}{0.400pt}}%
\begin{picture}(1500,900)(0,0)
\font\gnuplot=cmr10 at 10pt
\gnuplot
\sbox{\plotpoint}{\rule[-0.200pt]{0.400pt}{0.400pt}}%
\put(180.0,82.0){\rule[-0.200pt]{4.818pt}{0.400pt}}
\put(160,82){\makebox(0,0)[r]{ 0.001}}
\put(1419.0,82.0){\rule[-0.200pt]{4.818pt}{0.400pt}}
\put(180.0,152.0){\rule[-0.200pt]{2.409pt}{0.400pt}}
\put(1429.0,152.0){\rule[-0.200pt]{2.409pt}{0.400pt}}
\put(180.0,193.0){\rule[-0.200pt]{2.409pt}{0.400pt}}
\put(1429.0,193.0){\rule[-0.200pt]{2.409pt}{0.400pt}}
\put(180.0,221.0){\rule[-0.200pt]{2.409pt}{0.400pt}}
\put(1429.0,221.0){\rule[-0.200pt]{2.409pt}{0.400pt}}
\put(180.0,244.0){\rule[-0.200pt]{2.409pt}{0.400pt}}
\put(1429.0,244.0){\rule[-0.200pt]{2.409pt}{0.400pt}}
\put(180.0,262.0){\rule[-0.200pt]{2.409pt}{0.400pt}}
\put(1429.0,262.0){\rule[-0.200pt]{2.409pt}{0.400pt}}
\put(180.0,278.0){\rule[-0.200pt]{2.409pt}{0.400pt}}
\put(1429.0,278.0){\rule[-0.200pt]{2.409pt}{0.400pt}}
\put(180.0,291.0){\rule[-0.200pt]{2.409pt}{0.400pt}}
\put(1429.0,291.0){\rule[-0.200pt]{2.409pt}{0.400pt}}
\put(180.0,303.0){\rule[-0.200pt]{2.409pt}{0.400pt}}
\put(1429.0,303.0){\rule[-0.200pt]{2.409pt}{0.400pt}}
\put(180.0,314.0){\rule[-0.200pt]{4.818pt}{0.400pt}}
\put(160,314){\makebox(0,0)[r]{ 0.01}}
\put(1419.0,314.0){\rule[-0.200pt]{4.818pt}{0.400pt}}
\put(180.0,383.0){\rule[-0.200pt]{2.409pt}{0.400pt}}
\put(1429.0,383.0){\rule[-0.200pt]{2.409pt}{0.400pt}}
\put(180.0,424.0){\rule[-0.200pt]{2.409pt}{0.400pt}}
\put(1429.0,424.0){\rule[-0.200pt]{2.409pt}{0.400pt}}
\put(180.0,453.0){\rule[-0.200pt]{2.409pt}{0.400pt}}
\put(1429.0,453.0){\rule[-0.200pt]{2.409pt}{0.400pt}}
\put(180.0,476.0){\rule[-0.200pt]{2.409pt}{0.400pt}}
\put(1429.0,476.0){\rule[-0.200pt]{2.409pt}{0.400pt}}
\put(180.0,494.0){\rule[-0.200pt]{2.409pt}{0.400pt}}
\put(1429.0,494.0){\rule[-0.200pt]{2.409pt}{0.400pt}}
\put(180.0,509.0){\rule[-0.200pt]{2.409pt}{0.400pt}}
\put(1429.0,509.0){\rule[-0.200pt]{2.409pt}{0.400pt}}
\put(180.0,523.0){\rule[-0.200pt]{2.409pt}{0.400pt}}
\put(1429.0,523.0){\rule[-0.200pt]{2.409pt}{0.400pt}}
\put(180.0,535.0){\rule[-0.200pt]{2.409pt}{0.400pt}}
\put(1429.0,535.0){\rule[-0.200pt]{2.409pt}{0.400pt}}
\put(180.0,545.0){\rule[-0.200pt]{4.818pt}{0.400pt}}
\put(160,545){\makebox(0,0)[r]{ 0.1}}
\put(1419.0,545.0){\rule[-0.200pt]{4.818pt}{0.400pt}}
\put(180.0,615.0){\rule[-0.200pt]{2.409pt}{0.400pt}}
\put(1429.0,615.0){\rule[-0.200pt]{2.409pt}{0.400pt}}
\put(180.0,656.0){\rule[-0.200pt]{2.409pt}{0.400pt}}
\put(1429.0,656.0){\rule[-0.200pt]{2.409pt}{0.400pt}}
\put(180.0,685.0){\rule[-0.200pt]{2.409pt}{0.400pt}}
\put(1429.0,685.0){\rule[-0.200pt]{2.409pt}{0.400pt}}
\put(180.0,707.0){\rule[-0.200pt]{2.409pt}{0.400pt}}
\put(1429.0,707.0){\rule[-0.200pt]{2.409pt}{0.400pt}}
\put(180.0,726.0){\rule[-0.200pt]{2.409pt}{0.400pt}}
\put(1429.0,726.0){\rule[-0.200pt]{2.409pt}{0.400pt}}
\put(180.0,741.0){\rule[-0.200pt]{2.409pt}{0.400pt}}
\put(1429.0,741.0){\rule[-0.200pt]{2.409pt}{0.400pt}}
\put(180.0,755.0){\rule[-0.200pt]{2.409pt}{0.400pt}}
\put(1429.0,755.0){\rule[-0.200pt]{2.409pt}{0.400pt}}
\put(180.0,766.0){\rule[-0.200pt]{2.409pt}{0.400pt}}
\put(1429.0,766.0){\rule[-0.200pt]{2.409pt}{0.400pt}}
\put(180.0,777.0){\rule[-0.200pt]{4.818pt}{0.400pt}}
\put(160,777){\makebox(0,0)[r]{ 1}}
\put(1419.0,777.0){\rule[-0.200pt]{4.818pt}{0.400pt}}
\put(180.0,82.0){\rule[-0.200pt]{0.400pt}{4.818pt}}
\put(180,41){\makebox(0,0){ 1}}
\put(180.0,757.0){\rule[-0.200pt]{0.400pt}{4.818pt}}
\put(345.0,82.0){\rule[-0.200pt]{0.400pt}{2.409pt}}
\put(345.0,767.0){\rule[-0.200pt]{0.400pt}{2.409pt}}
\put(441.0,82.0){\rule[-0.200pt]{0.400pt}{2.409pt}}
\put(441.0,767.0){\rule[-0.200pt]{0.400pt}{2.409pt}}
\put(509.0,82.0){\rule[-0.200pt]{0.400pt}{2.409pt}}
\put(509.0,767.0){\rule[-0.200pt]{0.400pt}{2.409pt}}
\put(562.0,82.0){\rule[-0.200pt]{0.400pt}{2.409pt}}
\put(562.0,767.0){\rule[-0.200pt]{0.400pt}{2.409pt}}
\put(606.0,82.0){\rule[-0.200pt]{0.400pt}{2.409pt}}
\put(606.0,767.0){\rule[-0.200pt]{0.400pt}{2.409pt}}
\put(642.0,82.0){\rule[-0.200pt]{0.400pt}{2.409pt}}
\put(642.0,767.0){\rule[-0.200pt]{0.400pt}{2.409pt}}
\put(674.0,82.0){\rule[-0.200pt]{0.400pt}{2.409pt}}
\put(674.0,767.0){\rule[-0.200pt]{0.400pt}{2.409pt}}
\put(702.0,82.0){\rule[-0.200pt]{0.400pt}{2.409pt}}
\put(702.0,767.0){\rule[-0.200pt]{0.400pt}{2.409pt}}
\put(727.0,82.0){\rule[-0.200pt]{0.400pt}{4.818pt}}
\put(727,41){\makebox(0,0){ 10}}
\put(727.0,757.0){\rule[-0.200pt]{0.400pt}{4.818pt}}
\put(892.0,82.0){\rule[-0.200pt]{0.400pt}{2.409pt}}
\put(892.0,767.0){\rule[-0.200pt]{0.400pt}{2.409pt}}
\put(988.0,82.0){\rule[-0.200pt]{0.400pt}{2.409pt}}
\put(988.0,767.0){\rule[-0.200pt]{0.400pt}{2.409pt}}
\put(1057.0,82.0){\rule[-0.200pt]{0.400pt}{2.409pt}}
\put(1057.0,767.0){\rule[-0.200pt]{0.400pt}{2.409pt}}
\put(1110.0,82.0){\rule[-0.200pt]{0.400pt}{2.409pt}}
\put(1110.0,767.0){\rule[-0.200pt]{0.400pt}{2.409pt}}
\put(1153.0,82.0){\rule[-0.200pt]{0.400pt}{2.409pt}}
\put(1153.0,767.0){\rule[-0.200pt]{0.400pt}{2.409pt}}
\put(1190.0,82.0){\rule[-0.200pt]{0.400pt}{2.409pt}}
\put(1190.0,767.0){\rule[-0.200pt]{0.400pt}{2.409pt}}
\put(1221.0,82.0){\rule[-0.200pt]{0.400pt}{2.409pt}}
\put(1221.0,767.0){\rule[-0.200pt]{0.400pt}{2.409pt}}
\put(1249.0,82.0){\rule[-0.200pt]{0.400pt}{2.409pt}}
\put(1249.0,767.0){\rule[-0.200pt]{0.400pt}{2.409pt}}
\put(1274.0,82.0){\rule[-0.200pt]{0.400pt}{4.818pt}}
\put(1274,41){\makebox(0,0){ 100}}
\put(1274.0,757.0){\rule[-0.200pt]{0.400pt}{4.818pt}}
\put(1439.0,82.0){\rule[-0.200pt]{0.400pt}{2.409pt}}
\put(1439.0,767.0){\rule[-0.200pt]{0.400pt}{2.409pt}}
\put(180.0,82.0){\rule[-0.200pt]{303.293pt}{0.400pt}}
\put(1439.0,82.0){\rule[-0.200pt]{0.400pt}{167.425pt}}
\put(180.0,777.0){\rule[-0.200pt]{303.293pt}{0.400pt}}
\put(5,450){\makebox(0,0){Log$_{10}(\rho (i)-\rho_{max})$}}
\put(830,0){\makebox(0,0){Log$_{10}$(i)}}
\put(809,839){\makebox(0,0){Fig4}}
\put(727,615){\makebox(0,0){Slope=-1/2}}
\put(180.0,82.0){\rule[-0.200pt]{0.400pt}{167.425pt}}
\put(180,645){\usebox{\plotpoint}}
\put(180,643.17){\rule{2.700pt}{0.400pt}}
\multiput(180.00,644.17)(7.396,-2.000){2}{\rule{1.350pt}{0.400pt}}
\multiput(193.00,641.95)(2.472,-0.447){3}{\rule{1.700pt}{0.108pt}}
\multiput(193.00,642.17)(8.472,-3.000){2}{\rule{0.850pt}{0.400pt}}
\multiput(205.00,638.95)(2.695,-0.447){3}{\rule{1.833pt}{0.108pt}}
\multiput(205.00,639.17)(9.195,-3.000){2}{\rule{0.917pt}{0.400pt}}
\multiput(218.00,635.95)(2.695,-0.447){3}{\rule{1.833pt}{0.108pt}}
\multiput(218.00,636.17)(9.195,-3.000){2}{\rule{0.917pt}{0.400pt}}
\put(231,632.17){\rule{2.700pt}{0.400pt}}
\multiput(231.00,633.17)(7.396,-2.000){2}{\rule{1.350pt}{0.400pt}}
\multiput(244.00,630.95)(2.472,-0.447){3}{\rule{1.700pt}{0.108pt}}
\multiput(244.00,631.17)(8.472,-3.000){2}{\rule{0.850pt}{0.400pt}}
\multiput(256.00,627.95)(2.695,-0.447){3}{\rule{1.833pt}{0.108pt}}
\multiput(256.00,628.17)(9.195,-3.000){2}{\rule{0.917pt}{0.400pt}}
\put(269,624.17){\rule{2.700pt}{0.400pt}}
\multiput(269.00,625.17)(7.396,-2.000){2}{\rule{1.350pt}{0.400pt}}
\multiput(282.00,622.95)(2.472,-0.447){3}{\rule{1.700pt}{0.108pt}}
\multiput(282.00,623.17)(8.472,-3.000){2}{\rule{0.850pt}{0.400pt}}
\multiput(294.00,619.95)(2.695,-0.447){3}{\rule{1.833pt}{0.108pt}}
\multiput(294.00,620.17)(9.195,-3.000){2}{\rule{0.917pt}{0.400pt}}
\put(307,616.17){\rule{2.700pt}{0.400pt}}
\multiput(307.00,617.17)(7.396,-2.000){2}{\rule{1.350pt}{0.400pt}}
\multiput(320.00,614.95)(2.695,-0.447){3}{\rule{1.833pt}{0.108pt}}
\multiput(320.00,615.17)(9.195,-3.000){2}{\rule{0.917pt}{0.400pt}}
\multiput(333.00,611.95)(2.472,-0.447){3}{\rule{1.700pt}{0.108pt}}
\multiput(333.00,612.17)(8.472,-3.000){2}{\rule{0.850pt}{0.400pt}}
\put(345,608.17){\rule{2.700pt}{0.400pt}}
\multiput(345.00,609.17)(7.396,-2.000){2}{\rule{1.350pt}{0.400pt}}
\multiput(358.00,606.95)(2.695,-0.447){3}{\rule{1.833pt}{0.108pt}}
\multiput(358.00,607.17)(9.195,-3.000){2}{\rule{0.917pt}{0.400pt}}
\multiput(371.00,603.95)(2.472,-0.447){3}{\rule{1.700pt}{0.108pt}}
\multiput(371.00,604.17)(8.472,-3.000){2}{\rule{0.850pt}{0.400pt}}
\multiput(383.00,600.95)(2.695,-0.447){3}{\rule{1.833pt}{0.108pt}}
\multiput(383.00,601.17)(9.195,-3.000){2}{\rule{0.917pt}{0.400pt}}
\put(396,597.17){\rule{2.700pt}{0.400pt}}
\multiput(396.00,598.17)(7.396,-2.000){2}{\rule{1.350pt}{0.400pt}}
\multiput(409.00,595.95)(2.695,-0.447){3}{\rule{1.833pt}{0.108pt}}
\multiput(409.00,596.17)(9.195,-3.000){2}{\rule{0.917pt}{0.400pt}}
\multiput(422.00,592.95)(2.472,-0.447){3}{\rule{1.700pt}{0.108pt}}
\multiput(422.00,593.17)(8.472,-3.000){2}{\rule{0.850pt}{0.400pt}}
\put(434,589.17){\rule{2.700pt}{0.400pt}}
\multiput(434.00,590.17)(7.396,-2.000){2}{\rule{1.350pt}{0.400pt}}
\multiput(447.00,587.95)(2.695,-0.447){3}{\rule{1.833pt}{0.108pt}}
\multiput(447.00,588.17)(9.195,-3.000){2}{\rule{0.917pt}{0.400pt}}
\multiput(460.00,584.95)(2.472,-0.447){3}{\rule{1.700pt}{0.108pt}}
\multiput(460.00,585.17)(8.472,-3.000){2}{\rule{0.850pt}{0.400pt}}
\put(472,581.17){\rule{2.700pt}{0.400pt}}
\multiput(472.00,582.17)(7.396,-2.000){2}{\rule{1.350pt}{0.400pt}}
\multiput(485.00,579.95)(2.695,-0.447){3}{\rule{1.833pt}{0.108pt}}
\multiput(485.00,580.17)(9.195,-3.000){2}{\rule{0.917pt}{0.400pt}}
\multiput(498.00,576.95)(2.695,-0.447){3}{\rule{1.833pt}{0.108pt}}
\multiput(498.00,577.17)(9.195,-3.000){2}{\rule{0.917pt}{0.400pt}}
\put(511,573.17){\rule{2.500pt}{0.400pt}}
\multiput(511.00,574.17)(6.811,-2.000){2}{\rule{1.250pt}{0.400pt}}
\multiput(523.00,571.95)(2.695,-0.447){3}{\rule{1.833pt}{0.108pt}}
\multiput(523.00,572.17)(9.195,-3.000){2}{\rule{0.917pt}{0.400pt}}
\multiput(536.00,568.95)(2.695,-0.447){3}{\rule{1.833pt}{0.108pt}}
\multiput(536.00,569.17)(9.195,-3.000){2}{\rule{0.917pt}{0.400pt}}
\multiput(549.00,565.95)(2.695,-0.447){3}{\rule{1.833pt}{0.108pt}}
\multiput(549.00,566.17)(9.195,-3.000){2}{\rule{0.917pt}{0.400pt}}
\put(562,562.17){\rule{2.500pt}{0.400pt}}
\multiput(562.00,563.17)(6.811,-2.000){2}{\rule{1.250pt}{0.400pt}}
\multiput(574.00,560.95)(2.695,-0.447){3}{\rule{1.833pt}{0.108pt}}
\multiput(574.00,561.17)(9.195,-3.000){2}{\rule{0.917pt}{0.400pt}}
\multiput(587.00,557.95)(2.695,-0.447){3}{\rule{1.833pt}{0.108pt}}
\multiput(587.00,558.17)(9.195,-3.000){2}{\rule{0.917pt}{0.400pt}}
\put(600,554.17){\rule{2.500pt}{0.400pt}}
\multiput(600.00,555.17)(6.811,-2.000){2}{\rule{1.250pt}{0.400pt}}
\multiput(612.00,552.95)(2.695,-0.447){3}{\rule{1.833pt}{0.108pt}}
\multiput(612.00,553.17)(9.195,-3.000){2}{\rule{0.917pt}{0.400pt}}
\multiput(625.00,549.95)(2.695,-0.447){3}{\rule{1.833pt}{0.108pt}}
\multiput(625.00,550.17)(9.195,-3.000){2}{\rule{0.917pt}{0.400pt}}
\put(638,546.17){\rule{2.700pt}{0.400pt}}
\multiput(638.00,547.17)(7.396,-2.000){2}{\rule{1.350pt}{0.400pt}}
\multiput(651.00,544.95)(2.472,-0.447){3}{\rule{1.700pt}{0.108pt}}
\multiput(651.00,545.17)(8.472,-3.000){2}{\rule{0.850pt}{0.400pt}}
\multiput(663.00,541.95)(2.695,-0.447){3}{\rule{1.833pt}{0.108pt}}
\multiput(663.00,542.17)(9.195,-3.000){2}{\rule{0.917pt}{0.400pt}}
\put(676,538.17){\rule{2.700pt}{0.400pt}}
\multiput(676.00,539.17)(7.396,-2.000){2}{\rule{1.350pt}{0.400pt}}
\multiput(689.00,536.95)(2.472,-0.447){3}{\rule{1.700pt}{0.108pt}}
\multiput(689.00,537.17)(8.472,-3.000){2}{\rule{0.850pt}{0.400pt}}
\multiput(701.00,533.95)(2.695,-0.447){3}{\rule{1.833pt}{0.108pt}}
\multiput(701.00,534.17)(9.195,-3.000){2}{\rule{0.917pt}{0.400pt}}
\multiput(714.00,530.95)(2.695,-0.447){3}{\rule{1.833pt}{0.108pt}}
\multiput(714.00,531.17)(9.195,-3.000){2}{\rule{0.917pt}{0.400pt}}
\put(727,527.17){\rule{2.700pt}{0.400pt}}
\multiput(727.00,528.17)(7.396,-2.000){2}{\rule{1.350pt}{0.400pt}}
\multiput(740.00,525.95)(2.472,-0.447){3}{\rule{1.700pt}{0.108pt}}
\multiput(740.00,526.17)(8.472,-3.000){2}{\rule{0.850pt}{0.400pt}}
\multiput(752.00,522.95)(2.695,-0.447){3}{\rule{1.833pt}{0.108pt}}
\multiput(752.00,523.17)(9.195,-3.000){2}{\rule{0.917pt}{0.400pt}}
\put(765,519.17){\rule{2.700pt}{0.400pt}}
\multiput(765.00,520.17)(7.396,-2.000){2}{\rule{1.350pt}{0.400pt}}
\multiput(778.00,517.95)(2.472,-0.447){3}{\rule{1.700pt}{0.108pt}}
\multiput(778.00,518.17)(8.472,-3.000){2}{\rule{0.850pt}{0.400pt}}
\multiput(790.00,514.95)(2.695,-0.447){3}{\rule{1.833pt}{0.108pt}}
\multiput(790.00,515.17)(9.195,-3.000){2}{\rule{0.917pt}{0.400pt}}
\put(803,511.17){\rule{2.700pt}{0.400pt}}
\multiput(803.00,512.17)(7.396,-2.000){2}{\rule{1.350pt}{0.400pt}}
\multiput(816.00,509.95)(2.695,-0.447){3}{\rule{1.833pt}{0.108pt}}
\multiput(816.00,510.17)(9.195,-3.000){2}{\rule{0.917pt}{0.400pt}}
\multiput(829.00,506.95)(2.472,-0.447){3}{\rule{1.700pt}{0.108pt}}
\multiput(829.00,507.17)(8.472,-3.000){2}{\rule{0.850pt}{0.400pt}}
\put(841,503.17){\rule{2.700pt}{0.400pt}}
\multiput(841.00,504.17)(7.396,-2.000){2}{\rule{1.350pt}{0.400pt}}
\multiput(854.00,501.95)(2.695,-0.447){3}{\rule{1.833pt}{0.108pt}}
\multiput(854.00,502.17)(9.195,-3.000){2}{\rule{0.917pt}{0.400pt}}
\multiput(867.00,498.95)(2.472,-0.447){3}{\rule{1.700pt}{0.108pt}}
\multiput(867.00,499.17)(8.472,-3.000){2}{\rule{0.850pt}{0.400pt}}
\multiput(879.00,495.95)(2.695,-0.447){3}{\rule{1.833pt}{0.108pt}}
\multiput(879.00,496.17)(9.195,-3.000){2}{\rule{0.917pt}{0.400pt}}
\put(892,492.17){\rule{2.700pt}{0.400pt}}
\multiput(892.00,493.17)(7.396,-2.000){2}{\rule{1.350pt}{0.400pt}}
\multiput(905.00,490.95)(2.695,-0.447){3}{\rule{1.833pt}{0.108pt}}
\multiput(905.00,491.17)(9.195,-3.000){2}{\rule{0.917pt}{0.400pt}}
\multiput(918.00,487.95)(2.472,-0.447){3}{\rule{1.700pt}{0.108pt}}
\multiput(918.00,488.17)(8.472,-3.000){2}{\rule{0.850pt}{0.400pt}}
\put(930,484.17){\rule{2.700pt}{0.400pt}}
\multiput(930.00,485.17)(7.396,-2.000){2}{\rule{1.350pt}{0.400pt}}
\multiput(943.00,482.95)(2.695,-0.447){3}{\rule{1.833pt}{0.108pt}}
\multiput(943.00,483.17)(9.195,-3.000){2}{\rule{0.917pt}{0.400pt}}
\multiput(956.00,479.95)(2.472,-0.447){3}{\rule{1.700pt}{0.108pt}}
\multiput(956.00,480.17)(8.472,-3.000){2}{\rule{0.850pt}{0.400pt}}
\put(968,476.17){\rule{2.700pt}{0.400pt}}
\multiput(968.00,477.17)(7.396,-2.000){2}{\rule{1.350pt}{0.400pt}}
\multiput(981.00,474.95)(2.695,-0.447){3}{\rule{1.833pt}{0.108pt}}
\multiput(981.00,475.17)(9.195,-3.000){2}{\rule{0.917pt}{0.400pt}}
\multiput(994.00,471.95)(2.695,-0.447){3}{\rule{1.833pt}{0.108pt}}
\multiput(994.00,472.17)(9.195,-3.000){2}{\rule{0.917pt}{0.400pt}}
\put(1007,468.17){\rule{2.500pt}{0.400pt}}
\multiput(1007.00,469.17)(6.811,-2.000){2}{\rule{1.250pt}{0.400pt}}
\multiput(1019.00,466.95)(2.695,-0.447){3}{\rule{1.833pt}{0.108pt}}
\multiput(1019.00,467.17)(9.195,-3.000){2}{\rule{0.917pt}{0.400pt}}
\multiput(1032.00,463.95)(2.695,-0.447){3}{\rule{1.833pt}{0.108pt}}
\multiput(1032.00,464.17)(9.195,-3.000){2}{\rule{0.917pt}{0.400pt}}
\multiput(1045.00,460.95)(2.472,-0.447){3}{\rule{1.700pt}{0.108pt}}
\multiput(1045.00,461.17)(8.472,-3.000){2}{\rule{0.850pt}{0.400pt}}
\put(1057,457.17){\rule{2.700pt}{0.400pt}}
\multiput(1057.00,458.17)(7.396,-2.000){2}{\rule{1.350pt}{0.400pt}}
\multiput(1070.00,455.95)(2.695,-0.447){3}{\rule{1.833pt}{0.108pt}}
\multiput(1070.00,456.17)(9.195,-3.000){2}{\rule{0.917pt}{0.400pt}}
\multiput(1083.00,452.95)(2.695,-0.447){3}{\rule{1.833pt}{0.108pt}}
\multiput(1083.00,453.17)(9.195,-3.000){2}{\rule{0.917pt}{0.400pt}}
\put(1096,449.17){\rule{2.500pt}{0.400pt}}
\multiput(1096.00,450.17)(6.811,-2.000){2}{\rule{1.250pt}{0.400pt}}
\multiput(1108.00,447.95)(2.695,-0.447){3}{\rule{1.833pt}{0.108pt}}
\multiput(1108.00,448.17)(9.195,-3.000){2}{\rule{0.917pt}{0.400pt}}
\multiput(1121.00,444.95)(2.695,-0.447){3}{\rule{1.833pt}{0.108pt}}
\multiput(1121.00,445.17)(9.195,-3.000){2}{\rule{0.917pt}{0.400pt}}
\put(1134,441.17){\rule{2.700pt}{0.400pt}}
\multiput(1134.00,442.17)(7.396,-2.000){2}{\rule{1.350pt}{0.400pt}}
\multiput(1147.00,439.95)(2.472,-0.447){3}{\rule{1.700pt}{0.108pt}}
\multiput(1147.00,440.17)(8.472,-3.000){2}{\rule{0.850pt}{0.400pt}}
\multiput(1159.00,436.95)(2.695,-0.447){3}{\rule{1.833pt}{0.108pt}}
\multiput(1159.00,437.17)(9.195,-3.000){2}{\rule{0.917pt}{0.400pt}}
\put(1172,433.17){\rule{2.700pt}{0.400pt}}
\multiput(1172.00,434.17)(7.396,-2.000){2}{\rule{1.350pt}{0.400pt}}
\multiput(1185.00,431.95)(2.472,-0.447){3}{\rule{1.700pt}{0.108pt}}
\multiput(1185.00,432.17)(8.472,-3.000){2}{\rule{0.850pt}{0.400pt}}
\multiput(1197.00,428.95)(2.695,-0.447){3}{\rule{1.833pt}{0.108pt}}
\multiput(1197.00,429.17)(9.195,-3.000){2}{\rule{0.917pt}{0.400pt}}
\multiput(1210.00,425.95)(2.695,-0.447){3}{\rule{1.833pt}{0.108pt}}
\multiput(1210.00,426.17)(9.195,-3.000){2}{\rule{0.917pt}{0.400pt}}
\put(1223,422.17){\rule{2.700pt}{0.400pt}}
\multiput(1223.00,423.17)(7.396,-2.000){2}{\rule{1.350pt}{0.400pt}}
\multiput(1236.00,420.95)(2.472,-0.447){3}{\rule{1.700pt}{0.108pt}}
\multiput(1236.00,421.17)(8.472,-3.000){2}{\rule{0.850pt}{0.400pt}}
\multiput(1248.00,417.95)(2.695,-0.447){3}{\rule{1.833pt}{0.108pt}}
\multiput(1248.00,418.17)(9.195,-3.000){2}{\rule{0.917pt}{0.400pt}}
\put(1261,414.17){\rule{2.700pt}{0.400pt}}
\multiput(1261.00,415.17)(7.396,-2.000){2}{\rule{1.350pt}{0.400pt}}
\multiput(1274.00,412.95)(2.472,-0.447){3}{\rule{1.700pt}{0.108pt}}
\multiput(1274.00,413.17)(8.472,-3.000){2}{\rule{0.850pt}{0.400pt}}
\multiput(1286.00,409.95)(2.695,-0.447){3}{\rule{1.833pt}{0.108pt}}
\multiput(1286.00,410.17)(9.195,-3.000){2}{\rule{0.917pt}{0.400pt}}
\put(1299,406.17){\rule{2.700pt}{0.400pt}}
\multiput(1299.00,407.17)(7.396,-2.000){2}{\rule{1.350pt}{0.400pt}}
\multiput(1312.00,404.95)(2.695,-0.447){3}{\rule{1.833pt}{0.108pt}}
\multiput(1312.00,405.17)(9.195,-3.000){2}{\rule{0.917pt}{0.400pt}}
\multiput(1325.00,401.95)(2.472,-0.447){3}{\rule{1.700pt}{0.108pt}}
\multiput(1325.00,402.17)(8.472,-3.000){2}{\rule{0.850pt}{0.400pt}}
\put(1337,398.17){\rule{2.700pt}{0.400pt}}
\multiput(1337.00,399.17)(7.396,-2.000){2}{\rule{1.350pt}{0.400pt}}
\multiput(1350.00,396.95)(2.695,-0.447){3}{\rule{1.833pt}{0.108pt}}
\multiput(1350.00,397.17)(9.195,-3.000){2}{\rule{0.917pt}{0.400pt}}
\multiput(1363.00,393.95)(2.472,-0.447){3}{\rule{1.700pt}{0.108pt}}
\multiput(1363.00,394.17)(8.472,-3.000){2}{\rule{0.850pt}{0.400pt}}
\multiput(1375.00,390.95)(2.695,-0.447){3}{\rule{1.833pt}{0.108pt}}
\multiput(1375.00,391.17)(9.195,-3.000){2}{\rule{0.917pt}{0.400pt}}
\put(1388,387.17){\rule{2.700pt}{0.400pt}}
\multiput(1388.00,388.17)(7.396,-2.000){2}{\rule{1.350pt}{0.400pt}}
\multiput(1401.00,385.95)(2.695,-0.447){3}{\rule{1.833pt}{0.108pt}}
\multiput(1401.00,386.17)(9.195,-3.000){2}{\rule{0.917pt}{0.400pt}}
\multiput(1414.00,382.95)(2.472,-0.447){3}{\rule{1.700pt}{0.108pt}}
\multiput(1414.00,383.17)(8.472,-3.000){2}{\rule{0.850pt}{0.400pt}}
\put(1426,379.17){\rule{2.700pt}{0.400pt}}
\multiput(1426.00,380.17)(7.396,-2.000){2}{\rule{1.350pt}{0.400pt}}
\put(180,639){\raisebox{-.8pt}{\makebox(0,0){$\diamond$}}}
\put(345,609){\raisebox{-.8pt}{\makebox(0,0){$\diamond$}}}
\put(441,590){\raisebox{-.8pt}{\makebox(0,0){$\diamond$}}}
\put(509,577){\raisebox{-.8pt}{\makebox(0,0){$\diamond$}}}
\put(562,566){\raisebox{-.8pt}{\makebox(0,0){$\diamond$}}}
\put(606,556){\raisebox{-.8pt}{\makebox(0,0){$\diamond$}}}
\put(642,548){\raisebox{-.8pt}{\makebox(0,0){$\diamond$}}}
\put(674,541){\raisebox{-.8pt}{\makebox(0,0){$\diamond$}}}
\put(702,534){\raisebox{-.8pt}{\makebox(0,0){$\diamond$}}}
\put(727,530){\raisebox{-.8pt}{\makebox(0,0){$\diamond$}}}
\put(750,525){\raisebox{-.8pt}{\makebox(0,0){$\diamond$}}}
\put(770,520){\raisebox{-.8pt}{\makebox(0,0){$\diamond$}}}
\put(789,518){\raisebox{-.8pt}{\makebox(0,0){$\diamond$}}}
\put(807,514){\raisebox{-.8pt}{\makebox(0,0){$\diamond$}}}
\put(823,509){\raisebox{-.8pt}{\makebox(0,0){$\diamond$}}}
\put(839,508){\raisebox{-.8pt}{\makebox(0,0){$\diamond$}}}
\put(853,505){\raisebox{-.8pt}{\makebox(0,0){$\diamond$}}}
\put(867,500){\raisebox{-.8pt}{\makebox(0,0){$\diamond$}}}
\put(880,500){\raisebox{-.8pt}{\makebox(0,0){$\diamond$}}}
\put(892,498){\raisebox{-.8pt}{\makebox(0,0){$\diamond$}}}
\put(903,493){\raisebox{-.8pt}{\makebox(0,0){$\diamond$}}}
\put(915,495){\raisebox{-.8pt}{\makebox(0,0){$\diamond$}}}
\put(925,494){\raisebox{-.8pt}{\makebox(0,0){$\diamond$}}}
\put(935,490){\raisebox{-.8pt}{\makebox(0,0){$\diamond$}}}
\put(945,487){\raisebox{-.8pt}{\makebox(0,0){$\diamond$}}}
\put(954,483){\raisebox{-.8pt}{\makebox(0,0){$\diamond$}}}
\put(963,479){\raisebox{-.8pt}{\makebox(0,0){$\diamond$}}}
\put(972,473){\raisebox{-.8pt}{\makebox(0,0){$\diamond$}}}
\put(980,469){\raisebox{-.8pt}{\makebox(0,0){$\diamond$}}}
\put(988,471){\raisebox{-.8pt}{\makebox(0,0){$\diamond$}}}
\put(996,472){\raisebox{-.8pt}{\makebox(0,0){$\diamond$}}}
\put(1004,472){\raisebox{-.8pt}{\makebox(0,0){$\diamond$}}}
\put(1011,468){\raisebox{-.8pt}{\makebox(0,0){$\diamond$}}}
\put(1018,466){\raisebox{-.8pt}{\makebox(0,0){$\diamond$}}}
\put(1025,468){\raisebox{-.8pt}{\makebox(0,0){$\diamond$}}}
\put(1032,464){\raisebox{-.8pt}{\makebox(0,0){$\diamond$}}}
\put(1038,464){\raisebox{-.8pt}{\makebox(0,0){$\diamond$}}}
\put(1044,460){\raisebox{-.8pt}{\makebox(0,0){$\diamond$}}}
\put(1051,459){\raisebox{-.8pt}{\makebox(0,0){$\diamond$}}}
\put(1057,463){\raisebox{-.8pt}{\makebox(0,0){$\diamond$}}}
\put(1062,464){\raisebox{-.8pt}{\makebox(0,0){$\diamond$}}}
\put(1068,460){\raisebox{-.8pt}{\makebox(0,0){$\diamond$}}}
\put(1074,460){\raisebox{-.8pt}{\makebox(0,0){$\diamond$}}}
\put(1079,459){\raisebox{-.8pt}{\makebox(0,0){$\diamond$}}}
\put(1085,455){\raisebox{-.8pt}{\makebox(0,0){$\diamond$}}}
\put(1090,453){\raisebox{-.8pt}{\makebox(0,0){$\diamond$}}}
\put(1095,454){\raisebox{-.8pt}{\makebox(0,0){$\diamond$}}}
\put(1100,453){\raisebox{-.8pt}{\makebox(0,0){$\diamond$}}}
\put(1105,453){\raisebox{-.8pt}{\makebox(0,0){$\diamond$}}}
\put(1110,451){\raisebox{-.8pt}{\makebox(0,0){$\diamond$}}}
\put(1114,447){\raisebox{-.8pt}{\makebox(0,0){$\diamond$}}}
\put(1119,447){\raisebox{-.8pt}{\makebox(0,0){$\diamond$}}}
\put(1123,444){\raisebox{-.8pt}{\makebox(0,0){$\diamond$}}}
\put(1128,448){\raisebox{-.8pt}{\makebox(0,0){$\diamond$}}}
\put(1132,438){\raisebox{-.8pt}{\makebox(0,0){$\diamond$}}}
\put(1137,437){\raisebox{-.8pt}{\makebox(0,0){$\diamond$}}}
\put(1141,437){\raisebox{-.8pt}{\makebox(0,0){$\diamond$}}}
\put(1145,442){\raisebox{-.8pt}{\makebox(0,0){$\diamond$}}}
\put(1149,435){\raisebox{-.8pt}{\makebox(0,0){$\diamond$}}}
\put(1153,434){\raisebox{-.8pt}{\makebox(0,0){$\diamond$}}}
\put(1157,436){\raisebox{-.8pt}{\makebox(0,0){$\diamond$}}}
\put(1161,439){\raisebox{-.8pt}{\makebox(0,0){$\diamond$}}}
\put(1165,438){\raisebox{-.8pt}{\makebox(0,0){$\diamond$}}}
\put(1168,446){\raisebox{-.8pt}{\makebox(0,0){$\diamond$}}}
\put(1172,450){\raisebox{-.8pt}{\makebox(0,0){$\diamond$}}}
\put(1176,443){\raisebox{-.8pt}{\makebox(0,0){$\diamond$}}}
\put(1179,445){\raisebox{-.8pt}{\makebox(0,0){$\diamond$}}}
\put(1183,446){\raisebox{-.8pt}{\makebox(0,0){$\diamond$}}}
\put(1186,446){\raisebox{-.8pt}{\makebox(0,0){$\diamond$}}}
\put(1190,449){\raisebox{-.8pt}{\makebox(0,0){$\diamond$}}}
\put(1193,448){\raisebox{-.8pt}{\makebox(0,0){$\diamond$}}}
\put(1196,444){\raisebox{-.8pt}{\makebox(0,0){$\diamond$}}}
\put(1200,442){\raisebox{-.8pt}{\makebox(0,0){$\diamond$}}}
\put(1203,440){\raisebox{-.8pt}{\makebox(0,0){$\diamond$}}}
\put(1206,444){\raisebox{-.8pt}{\makebox(0,0){$\diamond$}}}
\put(1209,451){\raisebox{-.8pt}{\makebox(0,0){$\diamond$}}}
\put(1212,450){\raisebox{-.8pt}{\makebox(0,0){$\diamond$}}}
\put(1215,450){\raisebox{-.8pt}{\makebox(0,0){$\diamond$}}}
\put(1218,444){\raisebox{-.8pt}{\makebox(0,0){$\diamond$}}}
\put(1221,442){\raisebox{-.8pt}{\makebox(0,0){$\diamond$}}}
\put(1224,446){\raisebox{-.8pt}{\makebox(0,0){$\diamond$}}}
\put(1227,445){\raisebox{-.8pt}{\makebox(0,0){$\diamond$}}}
\put(1230,440){\raisebox{-.8pt}{\makebox(0,0){$\diamond$}}}
\put(1233,440){\raisebox{-.8pt}{\makebox(0,0){$\diamond$}}}
\put(1236,433){\raisebox{-.8pt}{\makebox(0,0){$\diamond$}}}
\put(1238,435){\raisebox{-.8pt}{\makebox(0,0){$\diamond$}}}
\put(1241,441){\raisebox{-.8pt}{\makebox(0,0){$\diamond$}}}
\put(1244,442){\raisebox{-.8pt}{\makebox(0,0){$\diamond$}}}
\put(1247,442){\raisebox{-.8pt}{\makebox(0,0){$\diamond$}}}
\put(1249,435){\raisebox{-.8pt}{\makebox(0,0){$\diamond$}}}
\put(1252,436){\raisebox{-.8pt}{\makebox(0,0){$\diamond$}}}
\put(1254,425){\raisebox{-.8pt}{\makebox(0,0){$\diamond$}}}
\put(1257,421){\raisebox{-.8pt}{\makebox(0,0){$\diamond$}}}
\put(1260,431){\raisebox{-.8pt}{\makebox(0,0){$\diamond$}}}
\put(1262,426){\raisebox{-.8pt}{\makebox(0,0){$\diamond$}}}
\put(1265,425){\raisebox{-.8pt}{\makebox(0,0){$\diamond$}}}
\put(1267,429){\raisebox{-.8pt}{\makebox(0,0){$\diamond$}}}
\put(1269,422){\raisebox{-.8pt}{\makebox(0,0){$\diamond$}}}
\put(1272,419){\raisebox{-.8pt}{\makebox(0,0){$\diamond$}}}
\put(1274,419){\raisebox{-.8pt}{\makebox(0,0){$\diamond$}}}
\put(1277,414){\raisebox{-.8pt}{\makebox(0,0){$\diamond$}}}
\put(1279,412){\raisebox{-.8pt}{\makebox(0,0){$\diamond$}}}
\put(1281,411){\raisebox{-.8pt}{\makebox(0,0){$\diamond$}}}
\put(1284,405){\raisebox{-.8pt}{\makebox(0,0){$\diamond$}}}
\put(1286,407){\raisebox{-.8pt}{\makebox(0,0){$\diamond$}}}
\put(1288,415){\raisebox{-.8pt}{\makebox(0,0){$\diamond$}}}
\put(1290,412){\raisebox{-.8pt}{\makebox(0,0){$\diamond$}}}
\put(1293,409){\raisebox{-.8pt}{\makebox(0,0){$\diamond$}}}
\put(1295,401){\raisebox{-.8pt}{\makebox(0,0){$\diamond$}}}
\put(1297,398){\raisebox{-.8pt}{\makebox(0,0){$\diamond$}}}
\put(1299,395){\raisebox{-.8pt}{\makebox(0,0){$\diamond$}}}
\put(1301,394){\raisebox{-.8pt}{\makebox(0,0){$\diamond$}}}
\put(1303,394){\raisebox{-.8pt}{\makebox(0,0){$\diamond$}}}
\put(1305,399){\raisebox{-.8pt}{\makebox(0,0){$\diamond$}}}
\put(1308,398){\raisebox{-.8pt}{\makebox(0,0){$\diamond$}}}
\put(1310,403){\raisebox{-.8pt}{\makebox(0,0){$\diamond$}}}
\put(1312,400){\raisebox{-.8pt}{\makebox(0,0){$\diamond$}}}
\put(1314,399){\raisebox{-.8pt}{\makebox(0,0){$\diamond$}}}
\put(1316,401){\raisebox{-.8pt}{\makebox(0,0){$\diamond$}}}
\put(1318,403){\raisebox{-.8pt}{\makebox(0,0){$\diamond$}}}
\put(1320,401){\raisebox{-.8pt}{\makebox(0,0){$\diamond$}}}
\put(1322,405){\raisebox{-.8pt}{\makebox(0,0){$\diamond$}}}
\put(1323,401){\raisebox{-.8pt}{\makebox(0,0){$\diamond$}}}
\put(1325,397){\raisebox{-.8pt}{\makebox(0,0){$\diamond$}}}
\put(1327,406){\raisebox{-.8pt}{\makebox(0,0){$\diamond$}}}
\put(1329,405){\raisebox{-.8pt}{\makebox(0,0){$\diamond$}}}
\put(1331,407){\raisebox{-.8pt}{\makebox(0,0){$\diamond$}}}
\put(1333,405){\raisebox{-.8pt}{\makebox(0,0){$\diamond$}}}
\put(1335,401){\raisebox{-.8pt}{\makebox(0,0){$\diamond$}}}
\put(1337,407){\raisebox{-.8pt}{\makebox(0,0){$\diamond$}}}
\put(1338,408){\raisebox{-.8pt}{\makebox(0,0){$\diamond$}}}
\put(1340,408){\raisebox{-.8pt}{\makebox(0,0){$\diamond$}}}
\put(1342,410){\raisebox{-.8pt}{\makebox(0,0){$\diamond$}}}
\put(1344,402){\raisebox{-.8pt}{\makebox(0,0){$\diamond$}}}
\put(1346,388){\raisebox{-.8pt}{\makebox(0,0){$\diamond$}}}
\put(1347,392){\raisebox{-.8pt}{\makebox(0,0){$\diamond$}}}
\put(1349,386){\raisebox{-.8pt}{\makebox(0,0){$\diamond$}}}
\put(1351,380){\raisebox{-.8pt}{\makebox(0,0){$\diamond$}}}
\put(1353,386){\raisebox{-.8pt}{\makebox(0,0){$\diamond$}}}
\put(1354,376){\raisebox{-.8pt}{\makebox(0,0){$\diamond$}}}
\put(1356,387){\raisebox{-.8pt}{\makebox(0,0){$\diamond$}}}
\put(1358,384){\raisebox{-.8pt}{\makebox(0,0){$\diamond$}}}
\put(1359,378){\raisebox{-.8pt}{\makebox(0,0){$\diamond$}}}
\put(1361,394){\raisebox{-.8pt}{\makebox(0,0){$\diamond$}}}
\put(1363,386){\raisebox{-.8pt}{\makebox(0,0){$\diamond$}}}
\put(1364,394){\raisebox{-.8pt}{\makebox(0,0){$\diamond$}}}
\put(1366,391){\raisebox{-.8pt}{\makebox(0,0){$\diamond$}}}
\put(1367,389){\raisebox{-.8pt}{\makebox(0,0){$\diamond$}}}
\put(1369,387){\raisebox{-.8pt}{\makebox(0,0){$\diamond$}}}
\put(1371,384){\raisebox{-.8pt}{\makebox(0,0){$\diamond$}}}
\put(1372,376){\raisebox{-.8pt}{\makebox(0,0){$\diamond$}}}
\put(1374,384){\raisebox{-.8pt}{\makebox(0,0){$\diamond$}}}
\put(1375,380){\raisebox{-.8pt}{\makebox(0,0){$\diamond$}}}
\put(1377,376){\raisebox{-.8pt}{\makebox(0,0){$\diamond$}}}
\put(1378,374){\raisebox{-.8pt}{\makebox(0,0){$\diamond$}}}
\put(1380,383){\raisebox{-.8pt}{\makebox(0,0){$\diamond$}}}
\put(1381,387){\raisebox{-.8pt}{\makebox(0,0){$\diamond$}}}
\put(1383,384){\raisebox{-.8pt}{\makebox(0,0){$\diamond$}}}
\put(1384,387){\raisebox{-.8pt}{\makebox(0,0){$\diamond$}}}
\put(1386,381){\raisebox{-.8pt}{\makebox(0,0){$\diamond$}}}
\put(1387,382){\raisebox{-.8pt}{\makebox(0,0){$\diamond$}}}
\put(1389,371){\raisebox{-.8pt}{\makebox(0,0){$\diamond$}}}
\put(1390,373){\raisebox{-.8pt}{\makebox(0,0){$\diamond$}}}
\put(1392,368){\raisebox{-.8pt}{\makebox(0,0){$\diamond$}}}
\put(1393,368){\raisebox{-.8pt}{\makebox(0,0){$\diamond$}}}
\put(1395,374){\raisebox{-.8pt}{\makebox(0,0){$\diamond$}}}
\put(1396,359){\raisebox{-.8pt}{\makebox(0,0){$\diamond$}}}
\put(1398,354){\raisebox{-.8pt}{\makebox(0,0){$\diamond$}}}
\put(1399,372){\raisebox{-.8pt}{\makebox(0,0){$\diamond$}}}
\put(1400,366){\raisebox{-.8pt}{\makebox(0,0){$\diamond$}}}
\put(1402,358){\raisebox{-.8pt}{\makebox(0,0){$\diamond$}}}
\put(1403,353){\raisebox{-.8pt}{\makebox(0,0){$\diamond$}}}
\put(1405,353){\raisebox{-.8pt}{\makebox(0,0){$\diamond$}}}
\put(1406,345){\raisebox{-.8pt}{\makebox(0,0){$\diamond$}}}
\put(1407,356){\raisebox{-.8pt}{\makebox(0,0){$\diamond$}}}
\put(1409,360){\raisebox{-.8pt}{\makebox(0,0){$\diamond$}}}
\put(1410,361){\raisebox{-.8pt}{\makebox(0,0){$\diamond$}}}
\put(1411,361){\raisebox{-.8pt}{\makebox(0,0){$\diamond$}}}
\put(1413,348){\raisebox{-.8pt}{\makebox(0,0){$\diamond$}}}
\put(1414,353){\raisebox{-.8pt}{\makebox(0,0){$\diamond$}}}
\put(1415,351){\raisebox{-.8pt}{\makebox(0,0){$\diamond$}}}
\put(1417,340){\raisebox{-.8pt}{\makebox(0,0){$\diamond$}}}
\put(1418,330){\raisebox{-.8pt}{\makebox(0,0){$\diamond$}}}
\put(1419,329){\raisebox{-.8pt}{\makebox(0,0){$\diamond$}}}
\put(1420,336){\raisebox{-.8pt}{\makebox(0,0){$\diamond$}}}
\put(1422,331){\raisebox{-.8pt}{\makebox(0,0){$\diamond$}}}
\put(1423,318){\raisebox{-.8pt}{\makebox(0,0){$\diamond$}}}
\put(1424,309){\raisebox{-.8pt}{\makebox(0,0){$\diamond$}}}
\put(1426,306){\raisebox{-.8pt}{\makebox(0,0){$\diamond$}}}
\put(1427,294){\raisebox{-.8pt}{\makebox(0,0){$\diamond$}}}
\put(1428,288){\raisebox{-.8pt}{\makebox(0,0){$\diamond$}}}
\put(1429,272){\raisebox{-.8pt}{\makebox(0,0){$\diamond$}}}
\put(1431,280){\raisebox{-.8pt}{\makebox(0,0){$\diamond$}}}
\put(1432,313){\raisebox{-.8pt}{\makebox(0,0){$\diamond$}}}
\put(1433,331){\raisebox{-.8pt}{\makebox(0,0){$\diamond$}}}
\put(1434,334){\raisebox{-.8pt}{\makebox(0,0){$\diamond$}}}
\put(1435,337){\raisebox{-.8pt}{\makebox(0,0){$\diamond$}}}
\put(1437,325){\raisebox{-.8pt}{\makebox(0,0){$\diamond$}}}
\put(1438,313){\raisebox{-.8pt}{\makebox(0,0){$\diamond$}}}
\put(1439,330){\raisebox{-.8pt}{\makebox(0,0){$\diamond$}}}
\end{picture}

\noindent Figure 4: A Log-Log plot of the density profile in the MC phase for $L=2000$, $n=1$, $c=0.5$, $\alpha=0.7$ and $\beta=0.9$. The straight line indicates the $1/\sqrt{\ell}$-decay obtained analytically.\\

In order to check the validity of our mean field-like theory and emphasize the analytical results obtained using this approach, we have shown in fig. 5 the variation of the critical value $\alpha^*$ of the platoon transition according to $n$ and $c$. It can be located by monitoring the variance of the headways[17,19]
\begin{equation}
\Delta^2=\overline{<u^2_{\mu}>_{\mu}}-\overline{<u_{\mu}>^2_{\mu}}
\end{equation}
where $<.>_{\mu}$ means the average over all the values of the gap $u_{\mu}=x_{\mu+1}-x_{\mu}-1$ ($x_{\mu}$ is the position of the particle $\mu$), as function of the injection rate $\alpha$, i.e $\rho$, and system size. For $\alpha > \alpha^*$($\rho > \rho^*$), $\Delta^2$ is independent of system size, while for $\alpha < \alpha^*$ ($\rho < \rho^*$) it is dominated by the macroscopic gaps and acquires an L-dependence.\\
In order to determine the value of $\alpha^*$, we compute the headway $\Delta^2$ for two system sizes namely $L=2000$ and $L=500$. The data for the two system sizes coincide for $\alpha > \alpha^*$ but differ for $\alpha < \alpha^*$. Since we restricted ourselves to high values of $c$, i.e $c\ge 1/2$, we remark, that the numerical data and the analytical expressions are in good agreement. The slight difference they may present for low values of $c$ results from the fact that for such values the system presents high correlations which are neglected in our analytical approach.

\setlength{\unitlength}{0.240900pt}
\ifx\plotpoint\undefined\newsavebox{\plotpoint}\fi
\sbox{\plotpoint}{\rule[-0.200pt]{0.400pt}{0.400pt}}%
\begin{picture}(1500,900)(0,0)
\font\gnuplot=cmr10 at 10pt
\gnuplot
\sbox{\plotpoint}{\rule[-0.200pt]{0.400pt}{0.400pt}}%
\put(201.0,123.0){\rule[-0.200pt]{4.818pt}{0.400pt}}
\put(181,123){\makebox(0,0)[r]{ 0.2}}
\put(1419.0,123.0){\rule[-0.200pt]{4.818pt}{0.400pt}}
\put(201.0,286.0){\rule[-0.200pt]{4.818pt}{0.400pt}}
\put(181,286){\makebox(0,0)[r]{ 0.25}}
\put(1419.0,286.0){\rule[-0.200pt]{4.818pt}{0.400pt}}
\put(201.0,450.0){\rule[-0.200pt]{4.818pt}{0.400pt}}
\put(181,450){\makebox(0,0)[r]{ 0.3}}
\put(1419.0,450.0){\rule[-0.200pt]{4.818pt}{0.400pt}}
\put(201.0,613.0){\rule[-0.200pt]{4.818pt}{0.400pt}}
\put(181,613){\makebox(0,0)[r]{ 0.35}}
\put(1419.0,613.0){\rule[-0.200pt]{4.818pt}{0.400pt}}
\put(201.0,777.0){\rule[-0.200pt]{4.818pt}{0.400pt}}
\put(181,777){\makebox(0,0)[r]{ 0.4}}
\put(1419.0,777.0){\rule[-0.200pt]{4.818pt}{0.400pt}}
\put(201.0,123.0){\rule[-0.200pt]{0.400pt}{4.818pt}}
\put(201,82){\makebox(0,0){ 1}}
\put(201.0,757.0){\rule[-0.200pt]{0.400pt}{4.818pt}}
\put(356.0,123.0){\rule[-0.200pt]{0.400pt}{4.818pt}}
\put(356,82){\makebox(0,0){ 1.5}}
\put(356.0,757.0){\rule[-0.200pt]{0.400pt}{4.818pt}}
\put(511.0,123.0){\rule[-0.200pt]{0.400pt}{4.818pt}}
\put(511,82){\makebox(0,0){ 2}}
\put(511.0,757.0){\rule[-0.200pt]{0.400pt}{4.818pt}}
\put(665.0,123.0){\rule[-0.200pt]{0.400pt}{4.818pt}}
\put(665,82){\makebox(0,0){ 2.5}}
\put(665.0,757.0){\rule[-0.200pt]{0.400pt}{4.818pt}}
\put(820.0,123.0){\rule[-0.200pt]{0.400pt}{4.818pt}}
\put(820,82){\makebox(0,0){ 3}}
\put(820.0,757.0){\rule[-0.200pt]{0.400pt}{4.818pt}}
\put(975.0,123.0){\rule[-0.200pt]{0.400pt}{4.818pt}}
\put(975,82){\makebox(0,0){ 3.5}}
\put(975.0,757.0){\rule[-0.200pt]{0.400pt}{4.818pt}}
\put(1130.0,123.0){\rule[-0.200pt]{0.400pt}{4.818pt}}
\put(1130,82){\makebox(0,0){ 4}}
\put(1130.0,757.0){\rule[-0.200pt]{0.400pt}{4.818pt}}
\put(1284.0,123.0){\rule[-0.200pt]{0.400pt}{4.818pt}}
\put(1284,82){\makebox(0,0){ 4.5}}
\put(1284.0,757.0){\rule[-0.200pt]{0.400pt}{4.818pt}}
\put(1439.0,123.0){\rule[-0.200pt]{0.400pt}{4.818pt}}
\put(1439,82){\makebox(0,0){ 5}}
\put(1439.0,757.0){\rule[-0.200pt]{0.400pt}{4.818pt}}
\put(201.0,123.0){\rule[-0.200pt]{298.234pt}{0.400pt}}
\put(1439.0,123.0){\rule[-0.200pt]{0.400pt}{157.549pt}}
\put(201.0,777.0){\rule[-0.200pt]{298.234pt}{0.400pt}}
\put(40,455){\makebox(0,0){$\alpha^*$}}
\put(820,21){\makebox(0,0){n}}
\put(820,839){\makebox(0,0){Fig 5a}}
\put(201.0,123.0){\rule[-0.200pt]{0.400pt}{157.549pt}}
\put(201,225){\usebox{\plotpoint}}
\multiput(201.00,225.58)(0.539,0.492){21}{\rule{0.533pt}{0.119pt}}
\multiput(201.00,224.17)(11.893,12.000){2}{\rule{0.267pt}{0.400pt}}
\multiput(214.00,237.58)(0.496,0.492){21}{\rule{0.500pt}{0.119pt}}
\multiput(214.00,236.17)(10.962,12.000){2}{\rule{0.250pt}{0.400pt}}
\multiput(226.00,249.58)(0.590,0.492){19}{\rule{0.573pt}{0.118pt}}
\multiput(226.00,248.17)(11.811,11.000){2}{\rule{0.286pt}{0.400pt}}
\multiput(239.00,260.58)(0.543,0.492){19}{\rule{0.536pt}{0.118pt}}
\multiput(239.00,259.17)(10.887,11.000){2}{\rule{0.268pt}{0.400pt}}
\multiput(251.00,271.58)(0.539,0.492){21}{\rule{0.533pt}{0.119pt}}
\multiput(251.00,270.17)(11.893,12.000){2}{\rule{0.267pt}{0.400pt}}
\multiput(264.00,283.58)(0.543,0.492){19}{\rule{0.536pt}{0.118pt}}
\multiput(264.00,282.17)(10.887,11.000){2}{\rule{0.268pt}{0.400pt}}
\multiput(276.00,294.58)(0.652,0.491){17}{\rule{0.620pt}{0.118pt}}
\multiput(276.00,293.17)(11.713,10.000){2}{\rule{0.310pt}{0.400pt}}
\multiput(289.00,304.58)(0.543,0.492){19}{\rule{0.536pt}{0.118pt}}
\multiput(289.00,303.17)(10.887,11.000){2}{\rule{0.268pt}{0.400pt}}
\multiput(301.00,315.58)(0.652,0.491){17}{\rule{0.620pt}{0.118pt}}
\multiput(301.00,314.17)(11.713,10.000){2}{\rule{0.310pt}{0.400pt}}
\multiput(314.00,325.58)(0.600,0.491){17}{\rule{0.580pt}{0.118pt}}
\multiput(314.00,324.17)(10.796,10.000){2}{\rule{0.290pt}{0.400pt}}
\multiput(326.00,335.58)(0.652,0.491){17}{\rule{0.620pt}{0.118pt}}
\multiput(326.00,334.17)(11.713,10.000){2}{\rule{0.310pt}{0.400pt}}
\multiput(339.00,345.58)(0.600,0.491){17}{\rule{0.580pt}{0.118pt}}
\multiput(339.00,344.17)(10.796,10.000){2}{\rule{0.290pt}{0.400pt}}
\multiput(351.00,355.59)(0.728,0.489){15}{\rule{0.678pt}{0.118pt}}
\multiput(351.00,354.17)(11.593,9.000){2}{\rule{0.339pt}{0.400pt}}
\multiput(364.00,364.58)(0.600,0.491){17}{\rule{0.580pt}{0.118pt}}
\multiput(364.00,363.17)(10.796,10.000){2}{\rule{0.290pt}{0.400pt}}
\multiput(376.00,374.59)(0.728,0.489){15}{\rule{0.678pt}{0.118pt}}
\multiput(376.00,373.17)(11.593,9.000){2}{\rule{0.339pt}{0.400pt}}
\multiput(389.00,383.59)(0.669,0.489){15}{\rule{0.633pt}{0.118pt}}
\multiput(389.00,382.17)(10.685,9.000){2}{\rule{0.317pt}{0.400pt}}
\multiput(401.00,392.59)(0.824,0.488){13}{\rule{0.750pt}{0.117pt}}
\multiput(401.00,391.17)(11.443,8.000){2}{\rule{0.375pt}{0.400pt}}
\multiput(414.00,400.59)(0.669,0.489){15}{\rule{0.633pt}{0.118pt}}
\multiput(414.00,399.17)(10.685,9.000){2}{\rule{0.317pt}{0.400pt}}
\multiput(426.00,409.59)(0.824,0.488){13}{\rule{0.750pt}{0.117pt}}
\multiput(426.00,408.17)(11.443,8.000){2}{\rule{0.375pt}{0.400pt}}
\multiput(439.00,417.59)(0.669,0.489){15}{\rule{0.633pt}{0.118pt}}
\multiput(439.00,416.17)(10.685,9.000){2}{\rule{0.317pt}{0.400pt}}
\multiput(451.00,426.59)(0.824,0.488){13}{\rule{0.750pt}{0.117pt}}
\multiput(451.00,425.17)(11.443,8.000){2}{\rule{0.375pt}{0.400pt}}
\multiput(464.00,434.59)(0.758,0.488){13}{\rule{0.700pt}{0.117pt}}
\multiput(464.00,433.17)(10.547,8.000){2}{\rule{0.350pt}{0.400pt}}
\multiput(476.00,442.59)(0.950,0.485){11}{\rule{0.843pt}{0.117pt}}
\multiput(476.00,441.17)(11.251,7.000){2}{\rule{0.421pt}{0.400pt}}
\multiput(489.00,449.59)(0.758,0.488){13}{\rule{0.700pt}{0.117pt}}
\multiput(489.00,448.17)(10.547,8.000){2}{\rule{0.350pt}{0.400pt}}
\multiput(501.00,457.59)(0.950,0.485){11}{\rule{0.843pt}{0.117pt}}
\multiput(501.00,456.17)(11.251,7.000){2}{\rule{0.421pt}{0.400pt}}
\multiput(514.00,464.59)(0.874,0.485){11}{\rule{0.786pt}{0.117pt}}
\multiput(514.00,463.17)(10.369,7.000){2}{\rule{0.393pt}{0.400pt}}
\multiput(526.00,471.59)(0.824,0.488){13}{\rule{0.750pt}{0.117pt}}
\multiput(526.00,470.17)(11.443,8.000){2}{\rule{0.375pt}{0.400pt}}
\multiput(539.00,479.59)(1.033,0.482){9}{\rule{0.900pt}{0.116pt}}
\multiput(539.00,478.17)(10.132,6.000){2}{\rule{0.450pt}{0.400pt}}
\multiput(551.00,485.59)(0.950,0.485){11}{\rule{0.843pt}{0.117pt}}
\multiput(551.00,484.17)(11.251,7.000){2}{\rule{0.421pt}{0.400pt}}
\multiput(564.00,492.59)(0.874,0.485){11}{\rule{0.786pt}{0.117pt}}
\multiput(564.00,491.17)(10.369,7.000){2}{\rule{0.393pt}{0.400pt}}
\multiput(576.00,499.59)(1.123,0.482){9}{\rule{0.967pt}{0.116pt}}
\multiput(576.00,498.17)(10.994,6.000){2}{\rule{0.483pt}{0.400pt}}
\multiput(589.00,505.59)(0.874,0.485){11}{\rule{0.786pt}{0.117pt}}
\multiput(589.00,504.17)(10.369,7.000){2}{\rule{0.393pt}{0.400pt}}
\multiput(601.00,512.59)(1.123,0.482){9}{\rule{0.967pt}{0.116pt}}
\multiput(601.00,511.17)(10.994,6.000){2}{\rule{0.483pt}{0.400pt}}
\multiput(614.00,518.59)(1.033,0.482){9}{\rule{0.900pt}{0.116pt}}
\multiput(614.00,517.17)(10.132,6.000){2}{\rule{0.450pt}{0.400pt}}
\multiput(626.00,524.59)(1.123,0.482){9}{\rule{0.967pt}{0.116pt}}
\multiput(626.00,523.17)(10.994,6.000){2}{\rule{0.483pt}{0.400pt}}
\multiput(639.00,530.59)(1.267,0.477){7}{\rule{1.060pt}{0.115pt}}
\multiput(639.00,529.17)(9.800,5.000){2}{\rule{0.530pt}{0.400pt}}
\multiput(651.00,535.59)(1.123,0.482){9}{\rule{0.967pt}{0.116pt}}
\multiput(651.00,534.17)(10.994,6.000){2}{\rule{0.483pt}{0.400pt}}
\multiput(664.00,541.59)(1.033,0.482){9}{\rule{0.900pt}{0.116pt}}
\multiput(664.00,540.17)(10.132,6.000){2}{\rule{0.450pt}{0.400pt}}
\multiput(676.00,547.59)(1.378,0.477){7}{\rule{1.140pt}{0.115pt}}
\multiput(676.00,546.17)(10.634,5.000){2}{\rule{0.570pt}{0.400pt}}
\multiput(689.00,552.59)(1.267,0.477){7}{\rule{1.060pt}{0.115pt}}
\multiput(689.00,551.17)(9.800,5.000){2}{\rule{0.530pt}{0.400pt}}
\multiput(701.00,557.59)(1.378,0.477){7}{\rule{1.140pt}{0.115pt}}
\multiput(701.00,556.17)(10.634,5.000){2}{\rule{0.570pt}{0.400pt}}
\multiput(714.00,562.59)(1.267,0.477){7}{\rule{1.060pt}{0.115pt}}
\multiput(714.00,561.17)(9.800,5.000){2}{\rule{0.530pt}{0.400pt}}
\multiput(726.00,567.59)(1.378,0.477){7}{\rule{1.140pt}{0.115pt}}
\multiput(726.00,566.17)(10.634,5.000){2}{\rule{0.570pt}{0.400pt}}
\multiput(739.00,572.59)(1.267,0.477){7}{\rule{1.060pt}{0.115pt}}
\multiput(739.00,571.17)(9.800,5.000){2}{\rule{0.530pt}{0.400pt}}
\multiput(751.00,577.59)(1.378,0.477){7}{\rule{1.140pt}{0.115pt}}
\multiput(751.00,576.17)(10.634,5.000){2}{\rule{0.570pt}{0.400pt}}
\multiput(764.00,582.60)(1.651,0.468){5}{\rule{1.300pt}{0.113pt}}
\multiput(764.00,581.17)(9.302,4.000){2}{\rule{0.650pt}{0.400pt}}
\multiput(776.00,586.59)(1.378,0.477){7}{\rule{1.140pt}{0.115pt}}
\multiput(776.00,585.17)(10.634,5.000){2}{\rule{0.570pt}{0.400pt}}
\multiput(789.00,591.60)(1.651,0.468){5}{\rule{1.300pt}{0.113pt}}
\multiput(789.00,590.17)(9.302,4.000){2}{\rule{0.650pt}{0.400pt}}
\multiput(801.00,595.60)(1.797,0.468){5}{\rule{1.400pt}{0.113pt}}
\multiput(801.00,594.17)(10.094,4.000){2}{\rule{0.700pt}{0.400pt}}
\multiput(814.00,599.60)(1.651,0.468){5}{\rule{1.300pt}{0.113pt}}
\multiput(814.00,598.17)(9.302,4.000){2}{\rule{0.650pt}{0.400pt}}
\multiput(826.00,603.59)(1.378,0.477){7}{\rule{1.140pt}{0.115pt}}
\multiput(826.00,602.17)(10.634,5.000){2}{\rule{0.570pt}{0.400pt}}
\multiput(839.00,608.61)(2.472,0.447){3}{\rule{1.700pt}{0.108pt}}
\multiput(839.00,607.17)(8.472,3.000){2}{\rule{0.850pt}{0.400pt}}
\multiput(851.00,611.60)(1.797,0.468){5}{\rule{1.400pt}{0.113pt}}
\multiput(851.00,610.17)(10.094,4.000){2}{\rule{0.700pt}{0.400pt}}
\multiput(864.00,615.60)(1.651,0.468){5}{\rule{1.300pt}{0.113pt}}
\multiput(864.00,614.17)(9.302,4.000){2}{\rule{0.650pt}{0.400pt}}
\multiput(876.00,619.60)(1.797,0.468){5}{\rule{1.400pt}{0.113pt}}
\multiput(876.00,618.17)(10.094,4.000){2}{\rule{0.700pt}{0.400pt}}
\multiput(889.00,623.61)(2.472,0.447){3}{\rule{1.700pt}{0.108pt}}
\multiput(889.00,622.17)(8.472,3.000){2}{\rule{0.850pt}{0.400pt}}
\multiput(901.00,626.60)(1.797,0.468){5}{\rule{1.400pt}{0.113pt}}
\multiput(901.00,625.17)(10.094,4.000){2}{\rule{0.700pt}{0.400pt}}
\multiput(914.00,630.61)(2.472,0.447){3}{\rule{1.700pt}{0.108pt}}
\multiput(914.00,629.17)(8.472,3.000){2}{\rule{0.850pt}{0.400pt}}
\multiput(926.00,633.60)(1.797,0.468){5}{\rule{1.400pt}{0.113pt}}
\multiput(926.00,632.17)(10.094,4.000){2}{\rule{0.700pt}{0.400pt}}
\multiput(939.00,637.61)(2.472,0.447){3}{\rule{1.700pt}{0.108pt}}
\multiput(939.00,636.17)(8.472,3.000){2}{\rule{0.850pt}{0.400pt}}
\multiput(951.00,640.61)(2.695,0.447){3}{\rule{1.833pt}{0.108pt}}
\multiput(951.00,639.17)(9.195,3.000){2}{\rule{0.917pt}{0.400pt}}
\multiput(964.00,643.60)(1.651,0.468){5}{\rule{1.300pt}{0.113pt}}
\multiput(964.00,642.17)(9.302,4.000){2}{\rule{0.650pt}{0.400pt}}
\multiput(976.00,647.61)(2.695,0.447){3}{\rule{1.833pt}{0.108pt}}
\multiput(976.00,646.17)(9.195,3.000){2}{\rule{0.917pt}{0.400pt}}
\multiput(989.00,650.61)(2.472,0.447){3}{\rule{1.700pt}{0.108pt}}
\multiput(989.00,649.17)(8.472,3.000){2}{\rule{0.850pt}{0.400pt}}
\multiput(1001.00,653.61)(2.695,0.447){3}{\rule{1.833pt}{0.108pt}}
\multiput(1001.00,652.17)(9.195,3.000){2}{\rule{0.917pt}{0.400pt}}
\multiput(1014.00,656.61)(2.472,0.447){3}{\rule{1.700pt}{0.108pt}}
\multiput(1014.00,655.17)(8.472,3.000){2}{\rule{0.850pt}{0.400pt}}
\put(1026,659.17){\rule{2.700pt}{0.400pt}}
\multiput(1026.00,658.17)(7.396,2.000){2}{\rule{1.350pt}{0.400pt}}
\multiput(1039.00,661.61)(2.472,0.447){3}{\rule{1.700pt}{0.108pt}}
\multiput(1039.00,660.17)(8.472,3.000){2}{\rule{0.850pt}{0.400pt}}
\multiput(1051.00,664.61)(2.695,0.447){3}{\rule{1.833pt}{0.108pt}}
\multiput(1051.00,663.17)(9.195,3.000){2}{\rule{0.917pt}{0.400pt}}
\multiput(1064.00,667.61)(2.472,0.447){3}{\rule{1.700pt}{0.108pt}}
\multiput(1064.00,666.17)(8.472,3.000){2}{\rule{0.850pt}{0.400pt}}
\put(1076,670.17){\rule{2.700pt}{0.400pt}}
\multiput(1076.00,669.17)(7.396,2.000){2}{\rule{1.350pt}{0.400pt}}
\multiput(1089.00,672.61)(2.472,0.447){3}{\rule{1.700pt}{0.108pt}}
\multiput(1089.00,671.17)(8.472,3.000){2}{\rule{0.850pt}{0.400pt}}
\multiput(1101.00,675.61)(2.695,0.447){3}{\rule{1.833pt}{0.108pt}}
\multiput(1101.00,674.17)(9.195,3.000){2}{\rule{0.917pt}{0.400pt}}
\put(1114,678.17){\rule{2.500pt}{0.400pt}}
\multiput(1114.00,677.17)(6.811,2.000){2}{\rule{1.250pt}{0.400pt}}
\multiput(1126.00,680.61)(2.695,0.447){3}{\rule{1.833pt}{0.108pt}}
\multiput(1126.00,679.17)(9.195,3.000){2}{\rule{0.917pt}{0.400pt}}
\put(1139,683.17){\rule{2.500pt}{0.400pt}}
\multiput(1139.00,682.17)(6.811,2.000){2}{\rule{1.250pt}{0.400pt}}
\multiput(1151.00,685.61)(2.695,0.447){3}{\rule{1.833pt}{0.108pt}}
\multiput(1151.00,684.17)(9.195,3.000){2}{\rule{0.917pt}{0.400pt}}
\put(1164,688.17){\rule{2.500pt}{0.400pt}}
\multiput(1164.00,687.17)(6.811,2.000){2}{\rule{1.250pt}{0.400pt}}
\multiput(1176.00,690.61)(2.695,0.447){3}{\rule{1.833pt}{0.108pt}}
\multiput(1176.00,689.17)(9.195,3.000){2}{\rule{0.917pt}{0.400pt}}
\put(1189,693.17){\rule{2.500pt}{0.400pt}}
\multiput(1189.00,692.17)(6.811,2.000){2}{\rule{1.250pt}{0.400pt}}
\put(1201,695.17){\rule{2.700pt}{0.400pt}}
\multiput(1201.00,694.17)(7.396,2.000){2}{\rule{1.350pt}{0.400pt}}
\multiput(1214.00,697.61)(2.472,0.447){3}{\rule{1.700pt}{0.108pt}}
\multiput(1214.00,696.17)(8.472,3.000){2}{\rule{0.850pt}{0.400pt}}
\put(1226,700.17){\rule{2.700pt}{0.400pt}}
\multiput(1226.00,699.17)(7.396,2.000){2}{\rule{1.350pt}{0.400pt}}
\put(1239,702.17){\rule{2.500pt}{0.400pt}}
\multiput(1239.00,701.17)(6.811,2.000){2}{\rule{1.250pt}{0.400pt}}
\multiput(1251.00,704.61)(2.695,0.447){3}{\rule{1.833pt}{0.108pt}}
\multiput(1251.00,703.17)(9.195,3.000){2}{\rule{0.917pt}{0.400pt}}
\put(1264,707.17){\rule{2.500pt}{0.400pt}}
\multiput(1264.00,706.17)(6.811,2.000){2}{\rule{1.250pt}{0.400pt}}
\put(1276,709.17){\rule{2.700pt}{0.400pt}}
\multiput(1276.00,708.17)(7.396,2.000){2}{\rule{1.350pt}{0.400pt}}
\multiput(1289.00,711.61)(2.472,0.447){3}{\rule{1.700pt}{0.108pt}}
\multiput(1289.00,710.17)(8.472,3.000){2}{\rule{0.850pt}{0.400pt}}
\put(1301,714.17){\rule{2.700pt}{0.400pt}}
\multiput(1301.00,713.17)(7.396,2.000){2}{\rule{1.350pt}{0.400pt}}
\put(1314,716.17){\rule{2.500pt}{0.400pt}}
\multiput(1314.00,715.17)(6.811,2.000){2}{\rule{1.250pt}{0.400pt}}
\multiput(1326.00,718.61)(2.695,0.447){3}{\rule{1.833pt}{0.108pt}}
\multiput(1326.00,717.17)(9.195,3.000){2}{\rule{0.917pt}{0.400pt}}
\put(1339,721.17){\rule{2.500pt}{0.400pt}}
\multiput(1339.00,720.17)(6.811,2.000){2}{\rule{1.250pt}{0.400pt}}
\put(1351,723.17){\rule{2.700pt}{0.400pt}}
\multiput(1351.00,722.17)(7.396,2.000){2}{\rule{1.350pt}{0.400pt}}
\multiput(1364.00,725.61)(2.472,0.447){3}{\rule{1.700pt}{0.108pt}}
\multiput(1364.00,724.17)(8.472,3.000){2}{\rule{0.850pt}{0.400pt}}
\put(1376,728.17){\rule{2.700pt}{0.400pt}}
\multiput(1376.00,727.17)(7.396,2.000){2}{\rule{1.350pt}{0.400pt}}
\multiput(1389.00,730.61)(2.472,0.447){3}{\rule{1.700pt}{0.108pt}}
\multiput(1389.00,729.17)(8.472,3.000){2}{\rule{0.850pt}{0.400pt}}
\put(1401,733.17){\rule{2.700pt}{0.400pt}}
\multiput(1401.00,732.17)(7.396,2.000){2}{\rule{1.350pt}{0.400pt}}
\multiput(1414.00,735.61)(2.472,0.447){3}{\rule{1.700pt}{0.108pt}}
\multiput(1414.00,734.17)(8.472,3.000){2}{\rule{0.850pt}{0.400pt}}
\put(1426,738.17){\rule{2.700pt}{0.400pt}}
\multiput(1426.00,737.17)(7.396,2.000){2}{\rule{1.350pt}{0.400pt}}
\put(201,222){\raisebox{-.8pt}{\makebox(0,0){$\diamond$}}}
\put(511,461){\raisebox{-.8pt}{\makebox(0,0){$\diamond$}}}
\put(820,593){\raisebox{-.8pt}{\makebox(0,0){$\diamond$}}}
\put(1130,679){\raisebox{-.8pt}{\makebox(0,0){$\diamond$}}}
\put(1439,740){\raisebox{-.8pt}{\makebox(0,0){$\diamond$}}}
\sbox{\plotpoint}{\rule[-0.400pt]{0.800pt}{0.800pt}}%
\put(201,237){\makebox(0,0){$+$}}
\put(511,450){\makebox(0,0){$+$}}
\put(820,613){\makebox(0,0){$+$}}
\put(1130,663){\makebox(0,0){$+$}}
\put(1439,728){\makebox(0,0){$+$}}
\end{picture}

\setlength{\unitlength}{0.240900pt}
\ifx\plotpoint\undefined\newsavebox{\plotpoint}\fi
\sbox{\plotpoint}{\rule[-0.200pt]{0.400pt}{0.400pt}}%
\begin{picture}(1500,900)(0,0)
\font\gnuplot=cmr10 at 10pt
\gnuplot
\sbox{\plotpoint}{\rule[-0.200pt]{0.400pt}{0.400pt}}%
\put(201.0,123.0){\rule[-0.200pt]{4.818pt}{0.400pt}}
\put(181,123){\makebox(0,0)[r]{-0.05}}
\put(1419.0,123.0){\rule[-0.200pt]{4.818pt}{0.400pt}}
\put(201.0,232.0){\rule[-0.200pt]{4.818pt}{0.400pt}}
\put(181,232){\makebox(0,0)[r]{ 0}}
\put(1419.0,232.0){\rule[-0.200pt]{4.818pt}{0.400pt}}
\put(201.0,341.0){\rule[-0.200pt]{4.818pt}{0.400pt}}
\put(181,341){\makebox(0,0)[r]{ 0.05}}
\put(1419.0,341.0){\rule[-0.200pt]{4.818pt}{0.400pt}}
\put(201.0,450.0){\rule[-0.200pt]{4.818pt}{0.400pt}}
\put(181,450){\makebox(0,0)[r]{ 0.1}}
\put(1419.0,450.0){\rule[-0.200pt]{4.818pt}{0.400pt}}
\put(201.0,559.0){\rule[-0.200pt]{4.818pt}{0.400pt}}
\put(181,559){\makebox(0,0)[r]{ 0.15}}
\put(1419.0,559.0){\rule[-0.200pt]{4.818pt}{0.400pt}}
\put(201.0,668.0){\rule[-0.200pt]{4.818pt}{0.400pt}}
\put(181,668){\makebox(0,0)[r]{ 0.2}}
\put(1419.0,668.0){\rule[-0.200pt]{4.818pt}{0.400pt}}
\put(201.0,777.0){\rule[-0.200pt]{4.818pt}{0.400pt}}
\put(181,777){\makebox(0,0)[r]{ 0.25}}
\put(1419.0,777.0){\rule[-0.200pt]{4.818pt}{0.400pt}}
\put(201.0,123.0){\rule[-0.200pt]{0.400pt}{4.818pt}}
\put(201,82){\makebox(0,0){ 0.5}}
\put(201.0,757.0){\rule[-0.200pt]{0.400pt}{4.818pt}}
\put(325.0,123.0){\rule[-0.200pt]{0.400pt}{4.818pt}}
\put(325,82){\makebox(0,0){ 0.55}}
\put(325.0,757.0){\rule[-0.200pt]{0.400pt}{4.818pt}}
\put(449.0,123.0){\rule[-0.200pt]{0.400pt}{4.818pt}}
\put(449,82){\makebox(0,0){ 0.6}}
\put(449.0,757.0){\rule[-0.200pt]{0.400pt}{4.818pt}}
\put(572.0,123.0){\rule[-0.200pt]{0.400pt}{4.818pt}}
\put(572,82){\makebox(0,0){ 0.65}}
\put(572.0,757.0){\rule[-0.200pt]{0.400pt}{4.818pt}}
\put(696.0,123.0){\rule[-0.200pt]{0.400pt}{4.818pt}}
\put(696,82){\makebox(0,0){ 0.7}}
\put(696.0,757.0){\rule[-0.200pt]{0.400pt}{4.818pt}}
\put(820.0,123.0){\rule[-0.200pt]{0.400pt}{4.818pt}}
\put(820,82){\makebox(0,0){ 0.75}}
\put(820.0,757.0){\rule[-0.200pt]{0.400pt}{4.818pt}}
\put(944.0,123.0){\rule[-0.200pt]{0.400pt}{4.818pt}}
\put(944,82){\makebox(0,0){ 0.8}}
\put(944.0,757.0){\rule[-0.200pt]{0.400pt}{4.818pt}}
\put(1068.0,123.0){\rule[-0.200pt]{0.400pt}{4.818pt}}
\put(1068,82){\makebox(0,0){ 0.85}}
\put(1068.0,757.0){\rule[-0.200pt]{0.400pt}{4.818pt}}
\put(1191.0,123.0){\rule[-0.200pt]{0.400pt}{4.818pt}}
\put(1191,82){\makebox(0,0){ 0.9}}
\put(1191.0,757.0){\rule[-0.200pt]{0.400pt}{4.818pt}}
\put(1315.0,123.0){\rule[-0.200pt]{0.400pt}{4.818pt}}
\put(1315,82){\makebox(0,0){ 0.95}}
\put(1315.0,757.0){\rule[-0.200pt]{0.400pt}{4.818pt}}
\put(1439.0,123.0){\rule[-0.200pt]{0.400pt}{4.818pt}}
\put(1439,82){\makebox(0,0){ 1}}
\put(1439.0,757.0){\rule[-0.200pt]{0.400pt}{4.818pt}}
\put(201.0,123.0){\rule[-0.200pt]{298.234pt}{0.400pt}}
\put(1439.0,123.0){\rule[-0.200pt]{0.400pt}{157.549pt}}
\put(201.0,777.0){\rule[-0.200pt]{298.234pt}{0.400pt}}
\put(40,450){\makebox(0,0){$\alpha^*$}}
\put(820,21){\makebox(0,0){c}}
\put(820,839){\makebox(0,0){Fig 5b}}
\put(201.0,123.0){\rule[-0.200pt]{0.400pt}{157.549pt}}
\put(201,734){\usebox{\plotpoint}}
\multiput(201.00,732.94)(1.797,-0.468){5}{\rule{1.400pt}{0.113pt}}
\multiput(201.00,733.17)(10.094,-4.000){2}{\rule{0.700pt}{0.400pt}}
\multiput(214.00,728.93)(1.267,-0.477){7}{\rule{1.060pt}{0.115pt}}
\multiput(214.00,729.17)(9.800,-5.000){2}{\rule{0.530pt}{0.400pt}}
\multiput(226.00,723.93)(1.378,-0.477){7}{\rule{1.140pt}{0.115pt}}
\multiput(226.00,724.17)(10.634,-5.000){2}{\rule{0.570pt}{0.400pt}}
\multiput(239.00,718.94)(1.651,-0.468){5}{\rule{1.300pt}{0.113pt}}
\multiput(239.00,719.17)(9.302,-4.000){2}{\rule{0.650pt}{0.400pt}}
\multiput(251.00,714.93)(1.378,-0.477){7}{\rule{1.140pt}{0.115pt}}
\multiput(251.00,715.17)(10.634,-5.000){2}{\rule{0.570pt}{0.400pt}}
\multiput(264.00,709.93)(1.267,-0.477){7}{\rule{1.060pt}{0.115pt}}
\multiput(264.00,710.17)(9.800,-5.000){2}{\rule{0.530pt}{0.400pt}}
\multiput(276.00,704.94)(1.797,-0.468){5}{\rule{1.400pt}{0.113pt}}
\multiput(276.00,705.17)(10.094,-4.000){2}{\rule{0.700pt}{0.400pt}}
\multiput(289.00,700.93)(1.267,-0.477){7}{\rule{1.060pt}{0.115pt}}
\multiput(289.00,701.17)(9.800,-5.000){2}{\rule{0.530pt}{0.400pt}}
\multiput(301.00,695.93)(1.378,-0.477){7}{\rule{1.140pt}{0.115pt}}
\multiput(301.00,696.17)(10.634,-5.000){2}{\rule{0.570pt}{0.400pt}}
\multiput(314.00,690.94)(1.651,-0.468){5}{\rule{1.300pt}{0.113pt}}
\multiput(314.00,691.17)(9.302,-4.000){2}{\rule{0.650pt}{0.400pt}}
\multiput(326.00,686.93)(1.378,-0.477){7}{\rule{1.140pt}{0.115pt}}
\multiput(326.00,687.17)(10.634,-5.000){2}{\rule{0.570pt}{0.400pt}}
\multiput(339.00,681.93)(1.267,-0.477){7}{\rule{1.060pt}{0.115pt}}
\multiput(339.00,682.17)(9.800,-5.000){2}{\rule{0.530pt}{0.400pt}}
\multiput(351.00,676.93)(1.378,-0.477){7}{\rule{1.140pt}{0.115pt}}
\multiput(351.00,677.17)(10.634,-5.000){2}{\rule{0.570pt}{0.400pt}}
\multiput(364.00,671.94)(1.651,-0.468){5}{\rule{1.300pt}{0.113pt}}
\multiput(364.00,672.17)(9.302,-4.000){2}{\rule{0.650pt}{0.400pt}}
\multiput(376.00,667.93)(1.378,-0.477){7}{\rule{1.140pt}{0.115pt}}
\multiput(376.00,668.17)(10.634,-5.000){2}{\rule{0.570pt}{0.400pt}}
\multiput(389.00,662.93)(1.267,-0.477){7}{\rule{1.060pt}{0.115pt}}
\multiput(389.00,663.17)(9.800,-5.000){2}{\rule{0.530pt}{0.400pt}}
\multiput(401.00,657.93)(1.378,-0.477){7}{\rule{1.140pt}{0.115pt}}
\multiput(401.00,658.17)(10.634,-5.000){2}{\rule{0.570pt}{0.400pt}}
\multiput(414.00,652.94)(1.651,-0.468){5}{\rule{1.300pt}{0.113pt}}
\multiput(414.00,653.17)(9.302,-4.000){2}{\rule{0.650pt}{0.400pt}}
\multiput(426.00,648.93)(1.378,-0.477){7}{\rule{1.140pt}{0.115pt}}
\multiput(426.00,649.17)(10.634,-5.000){2}{\rule{0.570pt}{0.400pt}}
\multiput(439.00,643.93)(1.267,-0.477){7}{\rule{1.060pt}{0.115pt}}
\multiput(439.00,644.17)(9.800,-5.000){2}{\rule{0.530pt}{0.400pt}}
\multiput(451.00,638.93)(1.378,-0.477){7}{\rule{1.140pt}{0.115pt}}
\multiput(451.00,639.17)(10.634,-5.000){2}{\rule{0.570pt}{0.400pt}}
\multiput(464.00,633.93)(1.267,-0.477){7}{\rule{1.060pt}{0.115pt}}
\multiput(464.00,634.17)(9.800,-5.000){2}{\rule{0.530pt}{0.400pt}}
\multiput(476.00,628.94)(1.797,-0.468){5}{\rule{1.400pt}{0.113pt}}
\multiput(476.00,629.17)(10.094,-4.000){2}{\rule{0.700pt}{0.400pt}}
\multiput(489.00,624.93)(1.267,-0.477){7}{\rule{1.060pt}{0.115pt}}
\multiput(489.00,625.17)(9.800,-5.000){2}{\rule{0.530pt}{0.400pt}}
\multiput(501.00,619.93)(1.378,-0.477){7}{\rule{1.140pt}{0.115pt}}
\multiput(501.00,620.17)(10.634,-5.000){2}{\rule{0.570pt}{0.400pt}}
\multiput(514.00,614.93)(1.267,-0.477){7}{\rule{1.060pt}{0.115pt}}
\multiput(514.00,615.17)(9.800,-5.000){2}{\rule{0.530pt}{0.400pt}}
\multiput(526.00,609.93)(1.378,-0.477){7}{\rule{1.140pt}{0.115pt}}
\multiput(526.00,610.17)(10.634,-5.000){2}{\rule{0.570pt}{0.400pt}}
\multiput(539.00,604.93)(1.267,-0.477){7}{\rule{1.060pt}{0.115pt}}
\multiput(539.00,605.17)(9.800,-5.000){2}{\rule{0.530pt}{0.400pt}}
\multiput(551.00,599.93)(1.378,-0.477){7}{\rule{1.140pt}{0.115pt}}
\multiput(551.00,600.17)(10.634,-5.000){2}{\rule{0.570pt}{0.400pt}}
\multiput(564.00,594.94)(1.651,-0.468){5}{\rule{1.300pt}{0.113pt}}
\multiput(564.00,595.17)(9.302,-4.000){2}{\rule{0.650pt}{0.400pt}}
\multiput(576.00,590.93)(1.378,-0.477){7}{\rule{1.140pt}{0.115pt}}
\multiput(576.00,591.17)(10.634,-5.000){2}{\rule{0.570pt}{0.400pt}}
\multiput(589.00,585.93)(1.267,-0.477){7}{\rule{1.060pt}{0.115pt}}
\multiput(589.00,586.17)(9.800,-5.000){2}{\rule{0.530pt}{0.400pt}}
\multiput(601.00,580.93)(1.378,-0.477){7}{\rule{1.140pt}{0.115pt}}
\multiput(601.00,581.17)(10.634,-5.000){2}{\rule{0.570pt}{0.400pt}}
\multiput(614.00,575.93)(1.267,-0.477){7}{\rule{1.060pt}{0.115pt}}
\multiput(614.00,576.17)(9.800,-5.000){2}{\rule{0.530pt}{0.400pt}}
\multiput(626.00,570.93)(1.378,-0.477){7}{\rule{1.140pt}{0.115pt}}
\multiput(626.00,571.17)(10.634,-5.000){2}{\rule{0.570pt}{0.400pt}}
\multiput(639.00,565.93)(1.267,-0.477){7}{\rule{1.060pt}{0.115pt}}
\multiput(639.00,566.17)(9.800,-5.000){2}{\rule{0.530pt}{0.400pt}}
\multiput(651.00,560.93)(1.378,-0.477){7}{\rule{1.140pt}{0.115pt}}
\multiput(651.00,561.17)(10.634,-5.000){2}{\rule{0.570pt}{0.400pt}}
\multiput(664.00,555.93)(1.267,-0.477){7}{\rule{1.060pt}{0.115pt}}
\multiput(664.00,556.17)(9.800,-5.000){2}{\rule{0.530pt}{0.400pt}}
\multiput(676.00,550.93)(1.378,-0.477){7}{\rule{1.140pt}{0.115pt}}
\multiput(676.00,551.17)(10.634,-5.000){2}{\rule{0.570pt}{0.400pt}}
\multiput(689.00,545.93)(1.267,-0.477){7}{\rule{1.060pt}{0.115pt}}
\multiput(689.00,546.17)(9.800,-5.000){2}{\rule{0.530pt}{0.400pt}}
\multiput(701.00,540.93)(1.378,-0.477){7}{\rule{1.140pt}{0.115pt}}
\multiput(701.00,541.17)(10.634,-5.000){2}{\rule{0.570pt}{0.400pt}}
\multiput(714.00,535.93)(1.267,-0.477){7}{\rule{1.060pt}{0.115pt}}
\multiput(714.00,536.17)(9.800,-5.000){2}{\rule{0.530pt}{0.400pt}}
\multiput(726.00,530.93)(1.378,-0.477){7}{\rule{1.140pt}{0.115pt}}
\multiput(726.00,531.17)(10.634,-5.000){2}{\rule{0.570pt}{0.400pt}}
\multiput(739.00,525.93)(1.267,-0.477){7}{\rule{1.060pt}{0.115pt}}
\multiput(739.00,526.17)(9.800,-5.000){2}{\rule{0.530pt}{0.400pt}}
\multiput(751.00,520.93)(1.378,-0.477){7}{\rule{1.140pt}{0.115pt}}
\multiput(751.00,521.17)(10.634,-5.000){2}{\rule{0.570pt}{0.400pt}}
\multiput(764.00,515.93)(1.267,-0.477){7}{\rule{1.060pt}{0.115pt}}
\multiput(764.00,516.17)(9.800,-5.000){2}{\rule{0.530pt}{0.400pt}}
\multiput(776.00,510.93)(1.378,-0.477){7}{\rule{1.140pt}{0.115pt}}
\multiput(776.00,511.17)(10.634,-5.000){2}{\rule{0.570pt}{0.400pt}}
\multiput(789.00,505.93)(1.267,-0.477){7}{\rule{1.060pt}{0.115pt}}
\multiput(789.00,506.17)(9.800,-5.000){2}{\rule{0.530pt}{0.400pt}}
\multiput(801.00,500.93)(1.378,-0.477){7}{\rule{1.140pt}{0.115pt}}
\multiput(801.00,501.17)(10.634,-5.000){2}{\rule{0.570pt}{0.400pt}}
\multiput(814.00,495.93)(1.267,-0.477){7}{\rule{1.060pt}{0.115pt}}
\multiput(814.00,496.17)(9.800,-5.000){2}{\rule{0.530pt}{0.400pt}}
\multiput(826.00,490.93)(1.378,-0.477){7}{\rule{1.140pt}{0.115pt}}
\multiput(826.00,491.17)(10.634,-5.000){2}{\rule{0.570pt}{0.400pt}}
\multiput(839.00,485.93)(1.267,-0.477){7}{\rule{1.060pt}{0.115pt}}
\multiput(839.00,486.17)(9.800,-5.000){2}{\rule{0.530pt}{0.400pt}}
\multiput(851.00,480.93)(1.123,-0.482){9}{\rule{0.967pt}{0.116pt}}
\multiput(851.00,481.17)(10.994,-6.000){2}{\rule{0.483pt}{0.400pt}}
\multiput(864.00,474.93)(1.267,-0.477){7}{\rule{1.060pt}{0.115pt}}
\multiput(864.00,475.17)(9.800,-5.000){2}{\rule{0.530pt}{0.400pt}}
\multiput(876.00,469.93)(1.378,-0.477){7}{\rule{1.140pt}{0.115pt}}
\multiput(876.00,470.17)(10.634,-5.000){2}{\rule{0.570pt}{0.400pt}}
\multiput(889.00,464.93)(1.267,-0.477){7}{\rule{1.060pt}{0.115pt}}
\multiput(889.00,465.17)(9.800,-5.000){2}{\rule{0.530pt}{0.400pt}}
\multiput(901.00,459.93)(1.378,-0.477){7}{\rule{1.140pt}{0.115pt}}
\multiput(901.00,460.17)(10.634,-5.000){2}{\rule{0.570pt}{0.400pt}}
\multiput(914.00,454.93)(1.267,-0.477){7}{\rule{1.060pt}{0.115pt}}
\multiput(914.00,455.17)(9.800,-5.000){2}{\rule{0.530pt}{0.400pt}}
\multiput(926.00,449.93)(1.378,-0.477){7}{\rule{1.140pt}{0.115pt}}
\multiput(926.00,450.17)(10.634,-5.000){2}{\rule{0.570pt}{0.400pt}}
\multiput(939.00,444.93)(1.267,-0.477){7}{\rule{1.060pt}{0.115pt}}
\multiput(939.00,445.17)(9.800,-5.000){2}{\rule{0.530pt}{0.400pt}}
\multiput(951.00,439.93)(1.123,-0.482){9}{\rule{0.967pt}{0.116pt}}
\multiput(951.00,440.17)(10.994,-6.000){2}{\rule{0.483pt}{0.400pt}}
\multiput(964.00,433.93)(1.267,-0.477){7}{\rule{1.060pt}{0.115pt}}
\multiput(964.00,434.17)(9.800,-5.000){2}{\rule{0.530pt}{0.400pt}}
\multiput(976.00,428.93)(1.378,-0.477){7}{\rule{1.140pt}{0.115pt}}
\multiput(976.00,429.17)(10.634,-5.000){2}{\rule{0.570pt}{0.400pt}}
\multiput(989.00,423.93)(1.267,-0.477){7}{\rule{1.060pt}{0.115pt}}
\multiput(989.00,424.17)(9.800,-5.000){2}{\rule{0.530pt}{0.400pt}}
\multiput(1001.00,418.93)(1.378,-0.477){7}{\rule{1.140pt}{0.115pt}}
\multiput(1001.00,419.17)(10.634,-5.000){2}{\rule{0.570pt}{0.400pt}}
\multiput(1014.00,413.93)(1.033,-0.482){9}{\rule{0.900pt}{0.116pt}}
\multiput(1014.00,414.17)(10.132,-6.000){2}{\rule{0.450pt}{0.400pt}}
\multiput(1026.00,407.93)(1.378,-0.477){7}{\rule{1.140pt}{0.115pt}}
\multiput(1026.00,408.17)(10.634,-5.000){2}{\rule{0.570pt}{0.400pt}}
\multiput(1039.00,402.93)(1.267,-0.477){7}{\rule{1.060pt}{0.115pt}}
\multiput(1039.00,403.17)(9.800,-5.000){2}{\rule{0.530pt}{0.400pt}}
\multiput(1051.00,397.93)(1.378,-0.477){7}{\rule{1.140pt}{0.115pt}}
\multiput(1051.00,398.17)(10.634,-5.000){2}{\rule{0.570pt}{0.400pt}}
\multiput(1064.00,392.93)(1.033,-0.482){9}{\rule{0.900pt}{0.116pt}}
\multiput(1064.00,393.17)(10.132,-6.000){2}{\rule{0.450pt}{0.400pt}}
\multiput(1076.00,386.93)(1.378,-0.477){7}{\rule{1.140pt}{0.115pt}}
\multiput(1076.00,387.17)(10.634,-5.000){2}{\rule{0.570pt}{0.400pt}}
\multiput(1089.00,381.93)(1.267,-0.477){7}{\rule{1.060pt}{0.115pt}}
\multiput(1089.00,382.17)(9.800,-5.000){2}{\rule{0.530pt}{0.400pt}}
\multiput(1101.00,376.93)(1.378,-0.477){7}{\rule{1.140pt}{0.115pt}}
\multiput(1101.00,377.17)(10.634,-5.000){2}{\rule{0.570pt}{0.400pt}}
\multiput(1114.00,371.93)(1.033,-0.482){9}{\rule{0.900pt}{0.116pt}}
\multiput(1114.00,372.17)(10.132,-6.000){2}{\rule{0.450pt}{0.400pt}}
\multiput(1126.00,365.93)(1.378,-0.477){7}{\rule{1.140pt}{0.115pt}}
\multiput(1126.00,366.17)(10.634,-5.000){2}{\rule{0.570pt}{0.400pt}}
\multiput(1139.00,360.93)(1.267,-0.477){7}{\rule{1.060pt}{0.115pt}}
\multiput(1139.00,361.17)(9.800,-5.000){2}{\rule{0.530pt}{0.400pt}}
\multiput(1151.00,355.93)(1.123,-0.482){9}{\rule{0.967pt}{0.116pt}}
\multiput(1151.00,356.17)(10.994,-6.000){2}{\rule{0.483pt}{0.400pt}}
\multiput(1164.00,349.93)(1.267,-0.477){7}{\rule{1.060pt}{0.115pt}}
\multiput(1164.00,350.17)(9.800,-5.000){2}{\rule{0.530pt}{0.400pt}}
\multiput(1176.00,344.93)(1.378,-0.477){7}{\rule{1.140pt}{0.115pt}}
\multiput(1176.00,345.17)(10.634,-5.000){2}{\rule{0.570pt}{0.400pt}}
\multiput(1189.00,339.93)(1.033,-0.482){9}{\rule{0.900pt}{0.116pt}}
\multiput(1189.00,340.17)(10.132,-6.000){2}{\rule{0.450pt}{0.400pt}}
\multiput(1201.00,333.93)(1.378,-0.477){7}{\rule{1.140pt}{0.115pt}}
\multiput(1201.00,334.17)(10.634,-5.000){2}{\rule{0.570pt}{0.400pt}}
\multiput(1214.00,328.93)(1.267,-0.477){7}{\rule{1.060pt}{0.115pt}}
\multiput(1214.00,329.17)(9.800,-5.000){2}{\rule{0.530pt}{0.400pt}}
\multiput(1226.00,323.93)(1.123,-0.482){9}{\rule{0.967pt}{0.116pt}}
\multiput(1226.00,324.17)(10.994,-6.000){2}{\rule{0.483pt}{0.400pt}}
\multiput(1239.00,317.93)(1.267,-0.477){7}{\rule{1.060pt}{0.115pt}}
\multiput(1239.00,318.17)(9.800,-5.000){2}{\rule{0.530pt}{0.400pt}}
\multiput(1251.00,312.93)(1.123,-0.482){9}{\rule{0.967pt}{0.116pt}}
\multiput(1251.00,313.17)(10.994,-6.000){2}{\rule{0.483pt}{0.400pt}}
\multiput(1264.00,306.93)(1.267,-0.477){7}{\rule{1.060pt}{0.115pt}}
\multiput(1264.00,307.17)(9.800,-5.000){2}{\rule{0.530pt}{0.400pt}}
\multiput(1276.00,301.93)(1.378,-0.477){7}{\rule{1.140pt}{0.115pt}}
\multiput(1276.00,302.17)(10.634,-5.000){2}{\rule{0.570pt}{0.400pt}}
\multiput(1289.00,296.93)(1.033,-0.482){9}{\rule{0.900pt}{0.116pt}}
\multiput(1289.00,297.17)(10.132,-6.000){2}{\rule{0.450pt}{0.400pt}}
\multiput(1301.00,290.93)(1.378,-0.477){7}{\rule{1.140pt}{0.115pt}}
\multiput(1301.00,291.17)(10.634,-5.000){2}{\rule{0.570pt}{0.400pt}}
\multiput(1314.00,285.93)(1.033,-0.482){9}{\rule{0.900pt}{0.116pt}}
\multiput(1314.00,286.17)(10.132,-6.000){2}{\rule{0.450pt}{0.400pt}}
\multiput(1326.00,279.93)(1.378,-0.477){7}{\rule{1.140pt}{0.115pt}}
\multiput(1326.00,280.17)(10.634,-5.000){2}{\rule{0.570pt}{0.400pt}}
\multiput(1339.00,274.93)(1.033,-0.482){9}{\rule{0.900pt}{0.116pt}}
\multiput(1339.00,275.17)(10.132,-6.000){2}{\rule{0.450pt}{0.400pt}}
\multiput(1351.00,268.93)(1.378,-0.477){7}{\rule{1.140pt}{0.115pt}}
\multiput(1351.00,269.17)(10.634,-5.000){2}{\rule{0.570pt}{0.400pt}}
\multiput(1364.00,263.93)(1.033,-0.482){9}{\rule{0.900pt}{0.116pt}}
\multiput(1364.00,264.17)(10.132,-6.000){2}{\rule{0.450pt}{0.400pt}}
\multiput(1376.00,257.93)(1.378,-0.477){7}{\rule{1.140pt}{0.115pt}}
\multiput(1376.00,258.17)(10.634,-5.000){2}{\rule{0.570pt}{0.400pt}}
\multiput(1389.00,252.93)(1.033,-0.482){9}{\rule{0.900pt}{0.116pt}}
\multiput(1389.00,253.17)(10.132,-6.000){2}{\rule{0.450pt}{0.400pt}}
\multiput(1401.00,246.93)(1.378,-0.477){7}{\rule{1.140pt}{0.115pt}}
\multiput(1401.00,247.17)(10.634,-5.000){2}{\rule{0.570pt}{0.400pt}}
\multiput(1414.00,241.93)(1.033,-0.482){9}{\rule{0.900pt}{0.116pt}}
\multiput(1414.00,242.17)(10.132,-6.000){2}{\rule{0.450pt}{0.400pt}}
\multiput(1426.00,235.93)(1.378,-0.477){7}{\rule{1.140pt}{0.115pt}}
\multiput(1426.00,236.17)(10.634,-5.000){2}{\rule{0.570pt}{0.400pt}}
\put(201,734){\raisebox{-.8pt}{\makebox(0,0){$\diamond$}}}
\put(449,641){\raisebox{-.8pt}{\makebox(0,0){$\diamond$}}}
\put(696,544){\raisebox{-.8pt}{\makebox(0,0){$\diamond$}}}
\put(944,444){\raisebox{-.8pt}{\makebox(0,0){$\diamond$}}}
\put(1191,339){\raisebox{-.8pt}{\makebox(0,0){$\diamond$}}}
\put(1439,232){\raisebox{-.8pt}{\makebox(0,0){$\diamond$}}}
\sbox{\plotpoint}{\rule[-0.400pt]{0.800pt}{0.800pt}}%
\put(201,744){\makebox(0,0){$+$}}
\put(449,646){\makebox(0,0){$+$}}
\put(696,537){\makebox(0,0){$+$}}
\put(944,450){\makebox(0,0){$+$}}
\put(1191,341){\makebox(0,0){$+$}}
\put(449,646){\makebox(0,0){$+$}}
\put(696,537){\makebox(0,0){$+$}}
\put(944,450){\makebox(0,0){$+$}}
\put(1191,341){\makebox(0,0){$+$}}
\put(449,646){\makebox(0,0){$+$}}
\put(696,537){\makebox(0,0){$+$}}
\put(944,450){\makebox(0,0){$+$}}
\put(1191,341){\makebox(0,0){$+$}}
\put(449,646){\makebox(0,0){$+$}}
\put(696,537){\makebox(0,0){$+$}}
\put(944,450){\makebox(0,0){$+$}}
\put(1191,341){\makebox(0,0){$+$}}
\put(449,646){\makebox(0,0){$+$}}
\put(696,537){\makebox(0,0){$+$}}
\put(944,450){\makebox(0,0){$+$}}
\put(1191,341){\makebox(0,0){$+$}}
\end{picture}

\noindent Figure 5: The variation of $\alpha^*$ vs (a) $n$ ($c=0.5$) and (b) $c$ ($n=1$).$\diamond$ present the analytical result while + denotes the numerical data. The smooth curve represents the fit of the analytical results.

\section{Conclusion}
\hspace*{0.5cm}The effect of particle wise disorder on the phase diagram of the ASEP with open boundaries is studied analytically for a random sequential dynamics. The steady state has been established and its properties as the density profile the current and the bulk density were solved using the exact results of the pure case combined with a mean field-like approach. The shape of the phase diagram was found to be universal. It doesn't depend on the form of the distribution probability of particle jumping rate $\varphi (p)$. The critical line of the phase transition between the high-density phase and the low-density phase, which is of first order for any disorder distribution, was determined analytically. Because of the particle-hole symmetry breaking, we have shown that this critical line is located at $\beta=\kappa(\alpha)\neq \alpha$ and its explicit form was given within our mean field-like approach. The phase transition between the high(low)-density phase and the maximum-current phase,which is of second order, is located at a critical value $\beta_c$($\alpha_c=\beta_c$). We have shown, within our analytical approach that the value of $\alpha_c$ depends only on the particle jumping rate distribution probability. Based on the fact that the transition between the maximal current phase and the high density phase(low density phase) doesn't depend on the injection(extraction)rate $\alpha$($\beta$), we have deduced that the system exhibits a phase transition between high(low)-density phases HD$_1$(LD$_1$) and HD$_2$(LD$_2$) at the same critical value $\alpha =\alpha_c$. The effect of the disorder doesn't change the nature of those critical lines. They remain of second order and they all meet at the critical point $(\alpha_c,\beta_c=\alpha_c)$ that depends only on $\varphi (p)$. As it was shown in a previous work[21], we have highlighted the universal behavior of the density profile in the maximum-current phase namely that the density of the system reaches its bulk value $\rho _{bulk}=1/2$ as $\ell^{-1/2}$ independently on the distribution $\varphi(p)$ neither on $\alpha$ nor $\beta$.\\
Assuming that the analytical results obtained for the platoon phase transition occurring in the low-density phase for systems with periodic boundary conditions hold for systems with open boundaries[19], we have calculated the value of the injection rate $\alpha^{*}$ at which the platoon phase transition occurs and we have shown that for a particular choice of the disorder distribution $\varphi(p)$ (eq. 35) the transition is of first order for all values $c<\frac{n}{2n+1}$. Using the exact current-density relation (eq. 37) we have shown that the disorder induces a cusp at maximum flow in a certain region of parameter space as it was found numerically[19]. From our approach we were able to establish the analytical expression of the variation of the slope at maximal current.\\
To check the validity of our analytical results obtained within our mean field-like theory, numerical simulations have been performed on a lattice of linear size. They show that the obtained results concerning the phase diagram and different quantities that are of interest are in good agreement with the analytical expressions.\\
We think that some results obtained for a disordered model within a random sequential dynamics hold for the ordered sequential dynamics and the parallel dynamics. This point may be clarified in a future work.

\newpage

\section*{Acknowledgment}
The author would thank Joachim Krug for his interest in this work and for his critical reading the paper. The most part of this work, which is supported by PROTARSIII grant N$^{0}$ D12/22, has been achieved during my visit within the association scheme at the Abdus- Salam International Center for Theoretical Physics (AS-ICTP). I acknowledge the financial support of the PROTARSIII and the UNESCO and the hospitality of the AS-ICTP.

\section*{References}
\begin{enumerate}
\item[1.]Van Beijeren, H., Kutner, R., Spohn, H.: Excess Noise for Driven Diffusive Systems. Phys. Rev. Lett. {\bf{45}}, 2026 (1985).\\
Schmittman, B., Zia, R.K.P.: Statistical Mechanics of Driven Diffusive Systems. Domb and Lebowitz, seri 17, London: Academic (1995).\\
Spohn, H.: Large Scale Dynamics of Interacting Particles. New York:Springer Verlag (1991).
\item[2.]Burgers, J.M.: The non Linear Diffusion Equation. Riedel, Boston (1974)
\item[3.]Meakin, P., Ramanlal, P., Sander, L., Ball, R.C.: Ballistic Deposition on Surfaces. Phys. Rev. A {\bf{34}}, 5091 (1986).
\item[4.]Kardar, M., Zhang Yi-Chen: Scaling Directed Polymers in Random Media. Phys. Rev. Lett. {\bf{58}}, 2087 (1987).
\item[5.]Liggett, T.M.: Interacting Particle Systems. NY: Springer Verlag (1991).
\item[6.]Kardar, M., Parisi, G., Zhang Yi-Chen: Dynamic Scaling of Growing Interfaces. Phys. Rev. Lett. {\bf{56}}, 889 (1986).
\item[7.]Wolf, D.E., Tang, H.L.: Inhomogeneous Growth Process. Phys. Rev. Lett. {\bf{65}}, 1591 (1990).\\
Kandel, D., Mukamel, D.: Defects, Interface Profile and Phase transitions in Growth Models. Europhys. Lett. {\bf{65}}, 325 (1992).
\item[8.]Krug, J.: Boundary-Induced Phase Transitions in Driven Diffusive Systems. Phys. Rev. Lett. {\bf{67}}, 1889 (1991).
\item[9.]Nagel, K., Schreckenberg, M.: A Cellular Automaton Model for Freeway Traffic. Physique I{\bf{2}}, 2221 (1992).\\
Schreckenberg, M., Schadschneider, A., Nagel, K., Ito, N.: Discrete Stochastic Models for Traffic Flow. Phys. Rev. E{\bf{51}}, 2939 (1995).\\
Nagatani, T.: Bunching of Cars in Asymmetric Exclusion Models for Free way Traffic. Phys. Rev. E{\bf{51}}, 922 (1995).: Self-Organized Criticality in 1D Traffic Flow Model with Parallel and Ordered Sequential Dynamics. J. Phys. A: Math. Gen. {\bf{28}} L119 (1995).\\
Migowsky, S., Wanschura, T., Rujan, P.: Competition and Cooperation on a Toy Autobahn Model. Z. Phys. B{\bf{95}}, 407 (1994).
\item[10.]Derrida, B., Domany, E., Mukamel, D.: An Exact Solution of a One-Dimensional Asymmetric Exclusion Model with Open Boundaries. J. Stat. Phys. {\bf{69}}, 667 (1992).
\item[11.]Sch\"{u}tz, G., Domany, E.: Phase Transitions in an Exactly Solvable One-Dimensional Exclusion Process. J. Stat. Phys. {\bf{72}}, 277 (1993).
\item[12.]Derrida, B., Evans, M.R., Hakim, V., Pasquier, V.: Exact Solution of 1d Asymmetric Exclusion Model using a Matrix Formulation. J. Phys. A: Math.\& Gen. {\bf{26}}, 1493 (1993).
\item[13.]Kolomeisky, A.B., Sch\"{u}tz, G.M., Kolomeisky, E.B., Straley, J.P.: Phase Diagram of One-Dimensional Driven Lattice Gases with Open Boundaries. J. Phys. A: Math.\& Gen. {\bf{31}}, 6911 (1998).
\item[14.]Rajewsky, N., Santen, L., Schadschneider, A., Schreckenberg, M.: The Asymmetric Exclusion Process: Comparison of Update Procedures. J. Stat. Phys. {\bf{92}}, 151 (1998).
\item[15.]Tilstra, L.G., Ernst, M.H.: Synchronous Asymmetric Exclusion Process. J. Phys. A: Math.\& Gen. {\bf{31}}, 5033 (1998).
\item[16.]Barma, M.: Driven diffusive systems with disorder. Physica A{\bf 372}, 22 (2006).\\
Tripathy, G.,Barma, M.: Driven Lattice Gases with Quenched Disorder: Exact Results and Different Macroscopic Regimes. Phys. Rev. E{\bf{58}}, 1911 (1998).
\item[17.]Krug, J., Ferrari, P.A.: Phase Transitions in Driven Diffusive Systems with Random Rates. J. Phys. A: Math.\& Gen. {\bf{29}}, L465 (1996).\\
Evans, M.R.: Bose-Einstein Condensation in Disordered Exclusion Models and Relation to Traffic Flow. Europhys. Lett. {\bf{36}}, 13 (1996).
\item[18.]Evans, M.R.: Exact Steady States of Disordered Hopping Particle Models with Parallel and Ordered Sequential Dynamics. J. Phys. A: Math.\& Gen. {\bf{30}}, 5669 (1997).
\item[19.]Bengrine, M., Benyoussef, A., Ez-Zahraouy, H., Krug, J., Loulidi, M., Mhirech, F.: A Simulation Study of an Asymmetric Exclusion Model with Open Boundaries and Random Rates. J. Phys. A: Math.\& Gen. {\bf{32}}, 2527 (1999).
\item[20.]Hager, J.S., Krug, J., Popkov, V., Sch$\ddot{u}$tz, G.M.: Minimal current phase and universal boundary layers in driven diffusive systems. Phys. Rev. E{\bf{63}}, 056110 (2001).
\item[21.]Krug, J.: Spontaneous Formation of Space-time Structures and Criticality. ed, Riste, T., Sherrington, D., Dordrecht: Kluwer, p 37 (1991).
\item[22.]Tilstra, L.G., Ernst, M.H.: Synchronous asymmetric exclusion processes. J. Phys. A: Math. Gen. {\bf{31}}, 5033 (1998)
\end{enumerate}

\end{document}